\documentclass[journal, comsoc]{IEEEtran}
\usepackage{amsmath,amsfonts}
\usepackage{algorithm}
\usepackage{algpseudocode} 
\usepackage{array}
\usepackage{textcomp}
\usepackage{stfloats}
\usepackage{url}
\usepackage{verbatim}
\usepackage{graphicx}
\usepackage{xcolor}
\usepackage{hyperref}
\usepackage{cite}
\usepackage{tikz}
\usepackage{subcaption}
\usepackage{booktabs}
\usepackage{tabularx}
\usepackage{pgfplots}
\usepackage{wrapfig}
\usepackage{etoolbox}
\usepackage{pgfplotstable}
\usepackage{makecell}
\usepackage{siunitx}
\usepackage{multirow}
\usepackage{orcidlink}
\usepackage[absolute,showboxes]{textpos}
\newcommand{\orcid}[1]{\raisebox{0.5ex}{\orcidlink{#1}}}
\DeclareSIUnit{\pkt}{pkt}

\makeatletter
\patchcmd{\ALG@step}{\addtocounter{ALG@line}{1}}{\refstepcounter{ALG@line}}{}{}
\newcommand{\ALG@lineautorefname}{Line}
\makeatother

\usepgfplotslibrary{groupplots,colormaps,colorbrewer,statistics,fillbetween}
\usetikzlibrary{automata,arrows,arrows.meta,positioning,calc,patterns,decorations, decorations.markings,external,shapes.geometric, fit, backgrounds}
\usepackage{pifont}

\usepackage{xspace}

\newcommand{\etal}{\textit{et al.\ }}
\newcommand{\eg}{\textit{e.g.,}~}
\newcommand{\ie}{\textit{i.e.,}~}

\newcommand{\one}{({\em i})\xspace}
\newcommand{\two}{({\em ii})\xspace}
\newcommand{\three}{({\em iii})\xspace}
\newcommand{\four}{({\em iv})\xspace}

\newcommand{\our}{LZn\xspace}

\newcommand{\PPColsSim}[5]{%
	\pgfkeysifdefined{/ppcolssim/#1/#2/#3/#4/#5}%
	{\pgfkeysvalueof{/ppcolssim/#1/#2/#3/#4/#5}}%
	{\PackageError{results}{Missing key '/ppcolssim/#1/#2/#3/#4'}{}}%
}%

\newcommand{\FetchKey}[1]{%
	\pgfkeysifdefined{#1}%
	{\pgfkeysvalueof{#1}}%
	{\PackageError{results}{Missing key '#1'}}%
}%

\newcommand{\UDColThresh}[1]{%
    \def\SF{#1}%
    \pgfmathsetmacro{\M}{2^\SF}%
    \pgfmathsetmacro{\result}{10*log10(8/\M)}%
    \pgfmathprintnumber[fixed,precision=2]{\result}%
}%

\newcommand{\PercColsSim}[4]{%
	\pgfkeysifdefined{/perccolssim/#1/#2/#3/#4}%
	{\pgfkeysvalueof{/perccolssim/#1/#2/#3/#4}\%}%
	{\PackageError{results}{Missing key '/perccolssim/#1/#2/#3/#4'}{}}%
}%

\newcommand{\PoissonSFSevenTXRSixty}{0.75}
\newcommand{\PoissonSFTwelveTXRSixty}{24.08}

\let\orgautoref\autoref
\renewcommand{\autoref}
{\def\sectionautorefname{Section}%
\def\subsectionautorefname{Section}%
\def\subsubsectionautorefname{Section}%
\orgautoref}

\makeatletter
\renewcommand{\paragraph}[1]{\vspace*{0.03in}\noindent{\bf #1.}\hspace{0.25ex \@plus1ex \@minus.2ex}}
\newcommand{\paragraphS}[1]{\vspace*{0.03in}\noindent{\bf #1}\hspace{0.25ex \@plus1ex \@minus.2ex}}
\makeatother

\pgfkeys{
/perccolssim/detection/7/30/CoRa/.initial=91.4,
/perccolssim/crcok/7/30/CoRa/.initial=77.0,
/perccolssim/detection/7/30/CIC/.initial=43.8,
/perccolssim/crcok/7/30/CIC/.initial=40.9,
/perccolssim/detection/7/10/TnB/.initial=91.8,
/perccolssim/crcok/7/10/TnB/.initial=85.7,
/perccolssim/detection/7/10/CoRa/.initial=96.7,
/perccolssim/crcok/7/10/CoRa/.initial=90.4,
/perccolssim/detection/7/20/CoRa/.initial=94.3,
/perccolssim/crcok/7/20/CoRa/.initial=84.1,
/perccolssim/detection/7/45/OpenLoRa/.initial=22.9,
/perccolssim/crcok/7/45/OpenLoRa/.initial=17.1,
/perccolssim/detection/12/15/OpenLoRa/.initial=3.3,
/perccolssim/crcok/12/15/OpenLoRa/.initial=1.9,
/perccolssim/detection/7/5/TnB/.initial=94.9,
/perccolssim/crcok/7/5/TnB/.initial=91.9,
/perccolssim/detection/12/15/TnB/.initial=63.4,
/perccolssim/crcok/12/15/TnB/.initial=59.3,
/perccolssim/detection/12/55/CIC/.initial=2.6,
/perccolssim/crcok/12/55/CIC/.initial=2.5,
/perccolssim/detection/7/25/TnB/.initial=82.6,
/perccolssim/crcok/7/25/TnB/.initial=73.2,
/perccolssim/detection/12/20/CoRa/.initial=88.0,
/perccolssim/crcok/12/20/CoRa/.initial=74.8,
/perccolssim/detection/12/25/Pyramid/.initial=1.5,
/perccolssim/crcok/12/25/Pyramid/.initial=0.1,
/perccolssim/detection/12/45/CIC/.initial=3.0,
/perccolssim/crcok/12/45/CIC/.initial=2.9,
/perccolssim/detection/12/5/OpenLoRa/.initial=10.0,
/perccolssim/crcok/12/5/OpenLoRa/.initial=8.1,
/perccolssim/detection/12/35/Pyramid/.initial=1.2,
/perccolssim/crcok/12/35/Pyramid/.initial=0.1,
/perccolssim/detection/7/30/Pyramid/.initial=26.8,
/perccolssim/crcok/7/30/Pyramid/.initial=12.6,
/perccolssim/detection/12/30/CoRa/.initial=84.9,
/perccolssim/crcok/12/30/CoRa/.initial=66.7,
/perccolssim/detection/7/10/CIC/.initial=58.1,
/perccolssim/crcok/7/10/CIC/.initial=56.2,
/perccolssim/detection/12/20/OpenLoRa/.initial=2.9,
/perccolssim/crcok/12/20/OpenLoRa/.initial=1.5,
/perccolssim/detection/7/10/Pyramid/.initial=38.4,
/perccolssim/crcok/7/10/Pyramid/.initial=22.5,
/perccolssim/detection/12/25/OpenLoRa/.initial=1.9,
/perccolssim/crcok/12/25/OpenLoRa/.initial=1.2,
/perccolssim/detection/7/25/OpenLoRa/.initial=35.4,
/perccolssim/crcok/7/25/OpenLoRa/.initial=28.2,
/perccolssim/detection/7/55/OpenLoRa/.initial=19.6,
/perccolssim/crcok/7/55/OpenLoRa/.initial=14.0,
/perccolssim/detection/12/60/Pyramid/.initial=5.6,
/perccolssim/crcok/12/60/Pyramid/.initial=0.1,
/perccolssim/detection/12/20/CIC/.initial=9.7,
/perccolssim/crcok/12/20/CIC/.initial=9.5,
/perccolssim/detection/12/30/TnB/.initial=51.8,
/perccolssim/crcok/12/30/TnB/.initial=46.5,
/perccolssim/detection/12/40/OpenLoRa/.initial=1.1,
/perccolssim/crcok/12/40/OpenLoRa/.initial=0.6,
/perccolssim/detection/12/45/TnB/.initial=44.0,
/perccolssim/crcok/12/45/TnB/.initial=40.1,
/perccolssim/detection/12/50/TnB/.initial=42.7,
/perccolssim/crcok/12/50/TnB/.initial=37.6,
/perccolssim/detection/7/45/CIC/.initial=37.3,
/perccolssim/crcok/7/45/CIC/.initial=34.3,
/perccolssim/detection/7/50/Pyramid/.initial=20.5,
/perccolssim/crcok/7/50/Pyramid/.initial=7.7,
/perccolssim/detection/7/40/CIC/.initial=38.4,
/perccolssim/crcok/7/40/CIC/.initial=34.9,
/perccolssim/detection/7/50/OpenLoRa/.initial=21.7,
/perccolssim/crcok/7/50/OpenLoRa/.initial=15.7,
/perccolssim/detection/12/10/Pyramid/.initial=1.7,
/perccolssim/crcok/12/10/Pyramid/.initial=0.6,
/perccolssim/detection/12/35/CIC/.initial=6.2,
/perccolssim/crcok/12/35/CIC/.initial=6.0,
/perccolssim/detection/12/50/CoRa/.initial=81.1,
/perccolssim/crcok/12/50/CoRa/.initial=55.7,
/perccolssim/detection/7/45/CoRa/.initial=86.5,
/perccolssim/crcok/7/45/CoRa/.initial=68.2,
/perccolssim/detection/7/50/TnB/.initial=70.4,
/perccolssim/crcok/7/50/TnB/.initial=58.1,
/perccolssim/detection/12/15/Pyramid/.initial=2.2,
/perccolssim/crcok/12/15/Pyramid/.initial=0.7,
/perccolssim/detection/7/45/TnB/.initial=71.9,
/perccolssim/crcok/7/45/TnB/.initial=60.7,
/perccolssim/detection/12/20/TnB/.initial=58.7,
/perccolssim/crcok/12/20/TnB/.initial=54.5,
/perccolssim/detection/12/30/OpenLoRa/.initial=1.5,
/perccolssim/crcok/12/30/OpenLoRa/.initial=0.8,
/perccolssim/detection/12/5/TnB/.initial=85.1,
/perccolssim/crcok/12/5/TnB/.initial=80.3,
/perccolssim/detection/7/15/OpenLoRa/.initial=47.6,
/perccolssim/crcok/7/15/OpenLoRa/.initial=41.7,
/perccolssim/detection/12/50/CIC/.initial=3.9,
/perccolssim/crcok/12/50/CIC/.initial=3.8,
/perccolssim/detection/12/35/TnB/.initial=49.2,
/perccolssim/crcok/12/35/TnB/.initial=44.5,
/perccolssim/detection/12/50/Pyramid/.initial=3.7,
/perccolssim/crcok/12/50/Pyramid/.initial=0.0,
/perccolssim/detection/7/20/TnB/.initial=85.4,
/perccolssim/crcok/7/20/TnB/.initial=77.5,
/perccolssim/detection/7/40/Pyramid/.initial=23.2,
/perccolssim/crcok/7/40/Pyramid/.initial=10.2,
/perccolssim/detection/7/60/Pyramid/.initial=17.9,
/perccolssim/crcok/7/60/Pyramid/.initial=6.8,
/perccolssim/detection/12/30/CIC/.initial=5.9,
/perccolssim/crcok/12/30/CIC/.initial=5.8,
/perccolssim/detection/7/5/CoRa/.initial=97.9,
/perccolssim/crcok/7/5/CoRa/.initial=95.5,
/perccolssim/detection/12/40/CoRa/.initial=83.2,
/perccolssim/crcok/12/40/CoRa/.initial=60.8,
/perccolssim/detection/7/30/TnB/.initial=79.4,
/perccolssim/crcok/7/30/TnB/.initial=69.3,
/perccolssim/detection/12/55/OpenLoRa/.initial=0.8,
/perccolssim/crcok/12/55/OpenLoRa/.initial=0.4,
/perccolssim/detection/12/35/CoRa/.initial=83.6,
/perccolssim/crcok/12/35/CoRa/.initial=63.4,
/perccolssim/detection/12/60/CIC/.initial=3.0,
/perccolssim/crcok/12/60/CIC/.initial=2.8,
/perccolssim/detection/7/15/Pyramid/.initial=33.8,
/perccolssim/crcok/7/15/Pyramid/.initial=18.9,
/perccolssim/detection/7/60/CIC/.initial=32.0,
/perccolssim/crcok/7/60/CIC/.initial=28.6,
/perccolssim/detection/7/35/OpenLoRa/.initial=28.3,
/perccolssim/crcok/7/35/OpenLoRa/.initial=21.8,
/perccolssim/detection/12/10/OpenLoRa/.initial=4.8,
/perccolssim/crcok/12/10/OpenLoRa/.initial=3.3,
/perccolssim/detection/12/35/OpenLoRa/.initial=1.7,
/perccolssim/crcok/12/35/OpenLoRa/.initial=0.7,
/perccolssim/detection/7/40/OpenLoRa/.initial=25.7,
/perccolssim/crcok/7/40/OpenLoRa/.initial=19.0,
/perccolssim/detection/12/45/OpenLoRa/.initial=1.3,
/perccolssim/crcok/12/45/OpenLoRa/.initial=0.6,
/perccolssim/detection/7/15/TnB/.initial=88.5,
/perccolssim/crcok/7/15/TnB/.initial=81.5,
/perccolssim/detection/12/40/Pyramid/.initial=2.2,
/perccolssim/crcok/12/40/Pyramid/.initial=0.1,
/perccolssim/detection/12/50/OpenLoRa/.initial=0.8,
/perccolssim/crcok/12/50/OpenLoRa/.initial=0.4,
/perccolssim/detection/12/10/CoRa/.initial=93.5,
/perccolssim/crcok/12/10/CoRa/.initial=88.4,
/perccolssim/detection/12/60/OpenLoRa/.initial=1.1,
/perccolssim/crcok/12/60/OpenLoRa/.initial=0.4,
/perccolssim/detection/12/45/CoRa/.initial=82.5,
/perccolssim/crcok/12/45/CoRa/.initial=58.8,
/perccolssim/detection/7/5/CIC/.initial=61.7,
/perccolssim/crcok/7/5/CIC/.initial=60.9,
/perccolssim/detection/7/45/Pyramid/.initial=21.8,
/perccolssim/crcok/7/45/Pyramid/.initial=9.0,
/perccolssim/detection/7/55/TnB/.initial=69.2,
/perccolssim/crcok/7/55/TnB/.initial=56.4,
/perccolssim/detection/7/35/CIC/.initial=40.6,
/perccolssim/crcok/7/35/CIC/.initial=37.5,
/perccolssim/detection/7/60/TnB/.initial=65.4,
/perccolssim/crcok/7/60/TnB/.initial=53.2,
/perccolssim/detection/12/30/Pyramid/.initial=2.1,
/perccolssim/crcok/12/30/Pyramid/.initial=0.4,
/perccolssim/detection/7/10/OpenLoRa/.initial=53.9,
/perccolssim/crcok/7/10/OpenLoRa/.initial=47.6,
/perccolssim/detection/12/15/CIC/.initial=10.3,
/perccolssim/crcok/12/15/CIC/.initial=10.0,
/perccolssim/detection/7/15/CoRa/.initial=95.5,
/perccolssim/crcok/7/15/CoRa/.initial=87.3,
/perccolssim/detection/12/55/CoRa/.initial=79.7,
/perccolssim/crcok/12/55/CoRa/.initial=54.1,
/perccolssim/detection/7/40/TnB/.initial=72.9,
/perccolssim/crcok/7/40/TnB/.initial=61.4,
/perccolssim/detection/7/55/CoRa/.initial=84.9,
/perccolssim/crcok/7/55/CoRa/.initial=63.3,
/perccolssim/detection/7/50/CoRa/.initial=85.1,
/perccolssim/crcok/7/50/CoRa/.initial=64.6,
/perccolssim/detection/12/15/CoRa/.initial=90.3,
/perccolssim/crcok/12/15/CoRa/.initial=80.4,
/perccolssim/detection/7/5/OpenLoRa/.initial=61.3,
/perccolssim/crcok/7/5/OpenLoRa/.initial=58.0,
/perccolssim/detection/7/50/CIC/.initial=36.1,
/perccolssim/crcok/7/50/CIC/.initial=32.4,
/perccolssim/detection/7/40/CoRa/.initial=87.6,
/perccolssim/crcok/7/40/CoRa/.initial=68.3,
/perccolssim/detection/7/25/CIC/.initial=46.9,
/perccolssim/crcok/7/25/CIC/.initial=44.1,
/perccolssim/detection/7/15/CIC/.initial=54.5,
/perccolssim/crcok/7/15/CIC/.initial=52.5,
/perccolssim/detection/7/20/Pyramid/.initial=31.1,
/perccolssim/crcok/7/20/Pyramid/.initial=17.2,
/perccolssim/detection/7/60/OpenLoRa/.initial=17.3,
/perccolssim/crcok/7/60/OpenLoRa/.initial=12.2,
/perccolssim/detection/7/25/Pyramid/.initial=28.4,
/perccolssim/crcok/7/25/Pyramid/.initial=14.1,
/perccolssim/detection/12/10/TnB/.initial=74.1,
/perccolssim/crcok/12/10/TnB/.initial=68.9,
/perccolssim/detection/12/20/Pyramid/.initial=3.0,
/perccolssim/crcok/12/20/Pyramid/.initial=0.6,
/perccolssim/detection/7/25/CoRa/.initial=92.4,
/perccolssim/crcok/7/25/CoRa/.initial=79.8,
/perccolssim/detection/12/5/CoRa/.initial=96.9,
/perccolssim/crcok/12/5/CoRa/.initial=95.7,
/perccolssim/detection/7/20/CIC/.initial=49.0,
/perccolssim/crcok/7/20/CIC/.initial=46.5,
/perccolssim/detection/12/25/CoRa/.initial=86.4,
/perccolssim/crcok/12/25/CoRa/.initial=70.0,
/perccolssim/detection/7/20/OpenLoRa/.initial=40.2,
/perccolssim/crcok/7/20/OpenLoRa/.initial=33.5,
/perccolssim/detection/12/10/CIC/.initial=13.8,
/perccolssim/crcok/12/10/CIC/.initial=13.6,
/perccolssim/detection/7/35/Pyramid/.initial=23.7,
/perccolssim/crcok/7/35/Pyramid/.initial=10.4,
/perccolssim/detection/12/5/Pyramid/.initial=5.4,
/perccolssim/crcok/12/5/Pyramid/.initial=3.6,
/perccolssim/detection/12/5/CIC/.initial=27.7,
/perccolssim/crcok/12/5/CIC/.initial=27.5,
/perccolssim/detection/12/40/TnB/.initial=46.2,
/perccolssim/crcok/12/40/TnB/.initial=41.5,
/perccolssim/detection/7/35/TnB/.initial=76.8,
/perccolssim/crcok/7/35/TnB/.initial=65.5,
/perccolssim/detection/12/40/CIC/.initial=5.3,
/perccolssim/crcok/12/40/CIC/.initial=5.1,
/perccolssim/detection/12/25/CIC/.initial=9.3,
/perccolssim/crcok/12/25/CIC/.initial=9.1,
/perccolssim/detection/12/60/TnB/.initial=38.7,
/perccolssim/crcok/12/60/TnB/.initial=33.5,
/perccolssim/detection/7/55/CIC/.initial=34.4,
/perccolssim/crcok/7/55/CIC/.initial=30.8,
/perccolssim/detection/12/60/CoRa/.initial=78.1,
/perccolssim/crcok/12/60/CoRa/.initial=50.7,
/perccolssim/detection/12/45/Pyramid/.initial=2.1,
/perccolssim/crcok/12/45/Pyramid/.initial=0.0,
/perccolssim/detection/7/30/OpenLoRa/.initial=32.5,
/perccolssim/crcok/7/30/OpenLoRa/.initial=24.9,
/perccolssim/detection/7/35/CoRa/.initial=89.3,
/perccolssim/crcok/7/35/CoRa/.initial=72.9,
/perccolssim/detection/7/55/Pyramid/.initial=19.9,
/perccolssim/crcok/7/55/Pyramid/.initial=8.3,
/perccolssim/detection/12/55/Pyramid/.initial=1.3,
/perccolssim/crcok/12/55/Pyramid/.initial=0.0,
/perccolssim/detection/7/5/Pyramid/.initial=41.4,
/perccolssim/crcok/7/5/Pyramid/.initial=26.2,
/perccolssim/detection/7/60/CoRa/.initial=82.2,
/perccolssim/crcok/7/60/CoRa/.initial=59.7,
/perccolssim/detection/12/25/TnB/.initial=55.7,
/perccolssim/crcok/12/25/TnB/.initial=50.5,
/perccolssim/detection/12/55/TnB/.initial=42.1,
/perccolssim/crcok/12/55/TnB/.initial=37.0,
/ppcolssim/detection/7/30/CoRa/OpenLoRa/.initial=59.0,
/ppcolssim/crcok/7/30/CoRa/OpenLoRa/.initial=52.1,
/ppcolssim/detection/7/30/CoRa/CIC/.initial=47.6,
/ppcolssim/crcok/7/30/CoRa/CIC/.initial=36.1,
/ppcolssim/detection/7/30/CoRa/Pyramid/.initial=64.6,
/ppcolssim/crcok/7/30/CoRa/Pyramid/.initial=64.3,
/ppcolssim/detection/7/30/CoRa/TnB/.initial=12.1,
/ppcolssim/crcok/7/30/CoRa/TnB/.initial=7.7,
/ppcolssim/detection/7/30/OpenLoRa/CoRa/.initial=-59.0,
/ppcolssim/crcok/7/30/OpenLoRa/CoRa/.initial=-52.1,
/ppcolssim/detection/7/30/OpenLoRa/CIC/.initial=-11.3,
/ppcolssim/crcok/7/30/OpenLoRa/CIC/.initial=-16.0,
/ppcolssim/detection/7/30/OpenLoRa/Pyramid/.initial=5.7,
/ppcolssim/crcok/7/30/OpenLoRa/Pyramid/.initial=12.3,
/ppcolssim/detection/7/30/OpenLoRa/TnB/.initial=-46.9,
/ppcolssim/crcok/7/30/OpenLoRa/TnB/.initial=-44.4,
/ppcolssim/detection/7/30/CIC/CoRa/.initial=-47.6,
/ppcolssim/crcok/7/30/CIC/CoRa/.initial=-36.1,
/ppcolssim/detection/7/30/CIC/OpenLoRa/.initial=11.3,
/ppcolssim/crcok/7/30/CIC/OpenLoRa/.initial=16.0,
/ppcolssim/detection/7/30/CIC/Pyramid/.initial=17.0,
/ppcolssim/crcok/7/30/CIC/Pyramid/.initial=28.3,
/ppcolssim/detection/7/30/CIC/TnB/.initial=-35.6,
/ppcolssim/crcok/7/30/CIC/TnB/.initial=-28.4,
/ppcolssim/detection/7/30/Pyramid/CoRa/.initial=-64.6,
/ppcolssim/crcok/7/30/Pyramid/CoRa/.initial=-64.3,
/ppcolssim/detection/7/30/Pyramid/OpenLoRa/.initial=-5.7,
/ppcolssim/crcok/7/30/Pyramid/OpenLoRa/.initial=-12.3,
/ppcolssim/detection/7/30/Pyramid/CIC/.initial=-17.0,
/ppcolssim/crcok/7/30/Pyramid/CIC/.initial=-28.3,
/ppcolssim/detection/7/30/Pyramid/TnB/.initial=-52.6,
/ppcolssim/crcok/7/30/Pyramid/TnB/.initial=-56.6,
/ppcolssim/detection/7/30/TnB/CoRa/.initial=-12.1,
/ppcolssim/crcok/7/30/TnB/CoRa/.initial=-7.7,
/ppcolssim/detection/7/30/TnB/OpenLoRa/.initial=46.9,
/ppcolssim/crcok/7/30/TnB/OpenLoRa/.initial=44.4,
/ppcolssim/detection/7/30/TnB/CIC/.initial=35.6,
/ppcolssim/crcok/7/30/TnB/CIC/.initial=28.4,
/ppcolssim/detection/7/30/TnB/Pyramid/.initial=52.6,
/ppcolssim/crcok/7/30/TnB/Pyramid/.initial=56.6,
/ppcolssim/detection/7/10/CoRa/OpenLoRa/.initial=42.8,
/ppcolssim/crcok/7/10/CoRa/OpenLoRa/.initial=42.8,
/ppcolssim/detection/7/10/CoRa/CIC/.initial=38.6,
/ppcolssim/crcok/7/10/CoRa/CIC/.initial=34.3,
/ppcolssim/detection/7/10/CoRa/Pyramid/.initial=58.3,
/ppcolssim/crcok/7/10/CoRa/Pyramid/.initial=68.0,
/ppcolssim/detection/7/10/CoRa/TnB/.initial=4.9,
/ppcolssim/crcok/7/10/CoRa/TnB/.initial=4.7,
/ppcolssim/detection/7/10/OpenLoRa/CoRa/.initial=-42.8,
/ppcolssim/crcok/7/10/OpenLoRa/CoRa/.initial=-42.8,
/ppcolssim/detection/7/10/OpenLoRa/CIC/.initial=-4.2,
/ppcolssim/crcok/7/10/OpenLoRa/CIC/.initial=-8.5,
/ppcolssim/detection/7/10/OpenLoRa/Pyramid/.initial=15.5,
/ppcolssim/crcok/7/10/OpenLoRa/Pyramid/.initial=25.2,
/ppcolssim/detection/7/10/OpenLoRa/TnB/.initial=-37.9,
/ppcolssim/crcok/7/10/OpenLoRa/TnB/.initial=-38.0,
/ppcolssim/detection/7/10/CIC/CoRa/.initial=-38.6,
/ppcolssim/crcok/7/10/CIC/CoRa/.initial=-34.3,
/ppcolssim/detection/7/10/CIC/OpenLoRa/.initial=4.2,
/ppcolssim/crcok/7/10/CIC/OpenLoRa/.initial=8.5,
/ppcolssim/detection/7/10/CIC/Pyramid/.initial=19.7,
/ppcolssim/crcok/7/10/CIC/Pyramid/.initial=33.7,
/ppcolssim/detection/7/10/CIC/TnB/.initial=-33.7,
/ppcolssim/crcok/7/10/CIC/TnB/.initial=-29.5,
/ppcolssim/detection/7/10/Pyramid/CoRa/.initial=-58.3,
/ppcolssim/crcok/7/10/Pyramid/CoRa/.initial=-68.0,
/ppcolssim/detection/7/10/Pyramid/OpenLoRa/.initial=-15.5,
/ppcolssim/crcok/7/10/Pyramid/OpenLoRa/.initial=-25.2,
/ppcolssim/detection/7/10/Pyramid/CIC/.initial=-19.7,
/ppcolssim/crcok/7/10/Pyramid/CIC/.initial=-33.7,
/ppcolssim/detection/7/10/Pyramid/TnB/.initial=-53.4,
/ppcolssim/crcok/7/10/Pyramid/TnB/.initial=-63.2,
/ppcolssim/detection/7/10/TnB/CoRa/.initial=-4.9,
/ppcolssim/crcok/7/10/TnB/CoRa/.initial=-4.7,
/ppcolssim/detection/7/10/TnB/OpenLoRa/.initial=37.9,
/ppcolssim/crcok/7/10/TnB/OpenLoRa/.initial=38.0,
/ppcolssim/detection/7/10/TnB/CIC/.initial=33.7,
/ppcolssim/crcok/7/10/TnB/CIC/.initial=29.5,
/ppcolssim/detection/7/10/TnB/Pyramid/.initial=53.4,
/ppcolssim/crcok/7/10/TnB/Pyramid/.initial=63.2,
/ppcolssim/detection/7/20/CoRa/OpenLoRa/.initial=54.1,
/ppcolssim/crcok/7/20/CoRa/OpenLoRa/.initial=50.6,
/ppcolssim/detection/7/20/CoRa/CIC/.initial=45.3,
/ppcolssim/crcok/7/20/CoRa/CIC/.initial=37.6,
/ppcolssim/detection/7/20/CoRa/Pyramid/.initial=63.2,
/ppcolssim/crcok/7/20/CoRa/Pyramid/.initial=66.9,
/ppcolssim/detection/7/20/CoRa/TnB/.initial=8.9,
/ppcolssim/crcok/7/20/CoRa/TnB/.initial=6.6,
/ppcolssim/detection/7/20/OpenLoRa/CoRa/.initial=-54.1,
/ppcolssim/crcok/7/20/OpenLoRa/CoRa/.initial=-50.6,
/ppcolssim/detection/7/20/OpenLoRa/CIC/.initial=-8.8,
/ppcolssim/crcok/7/20/OpenLoRa/CIC/.initial=-13.0,
/ppcolssim/detection/7/20/OpenLoRa/Pyramid/.initial=9.1,
/ppcolssim/crcok/7/20/OpenLoRa/Pyramid/.initial=16.3,
/ppcolssim/detection/7/20/OpenLoRa/TnB/.initial=-45.2,
/ppcolssim/crcok/7/20/OpenLoRa/TnB/.initial=-44.0,
/ppcolssim/detection/7/20/CIC/CoRa/.initial=-45.3,
/ppcolssim/crcok/7/20/CIC/CoRa/.initial=-37.6,
/ppcolssim/detection/7/20/CIC/OpenLoRa/.initial=8.8,
/ppcolssim/crcok/7/20/CIC/OpenLoRa/.initial=13.0,
/ppcolssim/detection/7/20/CIC/Pyramid/.initial=17.9,
/ppcolssim/crcok/7/20/CIC/Pyramid/.initial=29.3,
/ppcolssim/detection/7/20/CIC/TnB/.initial=-36.4,
/ppcolssim/crcok/7/20/CIC/TnB/.initial=-31.0,
/ppcolssim/detection/7/20/Pyramid/CoRa/.initial=-63.2,
/ppcolssim/crcok/7/20/Pyramid/CoRa/.initial=-66.9,
/ppcolssim/detection/7/20/Pyramid/OpenLoRa/.initial=-9.1,
/ppcolssim/crcok/7/20/Pyramid/OpenLoRa/.initial=-16.3,
/ppcolssim/detection/7/20/Pyramid/CIC/.initial=-17.9,
/ppcolssim/crcok/7/20/Pyramid/CIC/.initial=-29.3,
/ppcolssim/detection/7/20/Pyramid/TnB/.initial=-54.3,
/ppcolssim/crcok/7/20/Pyramid/TnB/.initial=-60.2,
/ppcolssim/detection/7/20/TnB/CoRa/.initial=-8.9,
/ppcolssim/crcok/7/20/TnB/CoRa/.initial=-6.6,
/ppcolssim/detection/7/20/TnB/OpenLoRa/.initial=45.2,
/ppcolssim/crcok/7/20/TnB/OpenLoRa/.initial=44.0,
/ppcolssim/detection/7/20/TnB/CIC/.initial=36.4,
/ppcolssim/crcok/7/20/TnB/CIC/.initial=31.0,
/ppcolssim/detection/7/20/TnB/Pyramid/.initial=54.3,
/ppcolssim/crcok/7/20/TnB/Pyramid/.initial=60.2,
/ppcolssim/detection/7/45/CoRa/OpenLoRa/.initial=63.7,
/ppcolssim/crcok/7/45/CoRa/OpenLoRa/.initial=51.1,
/ppcolssim/detection/7/45/CoRa/CIC/.initial=49.2,
/ppcolssim/crcok/7/45/CoRa/CIC/.initial=33.9,
/ppcolssim/detection/7/45/CoRa/Pyramid/.initial=64.8,
/ppcolssim/crcok/7/45/CoRa/Pyramid/.initial=59.2,
/ppcolssim/detection/7/45/CoRa/TnB/.initial=14.6,
/ppcolssim/crcok/7/45/CoRa/TnB/.initial=7.6,
/ppcolssim/detection/7/45/OpenLoRa/CoRa/.initial=-63.7,
/ppcolssim/crcok/7/45/OpenLoRa/CoRa/.initial=-51.1,
/ppcolssim/detection/7/45/OpenLoRa/CIC/.initial=-14.4,
/ppcolssim/crcok/7/45/OpenLoRa/CIC/.initial=-17.2,
/ppcolssim/detection/7/45/OpenLoRa/Pyramid/.initial=1.1,
/ppcolssim/crcok/7/45/OpenLoRa/Pyramid/.initial=8.1,
/ppcolssim/detection/7/45/OpenLoRa/TnB/.initial=-49.0,
/ppcolssim/crcok/7/45/OpenLoRa/TnB/.initial=-43.5,
/ppcolssim/detection/7/45/CIC/CoRa/.initial=-49.2,
/ppcolssim/crcok/7/45/CIC/CoRa/.initial=-33.9,
/ppcolssim/detection/7/45/CIC/OpenLoRa/.initial=14.4,
/ppcolssim/crcok/7/45/CIC/OpenLoRa/.initial=17.2,
/ppcolssim/detection/7/45/CIC/Pyramid/.initial=15.6,
/ppcolssim/crcok/7/45/CIC/Pyramid/.initial=25.3,
/ppcolssim/detection/7/45/CIC/TnB/.initial=-34.6,
/ppcolssim/crcok/7/45/CIC/TnB/.initial=-26.4,
/ppcolssim/detection/7/45/Pyramid/CoRa/.initial=-64.8,
/ppcolssim/crcok/7/45/Pyramid/CoRa/.initial=-59.2,
/ppcolssim/detection/7/45/Pyramid/OpenLoRa/.initial=-1.1,
/ppcolssim/crcok/7/45/Pyramid/OpenLoRa/.initial=-8.1,
/ppcolssim/detection/7/45/Pyramid/CIC/.initial=-15.6,
/ppcolssim/crcok/7/45/Pyramid/CIC/.initial=-25.3,
/ppcolssim/detection/7/45/Pyramid/TnB/.initial=-50.1,
/ppcolssim/crcok/7/45/Pyramid/TnB/.initial=-51.7,
/ppcolssim/detection/7/45/TnB/CoRa/.initial=-14.6,
/ppcolssim/crcok/7/45/TnB/CoRa/.initial=-7.6,
/ppcolssim/detection/7/45/TnB/OpenLoRa/.initial=49.0,
/ppcolssim/crcok/7/45/TnB/OpenLoRa/.initial=43.5,
/ppcolssim/detection/7/45/TnB/CIC/.initial=34.6,
/ppcolssim/crcok/7/45/TnB/CIC/.initial=26.4,
/ppcolssim/detection/7/45/TnB/Pyramid/.initial=50.1,
/ppcolssim/crcok/7/45/TnB/Pyramid/.initial=51.7,
/ppcolssim/detection/12/15/CoRa/OpenLoRa/.initial=87.0,
/ppcolssim/crcok/12/15/CoRa/OpenLoRa/.initial=78.5,
/ppcolssim/detection/12/15/CoRa/CIC/.initial=80.0,
/ppcolssim/crcok/12/15/CoRa/CIC/.initial=70.4,
/ppcolssim/detection/12/15/CoRa/Pyramid/.initial=88.1,
/ppcolssim/crcok/12/15/CoRa/Pyramid/.initial=79.7,
/ppcolssim/detection/12/15/CoRa/TnB/.initial=26.9,
/ppcolssim/crcok/12/15/CoRa/TnB/.initial=21.1,
/ppcolssim/detection/12/15/OpenLoRa/CoRa/.initial=-87.0,
/ppcolssim/crcok/12/15/OpenLoRa/CoRa/.initial=-78.5,
/ppcolssim/detection/12/15/OpenLoRa/CIC/.initial=-7.0,
/ppcolssim/crcok/12/15/OpenLoRa/CIC/.initial=-8.1,
/ppcolssim/detection/12/15/OpenLoRa/Pyramid/.initial=1.1,
/ppcolssim/crcok/12/15/OpenLoRa/Pyramid/.initial=1.2,
/ppcolssim/detection/12/15/OpenLoRa/TnB/.initial=-60.1,
/ppcolssim/crcok/12/15/OpenLoRa/TnB/.initial=-57.4,
/ppcolssim/detection/12/15/CIC/CoRa/.initial=-80.0,
/ppcolssim/crcok/12/15/CIC/CoRa/.initial=-70.4,
/ppcolssim/detection/12/15/CIC/OpenLoRa/.initial=7.0,
/ppcolssim/crcok/12/15/CIC/OpenLoRa/.initial=8.1,
/ppcolssim/detection/12/15/CIC/Pyramid/.initial=8.1,
/ppcolssim/crcok/12/15/CIC/Pyramid/.initial=9.3,
/ppcolssim/detection/12/15/CIC/TnB/.initial=-53.1,
/ppcolssim/crcok/12/15/CIC/TnB/.initial=-49.3,
/ppcolssim/detection/12/15/Pyramid/CoRa/.initial=-88.1,
/ppcolssim/crcok/12/15/Pyramid/CoRa/.initial=-79.7,
/ppcolssim/detection/12/15/Pyramid/OpenLoRa/.initial=-1.1,
/ppcolssim/crcok/12/15/Pyramid/OpenLoRa/.initial=-1.2,
/ppcolssim/detection/12/15/Pyramid/CIC/.initial=-8.1,
/ppcolssim/crcok/12/15/Pyramid/CIC/.initial=-9.3,
/ppcolssim/detection/12/15/Pyramid/TnB/.initial=-61.2,
/ppcolssim/crcok/12/15/Pyramid/TnB/.initial=-58.6,
/ppcolssim/detection/12/15/TnB/CoRa/.initial=-26.9,
/ppcolssim/crcok/12/15/TnB/CoRa/.initial=-21.1,
/ppcolssim/detection/12/15/TnB/OpenLoRa/.initial=60.1,
/ppcolssim/crcok/12/15/TnB/OpenLoRa/.initial=57.4,
/ppcolssim/detection/12/15/TnB/CIC/.initial=53.1,
/ppcolssim/crcok/12/15/TnB/CIC/.initial=49.3,
/ppcolssim/detection/12/15/TnB/Pyramid/.initial=61.2,
/ppcolssim/crcok/12/15/TnB/Pyramid/.initial=58.6,
/ppcolssim/detection/7/5/CoRa/OpenLoRa/.initial=36.7,
/ppcolssim/crcok/7/5/CoRa/OpenLoRa/.initial=37.5,
/ppcolssim/detection/7/5/CoRa/CIC/.initial=36.3,
/ppcolssim/crcok/7/5/CoRa/CIC/.initial=34.6,
/ppcolssim/detection/7/5/CoRa/Pyramid/.initial=56.5,
/ppcolssim/crcok/7/5/CoRa/Pyramid/.initial=69.3,
/ppcolssim/detection/7/5/CoRa/TnB/.initial=3.0,
/ppcolssim/crcok/7/5/CoRa/TnB/.initial=3.6,
/ppcolssim/detection/7/5/OpenLoRa/CoRa/.initial=-36.7,
/ppcolssim/crcok/7/5/OpenLoRa/CoRa/.initial=-37.5,
/ppcolssim/detection/7/5/OpenLoRa/CIC/.initial=-0.4,
/ppcolssim/crcok/7/5/OpenLoRa/CIC/.initial=-2.9,
/ppcolssim/detection/7/5/OpenLoRa/Pyramid/.initial=19.9,
/ppcolssim/crcok/7/5/OpenLoRa/Pyramid/.initial=31.8,
/ppcolssim/detection/7/5/OpenLoRa/TnB/.initial=-33.7,
/ppcolssim/crcok/7/5/OpenLoRa/TnB/.initial=-34.0,
/ppcolssim/detection/7/5/CIC/CoRa/.initial=-36.3,
/ppcolssim/crcok/7/5/CIC/CoRa/.initial=-34.6,
/ppcolssim/detection/7/5/CIC/OpenLoRa/.initial=0.4,
/ppcolssim/crcok/7/5/CIC/OpenLoRa/.initial=2.9,
/ppcolssim/detection/7/5/CIC/Pyramid/.initial=20.3,
/ppcolssim/crcok/7/5/CIC/Pyramid/.initial=34.7,
/ppcolssim/detection/7/5/CIC/TnB/.initial=-33.3,
/ppcolssim/crcok/7/5/CIC/TnB/.initial=-31.1,
/ppcolssim/detection/7/5/Pyramid/CoRa/.initial=-56.5,
/ppcolssim/crcok/7/5/Pyramid/CoRa/.initial=-69.3,
/ppcolssim/detection/7/5/Pyramid/OpenLoRa/.initial=-19.9,
/ppcolssim/crcok/7/5/Pyramid/OpenLoRa/.initial=-31.8,
/ppcolssim/detection/7/5/Pyramid/CIC/.initial=-20.3,
/ppcolssim/crcok/7/5/Pyramid/CIC/.initial=-34.7,
/ppcolssim/detection/7/5/Pyramid/TnB/.initial=-53.5,
/ppcolssim/crcok/7/5/Pyramid/TnB/.initial=-65.8,
/ppcolssim/detection/7/5/TnB/CoRa/.initial=-3.0,
/ppcolssim/crcok/7/5/TnB/CoRa/.initial=-3.6,
/ppcolssim/detection/7/5/TnB/OpenLoRa/.initial=33.7,
/ppcolssim/crcok/7/5/TnB/OpenLoRa/.initial=34.0,
/ppcolssim/detection/7/5/TnB/CIC/.initial=33.3,
/ppcolssim/crcok/7/5/TnB/CIC/.initial=31.1,
/ppcolssim/detection/7/5/TnB/Pyramid/.initial=53.5,
/ppcolssim/crcok/7/5/TnB/Pyramid/.initial=65.8,
/ppcolssim/detection/12/55/CoRa/OpenLoRa/.initial=78.9,
/ppcolssim/crcok/12/55/CoRa/OpenLoRa/.initial=53.7,
/ppcolssim/detection/12/55/CoRa/CIC/.initial=77.2,
/ppcolssim/crcok/12/55/CoRa/CIC/.initial=51.6,
/ppcolssim/detection/12/55/CoRa/Pyramid/.initial=78.4,
/ppcolssim/crcok/12/55/CoRa/Pyramid/.initial=54.1,
/ppcolssim/detection/12/55/CoRa/TnB/.initial=37.7,
/ppcolssim/crcok/12/55/CoRa/TnB/.initial=17.0,
/ppcolssim/detection/12/55/OpenLoRa/CoRa/.initial=-78.9,
/ppcolssim/crcok/12/55/OpenLoRa/CoRa/.initial=-53.7,
/ppcolssim/detection/12/55/OpenLoRa/CIC/.initial=-1.7,
/ppcolssim/crcok/12/55/OpenLoRa/CIC/.initial=-2.1,
/ppcolssim/detection/12/55/OpenLoRa/Pyramid/.initial=-0.5,
/ppcolssim/crcok/12/55/OpenLoRa/Pyramid/.initial=0.4,
/ppcolssim/detection/12/55/OpenLoRa/TnB/.initial=-41.2,
/ppcolssim/crcok/12/55/OpenLoRa/TnB/.initial=-36.7,
/ppcolssim/detection/12/55/CIC/CoRa/.initial=-77.2,
/ppcolssim/crcok/12/55/CIC/CoRa/.initial=-51.6,
/ppcolssim/detection/12/55/CIC/OpenLoRa/.initial=1.7,
/ppcolssim/crcok/12/55/CIC/OpenLoRa/.initial=2.1,
/ppcolssim/detection/12/55/CIC/Pyramid/.initial=1.2,
/ppcolssim/crcok/12/55/CIC/Pyramid/.initial=2.5,
/ppcolssim/detection/12/55/CIC/TnB/.initial=-39.5,
/ppcolssim/crcok/12/55/CIC/TnB/.initial=-34.6,
/ppcolssim/detection/12/55/Pyramid/CoRa/.initial=-78.4,
/ppcolssim/crcok/12/55/Pyramid/CoRa/.initial=-54.1,
/ppcolssim/detection/12/55/Pyramid/OpenLoRa/.initial=0.5,
/ppcolssim/crcok/12/55/Pyramid/OpenLoRa/.initial=-0.4,
/ppcolssim/detection/12/55/Pyramid/CIC/.initial=-1.2,
/ppcolssim/crcok/12/55/Pyramid/CIC/.initial=-2.5,
/ppcolssim/detection/12/55/Pyramid/TnB/.initial=-40.7,
/ppcolssim/crcok/12/55/Pyramid/TnB/.initial=-37.0,
/ppcolssim/detection/12/55/TnB/CoRa/.initial=-37.7,
/ppcolssim/crcok/12/55/TnB/CoRa/.initial=-17.0,
/ppcolssim/detection/12/55/TnB/OpenLoRa/.initial=41.2,
/ppcolssim/crcok/12/55/TnB/OpenLoRa/.initial=36.7,
/ppcolssim/detection/12/55/TnB/CIC/.initial=39.5,
/ppcolssim/crcok/12/55/TnB/CIC/.initial=34.6,
/ppcolssim/detection/12/55/TnB/Pyramid/.initial=40.7,
/ppcolssim/crcok/12/55/TnB/Pyramid/.initial=37.0,
/ppcolssim/detection/7/25/CoRa/OpenLoRa/.initial=57.0,
/ppcolssim/crcok/7/25/CoRa/OpenLoRa/.initial=51.6,
/ppcolssim/detection/7/25/CoRa/CIC/.initial=45.5,
/ppcolssim/crcok/7/25/CoRa/CIC/.initial=35.7,
/ppcolssim/detection/7/25/CoRa/Pyramid/.initial=64.0,
/ppcolssim/crcok/7/25/CoRa/Pyramid/.initial=65.7,
/ppcolssim/detection/7/25/CoRa/TnB/.initial=9.7,
/ppcolssim/crcok/7/25/CoRa/TnB/.initial=6.6,
/ppcolssim/detection/7/25/OpenLoRa/CoRa/.initial=-57.0,
/ppcolssim/crcok/7/25/OpenLoRa/CoRa/.initial=-51.6,
/ppcolssim/detection/7/25/OpenLoRa/CIC/.initial=-11.5,
/ppcolssim/crcok/7/25/OpenLoRa/CIC/.initial=-16.0,
/ppcolssim/detection/7/25/OpenLoRa/Pyramid/.initial=7.0,
/ppcolssim/crcok/7/25/OpenLoRa/Pyramid/.initial=14.1,
/ppcolssim/detection/7/25/OpenLoRa/TnB/.initial=-47.2,
/ppcolssim/crcok/7/25/OpenLoRa/TnB/.initial=-45.0,
/ppcolssim/detection/7/25/CIC/CoRa/.initial=-45.5,
/ppcolssim/crcok/7/25/CIC/CoRa/.initial=-35.7,
/ppcolssim/detection/7/25/CIC/OpenLoRa/.initial=11.5,
/ppcolssim/crcok/7/25/CIC/OpenLoRa/.initial=16.0,
/ppcolssim/detection/7/25/CIC/Pyramid/.initial=18.5,
/ppcolssim/crcok/7/25/CIC/Pyramid/.initial=30.0,
/ppcolssim/detection/7/25/CIC/TnB/.initial=-35.8,
/ppcolssim/crcok/7/25/CIC/TnB/.initial=-29.1,
/ppcolssim/detection/7/25/Pyramid/CoRa/.initial=-64.0,
/ppcolssim/crcok/7/25/Pyramid/CoRa/.initial=-65.7,
/ppcolssim/detection/7/25/Pyramid/OpenLoRa/.initial=-7.0,
/ppcolssim/crcok/7/25/Pyramid/OpenLoRa/.initial=-14.1,
/ppcolssim/detection/7/25/Pyramid/CIC/.initial=-18.5,
/ppcolssim/crcok/7/25/Pyramid/CIC/.initial=-30.0,
/ppcolssim/detection/7/25/Pyramid/TnB/.initial=-54.3,
/ppcolssim/crcok/7/25/Pyramid/TnB/.initial=-59.1,
/ppcolssim/detection/7/25/TnB/CoRa/.initial=-9.7,
/ppcolssim/crcok/7/25/TnB/CoRa/.initial=-6.6,
/ppcolssim/detection/7/25/TnB/OpenLoRa/.initial=47.2,
/ppcolssim/crcok/7/25/TnB/OpenLoRa/.initial=45.0,
/ppcolssim/detection/7/25/TnB/CIC/.initial=35.8,
/ppcolssim/crcok/7/25/TnB/CIC/.initial=29.1,
/ppcolssim/detection/7/25/TnB/Pyramid/.initial=54.3,
/ppcolssim/crcok/7/25/TnB/Pyramid/.initial=59.1,
/ppcolssim/detection/12/20/CoRa/OpenLoRa/.initial=85.1,
/ppcolssim/crcok/12/20/CoRa/OpenLoRa/.initial=73.2,
/ppcolssim/detection/12/20/CoRa/CIC/.initial=78.3,
/ppcolssim/crcok/12/20/CoRa/CIC/.initial=65.2,
/ppcolssim/detection/12/20/CoRa/Pyramid/.initial=85.0,
/ppcolssim/crcok/12/20/CoRa/Pyramid/.initial=74.1,
/ppcolssim/detection/12/20/CoRa/TnB/.initial=29.3,
/ppcolssim/crcok/12/20/CoRa/TnB/.initial=20.3,
/ppcolssim/detection/12/20/OpenLoRa/CoRa/.initial=-85.1,
/ppcolssim/crcok/12/20/OpenLoRa/CoRa/.initial=-73.2,
/ppcolssim/detection/12/20/OpenLoRa/CIC/.initial=-6.8,
/ppcolssim/crcok/12/20/OpenLoRa/CIC/.initial=-8.0,
/ppcolssim/detection/12/20/OpenLoRa/Pyramid/.initial=-0.1,
/ppcolssim/crcok/12/20/OpenLoRa/Pyramid/.initial=0.9,
/ppcolssim/detection/12/20/OpenLoRa/TnB/.initial=-55.8,
/ppcolssim/crcok/12/20/OpenLoRa/TnB/.initial=-52.9,
/ppcolssim/detection/12/20/CIC/CoRa/.initial=-78.3,
/ppcolssim/crcok/12/20/CIC/CoRa/.initial=-65.2,
/ppcolssim/detection/12/20/CIC/OpenLoRa/.initial=6.8,
/ppcolssim/crcok/12/20/CIC/OpenLoRa/.initial=8.0,
/ppcolssim/detection/12/20/CIC/Pyramid/.initial=6.7,
/ppcolssim/crcok/12/20/CIC/Pyramid/.initial=8.9,
/ppcolssim/detection/12/20/CIC/TnB/.initial=-49.0,
/ppcolssim/crcok/12/20/CIC/TnB/.initial=-44.9,
/ppcolssim/detection/12/20/Pyramid/CoRa/.initial=-85.0,
/ppcolssim/crcok/12/20/Pyramid/CoRa/.initial=-74.1,
/ppcolssim/detection/12/20/Pyramid/OpenLoRa/.initial=0.1,
/ppcolssim/crcok/12/20/Pyramid/OpenLoRa/.initial=-0.9,
/ppcolssim/detection/12/20/Pyramid/CIC/.initial=-6.7,
/ppcolssim/crcok/12/20/Pyramid/CIC/.initial=-8.9,
/ppcolssim/detection/12/20/Pyramid/TnB/.initial=-55.7,
/ppcolssim/crcok/12/20/Pyramid/TnB/.initial=-53.8,
/ppcolssim/detection/12/20/TnB/CoRa/.initial=-29.3,
/ppcolssim/crcok/12/20/TnB/CoRa/.initial=-20.3,
/ppcolssim/detection/12/20/TnB/OpenLoRa/.initial=55.8,
/ppcolssim/crcok/12/20/TnB/OpenLoRa/.initial=52.9,
/ppcolssim/detection/12/20/TnB/CIC/.initial=49.0,
/ppcolssim/crcok/12/20/TnB/CIC/.initial=44.9,
/ppcolssim/detection/12/20/TnB/Pyramid/.initial=55.7,
/ppcolssim/crcok/12/20/TnB/Pyramid/.initial=53.8,
/ppcolssim/detection/12/25/CoRa/OpenLoRa/.initial=84.5,
/ppcolssim/crcok/12/25/CoRa/OpenLoRa/.initial=68.9,
/ppcolssim/detection/12/25/CoRa/CIC/.initial=77.1,
/ppcolssim/crcok/12/25/CoRa/CIC/.initial=60.9,
/ppcolssim/detection/12/25/CoRa/Pyramid/.initial=84.9,
/ppcolssim/crcok/12/25/CoRa/Pyramid/.initial=69.9,
/ppcolssim/detection/12/25/CoRa/TnB/.initial=30.7,
/ppcolssim/crcok/12/25/CoRa/TnB/.initial=19.5,
/ppcolssim/detection/12/25/OpenLoRa/CoRa/.initial=-84.5,
/ppcolssim/crcok/12/25/OpenLoRa/CoRa/.initial=-68.9,
/ppcolssim/detection/12/25/OpenLoRa/CIC/.initial=-7.4,
/ppcolssim/crcok/12/25/OpenLoRa/CIC/.initial=-7.9,
/ppcolssim/detection/12/25/OpenLoRa/Pyramid/.initial=0.5,
/ppcolssim/crcok/12/25/OpenLoRa/Pyramid/.initial=1.1,
/ppcolssim/detection/12/25/OpenLoRa/TnB/.initial=-53.8,
/ppcolssim/crcok/12/25/OpenLoRa/TnB/.initial=-49.4,
/ppcolssim/detection/12/25/CIC/CoRa/.initial=-77.1,
/ppcolssim/crcok/12/25/CIC/CoRa/.initial=-60.9,
/ppcolssim/detection/12/25/CIC/OpenLoRa/.initial=7.4,
/ppcolssim/crcok/12/25/CIC/OpenLoRa/.initial=7.9,
/ppcolssim/detection/12/25/CIC/Pyramid/.initial=7.9,
/ppcolssim/crcok/12/25/CIC/Pyramid/.initial=9.0,
/ppcolssim/detection/12/25/CIC/TnB/.initial=-46.4,
/ppcolssim/crcok/12/25/CIC/TnB/.initial=-41.4,
/ppcolssim/detection/12/25/Pyramid/CoRa/.initial=-84.9,
/ppcolssim/crcok/12/25/Pyramid/CoRa/.initial=-69.9,
/ppcolssim/detection/12/25/Pyramid/OpenLoRa/.initial=-0.5,
/ppcolssim/crcok/12/25/Pyramid/OpenLoRa/.initial=-1.1,
/ppcolssim/detection/12/25/Pyramid/CIC/.initial=-7.9,
/ppcolssim/crcok/12/25/Pyramid/CIC/.initial=-9.0,
/ppcolssim/detection/12/25/Pyramid/TnB/.initial=-54.2,
/ppcolssim/crcok/12/25/Pyramid/TnB/.initial=-50.4,
/ppcolssim/detection/12/25/TnB/CoRa/.initial=-30.7,
/ppcolssim/crcok/12/25/TnB/CoRa/.initial=-19.5,
/ppcolssim/detection/12/25/TnB/OpenLoRa/.initial=53.8,
/ppcolssim/crcok/12/25/TnB/OpenLoRa/.initial=49.4,
/ppcolssim/detection/12/25/TnB/CIC/.initial=46.4,
/ppcolssim/crcok/12/25/TnB/CIC/.initial=41.4,
/ppcolssim/detection/12/25/TnB/Pyramid/.initial=54.2,
/ppcolssim/crcok/12/25/TnB/Pyramid/.initial=50.4,
/ppcolssim/detection/12/45/CoRa/OpenLoRa/.initial=81.3,
/ppcolssim/crcok/12/45/CoRa/OpenLoRa/.initial=58.2,
/ppcolssim/detection/12/45/CoRa/CIC/.initial=79.5,
/ppcolssim/crcok/12/45/CoRa/CIC/.initial=55.9,
/ppcolssim/detection/12/45/CoRa/Pyramid/.initial=80.5,
/ppcolssim/crcok/12/45/CoRa/Pyramid/.initial=58.7,
/ppcolssim/detection/12/45/CoRa/TnB/.initial=38.5,
/ppcolssim/crcok/12/45/CoRa/TnB/.initial=18.6,
/ppcolssim/detection/12/45/OpenLoRa/CoRa/.initial=-81.3,
/ppcolssim/crcok/12/45/OpenLoRa/CoRa/.initial=-58.2,
/ppcolssim/detection/12/45/OpenLoRa/CIC/.initial=-1.8,
/ppcolssim/crcok/12/45/OpenLoRa/CIC/.initial=-2.3,
/ppcolssim/detection/12/45/OpenLoRa/Pyramid/.initial=-0.8,
/ppcolssim/crcok/12/45/OpenLoRa/Pyramid/.initial=0.6,
/ppcolssim/detection/12/45/OpenLoRa/TnB/.initial=-42.8,
/ppcolssim/crcok/12/45/OpenLoRa/TnB/.initial=-39.5,
/ppcolssim/detection/12/45/CIC/CoRa/.initial=-79.5,
/ppcolssim/crcok/12/45/CIC/CoRa/.initial=-55.9,
/ppcolssim/detection/12/45/CIC/OpenLoRa/.initial=1.8,
/ppcolssim/crcok/12/45/CIC/OpenLoRa/.initial=2.3,
/ppcolssim/detection/12/45/CIC/Pyramid/.initial=1.0,
/ppcolssim/crcok/12/45/CIC/Pyramid/.initial=2.8,
/ppcolssim/detection/12/45/CIC/TnB/.initial=-41.0,
/ppcolssim/crcok/12/45/CIC/TnB/.initial=-37.3,
/ppcolssim/detection/12/45/Pyramid/CoRa/.initial=-80.5,
/ppcolssim/crcok/12/45/Pyramid/CoRa/.initial=-58.7,
/ppcolssim/detection/12/45/Pyramid/OpenLoRa/.initial=0.8,
/ppcolssim/crcok/12/45/Pyramid/OpenLoRa/.initial=-0.6,
/ppcolssim/detection/12/45/Pyramid/CIC/.initial=-1.0,
/ppcolssim/crcok/12/45/Pyramid/CIC/.initial=-2.8,
/ppcolssim/detection/12/45/Pyramid/TnB/.initial=-42.0,
/ppcolssim/crcok/12/45/Pyramid/TnB/.initial=-40.1,
/ppcolssim/detection/12/45/TnB/CoRa/.initial=-38.5,
/ppcolssim/crcok/12/45/TnB/CoRa/.initial=-18.6,
/ppcolssim/detection/12/45/TnB/OpenLoRa/.initial=42.8,
/ppcolssim/crcok/12/45/TnB/OpenLoRa/.initial=39.5,
/ppcolssim/detection/12/45/TnB/CIC/.initial=41.0,
/ppcolssim/crcok/12/45/TnB/CIC/.initial=37.3,
/ppcolssim/detection/12/45/TnB/Pyramid/.initial=42.0,
/ppcolssim/crcok/12/45/TnB/Pyramid/.initial=40.1,
/ppcolssim/detection/12/5/CoRa/OpenLoRa/.initial=86.9,
/ppcolssim/crcok/12/5/CoRa/OpenLoRa/.initial=87.6,
/ppcolssim/detection/12/5/CoRa/CIC/.initial=69.2,
/ppcolssim/crcok/12/5/CoRa/CIC/.initial=68.1,
/ppcolssim/detection/12/5/CoRa/Pyramid/.initial=91.5,
/ppcolssim/crcok/12/5/CoRa/Pyramid/.initial=92.1,
/ppcolssim/detection/12/5/CoRa/TnB/.initial=11.8,
/ppcolssim/crcok/12/5/CoRa/TnB/.initial=15.4,
/ppcolssim/detection/12/5/OpenLoRa/CoRa/.initial=-86.9,
/ppcolssim/crcok/12/5/OpenLoRa/CoRa/.initial=-87.6,
/ppcolssim/detection/12/5/OpenLoRa/CIC/.initial=-17.8,
/ppcolssim/crcok/12/5/OpenLoRa/CIC/.initial=-19.4,
/ppcolssim/detection/12/5/OpenLoRa/Pyramid/.initial=4.6,
/ppcolssim/crcok/12/5/OpenLoRa/Pyramid/.initial=4.5,
/ppcolssim/detection/12/5/OpenLoRa/TnB/.initial=-75.1,
/ppcolssim/crcok/12/5/OpenLoRa/TnB/.initial=-72.2,
/ppcolssim/detection/12/5/CIC/CoRa/.initial=-69.2,
/ppcolssim/crcok/12/5/CIC/CoRa/.initial=-68.1,
/ppcolssim/detection/12/5/CIC/OpenLoRa/.initial=17.8,
/ppcolssim/crcok/12/5/CIC/OpenLoRa/.initial=19.4,
/ppcolssim/detection/12/5/CIC/Pyramid/.initial=22.4,
/ppcolssim/crcok/12/5/CIC/Pyramid/.initial=23.9,
/ppcolssim/detection/12/5/CIC/TnB/.initial=-57.4,
/ppcolssim/crcok/12/5/CIC/TnB/.initial=-52.8,
/ppcolssim/detection/12/5/Pyramid/CoRa/.initial=-91.5,
/ppcolssim/crcok/12/5/Pyramid/CoRa/.initial=-92.1,
/ppcolssim/detection/12/5/Pyramid/OpenLoRa/.initial=-4.6,
/ppcolssim/crcok/12/5/Pyramid/OpenLoRa/.initial=-4.5,
/ppcolssim/detection/12/5/Pyramid/CIC/.initial=-22.4,
/ppcolssim/crcok/12/5/Pyramid/CIC/.initial=-23.9,
/ppcolssim/detection/12/5/Pyramid/TnB/.initial=-79.7,
/ppcolssim/crcok/12/5/Pyramid/TnB/.initial=-76.7,
/ppcolssim/detection/12/5/TnB/CoRa/.initial=-11.8,
/ppcolssim/crcok/12/5/TnB/CoRa/.initial=-15.4,
/ppcolssim/detection/12/5/TnB/OpenLoRa/.initial=75.1,
/ppcolssim/crcok/12/5/TnB/OpenLoRa/.initial=72.2,
/ppcolssim/detection/12/5/TnB/CIC/.initial=57.4,
/ppcolssim/crcok/12/5/TnB/CIC/.initial=52.8,
/ppcolssim/detection/12/5/TnB/Pyramid/.initial=79.7,
/ppcolssim/crcok/12/5/TnB/Pyramid/.initial=76.7,
/ppcolssim/detection/12/35/CoRa/OpenLoRa/.initial=81.9,
/ppcolssim/crcok/12/35/CoRa/OpenLoRa/.initial=62.7,
/ppcolssim/detection/12/35/CoRa/CIC/.initial=77.4,
/ppcolssim/crcok/12/35/CoRa/CIC/.initial=57.4,
/ppcolssim/detection/12/35/CoRa/Pyramid/.initial=82.4,
/ppcolssim/crcok/12/35/CoRa/Pyramid/.initial=63.4,
/ppcolssim/detection/12/35/CoRa/TnB/.initial=34.4,
/ppcolssim/crcok/12/35/CoRa/TnB/.initial=18.9,
/ppcolssim/detection/12/35/OpenLoRa/CoRa/.initial=-81.9,
/ppcolssim/crcok/12/35/OpenLoRa/CoRa/.initial=-62.7,
/ppcolssim/detection/12/35/OpenLoRa/CIC/.initial=-4.6,
/ppcolssim/crcok/12/35/OpenLoRa/CIC/.initial=-5.3,
/ppcolssim/detection/12/35/OpenLoRa/Pyramid/.initial=0.5,
/ppcolssim/crcok/12/35/OpenLoRa/Pyramid/.initial=0.7,
/ppcolssim/detection/12/35/OpenLoRa/TnB/.initial=-47.5,
/ppcolssim/crcok/12/35/OpenLoRa/TnB/.initial=-43.8,
/ppcolssim/detection/12/35/CIC/CoRa/.initial=-77.4,
/ppcolssim/crcok/12/35/CIC/CoRa/.initial=-57.4,
/ppcolssim/detection/12/35/CIC/OpenLoRa/.initial=4.6,
/ppcolssim/crcok/12/35/CIC/OpenLoRa/.initial=5.3,
/ppcolssim/detection/12/35/CIC/Pyramid/.initial=5.0,
/ppcolssim/crcok/12/35/CIC/Pyramid/.initial=6.0,
/ppcolssim/detection/12/35/CIC/TnB/.initial=-43.0,
/ppcolssim/crcok/12/35/CIC/TnB/.initial=-38.5,
/ppcolssim/detection/12/35/Pyramid/CoRa/.initial=-82.4,
/ppcolssim/crcok/12/35/Pyramid/CoRa/.initial=-63.4,
/ppcolssim/detection/12/35/Pyramid/OpenLoRa/.initial=-0.5,
/ppcolssim/crcok/12/35/Pyramid/OpenLoRa/.initial=-0.7,
/ppcolssim/detection/12/35/Pyramid/CIC/.initial=-5.0,
/ppcolssim/crcok/12/35/Pyramid/CIC/.initial=-6.0,
/ppcolssim/detection/12/35/Pyramid/TnB/.initial=-48.0,
/ppcolssim/crcok/12/35/Pyramid/TnB/.initial=-44.5,
/ppcolssim/detection/12/35/TnB/CoRa/.initial=-34.4,
/ppcolssim/crcok/12/35/TnB/CoRa/.initial=-18.9,
/ppcolssim/detection/12/35/TnB/OpenLoRa/.initial=47.5,
/ppcolssim/crcok/12/35/TnB/OpenLoRa/.initial=43.8,
/ppcolssim/detection/12/35/TnB/CIC/.initial=43.0,
/ppcolssim/crcok/12/35/TnB/CIC/.initial=38.5,
/ppcolssim/detection/12/35/TnB/Pyramid/.initial=48.0,
/ppcolssim/crcok/12/35/TnB/Pyramid/.initial=44.5,
/ppcolssim/detection/12/30/CoRa/OpenLoRa/.initial=83.4,
/ppcolssim/crcok/12/30/CoRa/OpenLoRa/.initial=66.0,
/ppcolssim/detection/12/30/CoRa/CIC/.initial=78.9,
/ppcolssim/crcok/12/30/CoRa/CIC/.initial=60.9,
/ppcolssim/detection/12/30/CoRa/Pyramid/.initial=82.8,
/ppcolssim/crcok/12/30/CoRa/Pyramid/.initial=66.4,
/ppcolssim/detection/12/30/CoRa/TnB/.initial=33.1,
/ppcolssim/crcok/12/30/CoRa/TnB/.initial=20.2,
/ppcolssim/detection/12/30/OpenLoRa/CoRa/.initial=-83.4,
/ppcolssim/crcok/12/30/OpenLoRa/CoRa/.initial=-66.0,
/ppcolssim/detection/12/30/OpenLoRa/CIC/.initial=-4.5,
/ppcolssim/crcok/12/30/OpenLoRa/CIC/.initial=-5.0,
/ppcolssim/detection/12/30/OpenLoRa/Pyramid/.initial=-0.6,
/ppcolssim/crcok/12/30/OpenLoRa/Pyramid/.initial=0.4,
/ppcolssim/detection/12/30/OpenLoRa/TnB/.initial=-50.3,
/ppcolssim/crcok/12/30/OpenLoRa/TnB/.initial=-45.7,
/ppcolssim/detection/12/30/CIC/CoRa/.initial=-78.9,
/ppcolssim/crcok/12/30/CIC/CoRa/.initial=-60.9,
/ppcolssim/detection/12/30/CIC/OpenLoRa/.initial=4.5,
/ppcolssim/crcok/12/30/CIC/OpenLoRa/.initial=5.0,
/ppcolssim/detection/12/30/CIC/Pyramid/.initial=3.9,
/ppcolssim/crcok/12/30/CIC/Pyramid/.initial=5.4,
/ppcolssim/detection/12/30/CIC/TnB/.initial=-45.9,
/ppcolssim/crcok/12/30/CIC/TnB/.initial=-40.7,
/ppcolssim/detection/12/30/Pyramid/CoRa/.initial=-82.8,
/ppcolssim/crcok/12/30/Pyramid/CoRa/.initial=-66.4,
/ppcolssim/detection/12/30/Pyramid/OpenLoRa/.initial=0.6,
/ppcolssim/crcok/12/30/Pyramid/OpenLoRa/.initial=-0.4,
/ppcolssim/detection/12/30/Pyramid/CIC/.initial=-3.9,
/ppcolssim/crcok/12/30/Pyramid/CIC/.initial=-5.4,
/ppcolssim/detection/12/30/Pyramid/TnB/.initial=-49.7,
/ppcolssim/crcok/12/30/Pyramid/TnB/.initial=-46.1,
/ppcolssim/detection/12/30/TnB/CoRa/.initial=-33.1,
/ppcolssim/crcok/12/30/TnB/CoRa/.initial=-20.2,
/ppcolssim/detection/12/30/TnB/OpenLoRa/.initial=50.3,
/ppcolssim/crcok/12/30/TnB/OpenLoRa/.initial=45.7,
/ppcolssim/detection/12/30/TnB/CIC/.initial=45.9,
/ppcolssim/crcok/12/30/TnB/CIC/.initial=40.7,
/ppcolssim/detection/12/30/TnB/Pyramid/.initial=49.7,
/ppcolssim/crcok/12/30/TnB/Pyramid/.initial=46.1,
/ppcolssim/detection/7/55/CoRa/OpenLoRa/.initial=65.3,
/ppcolssim/crcok/7/55/CoRa/OpenLoRa/.initial=49.2,
/ppcolssim/detection/7/55/CoRa/CIC/.initial=50.5,
/ppcolssim/crcok/7/55/CoRa/CIC/.initial=32.5,
/ppcolssim/detection/7/55/CoRa/Pyramid/.initial=65.0,
/ppcolssim/crcok/7/55/CoRa/Pyramid/.initial=55.0,
/ppcolssim/detection/7/55/CoRa/TnB/.initial=15.7,
/ppcolssim/crcok/7/55/CoRa/TnB/.initial=6.9,
/ppcolssim/detection/7/55/OpenLoRa/CoRa/.initial=-65.3,
/ppcolssim/crcok/7/55/OpenLoRa/CoRa/.initial=-49.2,
/ppcolssim/detection/7/55/OpenLoRa/CIC/.initial=-14.8,
/ppcolssim/crcok/7/55/OpenLoRa/CIC/.initial=-16.7,
/ppcolssim/detection/7/55/OpenLoRa/Pyramid/.initial=-0.3,
/ppcolssim/crcok/7/55/OpenLoRa/Pyramid/.initial=5.8,
/ppcolssim/detection/7/55/OpenLoRa/TnB/.initial=-49.6,
/ppcolssim/crcok/7/55/OpenLoRa/TnB/.initial=-42.3,
/ppcolssim/detection/7/55/CIC/CoRa/.initial=-50.5,
/ppcolssim/crcok/7/55/CIC/CoRa/.initial=-32.5,
/ppcolssim/detection/7/55/CIC/OpenLoRa/.initial=14.8,
/ppcolssim/crcok/7/55/CIC/OpenLoRa/.initial=16.7,
/ppcolssim/detection/7/55/CIC/Pyramid/.initial=14.5,
/ppcolssim/crcok/7/55/CIC/Pyramid/.initial=22.5,
/ppcolssim/detection/7/55/CIC/TnB/.initial=-34.8,
/ppcolssim/crcok/7/55/CIC/TnB/.initial=-25.6,
/ppcolssim/detection/7/55/Pyramid/CoRa/.initial=-65.0,
/ppcolssim/crcok/7/55/Pyramid/CoRa/.initial=-55.0,
/ppcolssim/detection/7/55/Pyramid/OpenLoRa/.initial=0.3,
/ppcolssim/crcok/7/55/Pyramid/OpenLoRa/.initial=-5.8,
/ppcolssim/detection/7/55/Pyramid/CIC/.initial=-14.5,
/ppcolssim/crcok/7/55/Pyramid/CIC/.initial=-22.5,
/ppcolssim/detection/7/55/Pyramid/TnB/.initial=-49.3,
/ppcolssim/crcok/7/55/Pyramid/TnB/.initial=-48.1,
/ppcolssim/detection/7/55/TnB/CoRa/.initial=-15.7,
/ppcolssim/crcok/7/55/TnB/CoRa/.initial=-6.9,
/ppcolssim/detection/7/55/TnB/OpenLoRa/.initial=49.6,
/ppcolssim/crcok/7/55/TnB/OpenLoRa/.initial=42.3,
/ppcolssim/detection/7/55/TnB/CIC/.initial=34.8,
/ppcolssim/crcok/7/55/TnB/CIC/.initial=25.6,
/ppcolssim/detection/7/55/TnB/Pyramid/.initial=49.3,
/ppcolssim/crcok/7/55/TnB/Pyramid/.initial=48.1,
/ppcolssim/detection/12/60/CoRa/OpenLoRa/.initial=77.1,
/ppcolssim/crcok/12/60/CoRa/OpenLoRa/.initial=50.3,
/ppcolssim/detection/12/60/CoRa/CIC/.initial=75.1,
/ppcolssim/crcok/12/60/CoRa/CIC/.initial=47.9,
/ppcolssim/detection/12/60/CoRa/Pyramid/.initial=72.5,
/ppcolssim/crcok/12/60/CoRa/Pyramid/.initial=50.6,
/ppcolssim/detection/12/60/CoRa/TnB/.initial=39.4,
/ppcolssim/crcok/12/60/CoRa/TnB/.initial=17.2,
/ppcolssim/detection/12/60/OpenLoRa/CoRa/.initial=-77.1,
/ppcolssim/crcok/12/60/OpenLoRa/CoRa/.initial=-50.3,
/ppcolssim/detection/12/60/OpenLoRa/CIC/.initial=-2.0,
/ppcolssim/crcok/12/60/OpenLoRa/CIC/.initial=-2.4,
/ppcolssim/detection/12/60/OpenLoRa/Pyramid/.initial=-4.6,
/ppcolssim/crcok/12/60/OpenLoRa/Pyramid/.initial=0.4,
/ppcolssim/detection/12/60/OpenLoRa/TnB/.initial=-37.6,
/ppcolssim/crcok/12/60/OpenLoRa/TnB/.initial=-33.1,
/ppcolssim/detection/12/60/CIC/CoRa/.initial=-75.1,
/ppcolssim/crcok/12/60/CIC/CoRa/.initial=-47.9,
/ppcolssim/detection/12/60/CIC/OpenLoRa/.initial=2.0,
/ppcolssim/crcok/12/60/CIC/OpenLoRa/.initial=2.4,
/ppcolssim/detection/12/60/CIC/Pyramid/.initial=-2.6,
/ppcolssim/crcok/12/60/CIC/Pyramid/.initial=2.7,
/ppcolssim/detection/12/60/CIC/TnB/.initial=-35.7,
/ppcolssim/crcok/12/60/CIC/TnB/.initial=-30.7,
/ppcolssim/detection/12/60/Pyramid/CoRa/.initial=-72.5,
/ppcolssim/crcok/12/60/Pyramid/CoRa/.initial=-50.6,
/ppcolssim/detection/12/60/Pyramid/OpenLoRa/.initial=4.6,
/ppcolssim/crcok/12/60/Pyramid/OpenLoRa/.initial=-0.4,
/ppcolssim/detection/12/60/Pyramid/CIC/.initial=2.6,
/ppcolssim/crcok/12/60/Pyramid/CIC/.initial=-2.7,
/ppcolssim/detection/12/60/Pyramid/TnB/.initial=-33.1,
/ppcolssim/crcok/12/60/Pyramid/TnB/.initial=-33.4,
/ppcolssim/detection/12/60/TnB/CoRa/.initial=-39.4,
/ppcolssim/crcok/12/60/TnB/CoRa/.initial=-17.2,
/ppcolssim/detection/12/60/TnB/OpenLoRa/.initial=37.6,
/ppcolssim/crcok/12/60/TnB/OpenLoRa/.initial=33.1,
/ppcolssim/detection/12/60/TnB/CIC/.initial=35.7,
/ppcolssim/crcok/12/60/TnB/CIC/.initial=30.7,
/ppcolssim/detection/12/60/TnB/Pyramid/.initial=33.1,
/ppcolssim/crcok/12/60/TnB/Pyramid/.initial=33.4,
/ppcolssim/detection/12/40/CoRa/OpenLoRa/.initial=82.0,
/ppcolssim/crcok/12/40/CoRa/OpenLoRa/.initial=60.1,
/ppcolssim/detection/12/40/CoRa/CIC/.initial=77.9,
/ppcolssim/crcok/12/40/CoRa/CIC/.initial=55.7,
/ppcolssim/detection/12/40/CoRa/Pyramid/.initial=80.9,
/ppcolssim/crcok/12/40/CoRa/Pyramid/.initial=60.7,
/ppcolssim/detection/12/40/CoRa/TnB/.initial=36.9,
/ppcolssim/crcok/12/40/CoRa/TnB/.initial=19.3,
/ppcolssim/detection/12/40/OpenLoRa/CoRa/.initial=-82.0,
/ppcolssim/crcok/12/40/OpenLoRa/CoRa/.initial=-60.1,
/ppcolssim/detection/12/40/OpenLoRa/CIC/.initial=-4.2,
/ppcolssim/crcok/12/40/OpenLoRa/CIC/.initial=-4.5,
/ppcolssim/detection/12/40/OpenLoRa/Pyramid/.initial=-1.1,
/ppcolssim/crcok/12/40/OpenLoRa/Pyramid/.initial=0.6,
/ppcolssim/detection/12/40/OpenLoRa/TnB/.initial=-45.1,
/ppcolssim/crcok/12/40/OpenLoRa/TnB/.initial=-40.8,
/ppcolssim/detection/12/40/CIC/CoRa/.initial=-77.9,
/ppcolssim/crcok/12/40/CIC/CoRa/.initial=-55.7,
/ppcolssim/detection/12/40/CIC/OpenLoRa/.initial=4.2,
/ppcolssim/crcok/12/40/CIC/OpenLoRa/.initial=4.5,
/ppcolssim/detection/12/40/CIC/Pyramid/.initial=3.1,
/ppcolssim/crcok/12/40/CIC/Pyramid/.initial=5.1,
/ppcolssim/detection/12/40/CIC/TnB/.initial=-40.9,
/ppcolssim/crcok/12/40/CIC/TnB/.initial=-36.4,
/ppcolssim/detection/12/40/Pyramid/CoRa/.initial=-80.9,
/ppcolssim/crcok/12/40/Pyramid/CoRa/.initial=-60.7,
/ppcolssim/detection/12/40/Pyramid/OpenLoRa/.initial=1.1,
/ppcolssim/crcok/12/40/Pyramid/OpenLoRa/.initial=-0.6,
/ppcolssim/detection/12/40/Pyramid/CIC/.initial=-3.1,
/ppcolssim/crcok/12/40/Pyramid/CIC/.initial=-5.1,
/ppcolssim/detection/12/40/Pyramid/TnB/.initial=-44.0,
/ppcolssim/crcok/12/40/Pyramid/TnB/.initial=-41.4,
/ppcolssim/detection/12/40/TnB/CoRa/.initial=-36.9,
/ppcolssim/crcok/12/40/TnB/CoRa/.initial=-19.3,
/ppcolssim/detection/12/40/TnB/OpenLoRa/.initial=45.1,
/ppcolssim/crcok/12/40/TnB/OpenLoRa/.initial=40.8,
/ppcolssim/detection/12/40/TnB/CIC/.initial=40.9,
/ppcolssim/crcok/12/40/TnB/CIC/.initial=36.4,
/ppcolssim/detection/12/40/TnB/Pyramid/.initial=44.0,
/ppcolssim/crcok/12/40/TnB/Pyramid/.initial=41.4,
/ppcolssim/detection/12/50/CoRa/OpenLoRa/.initial=80.3,
/ppcolssim/crcok/12/50/CoRa/OpenLoRa/.initial=55.3,
/ppcolssim/detection/12/50/CoRa/CIC/.initial=77.3,
/ppcolssim/crcok/12/50/CoRa/CIC/.initial=51.9,
/ppcolssim/detection/12/50/CoRa/Pyramid/.initial=77.5,
/ppcolssim/crcok/12/50/CoRa/Pyramid/.initial=55.6,
/ppcolssim/detection/12/50/CoRa/TnB/.initial=38.5,
/ppcolssim/crcok/12/50/CoRa/TnB/.initial=18.1,
/ppcolssim/detection/12/50/OpenLoRa/CoRa/.initial=-80.3,
/ppcolssim/crcok/12/50/OpenLoRa/CoRa/.initial=-55.3,
/ppcolssim/detection/12/50/OpenLoRa/CIC/.initial=-3.0,
/ppcolssim/crcok/12/50/OpenLoRa/CIC/.initial=-3.3,
/ppcolssim/detection/12/50/OpenLoRa/Pyramid/.initial=-2.8,
/ppcolssim/crcok/12/50/OpenLoRa/Pyramid/.initial=0.4,
/ppcolssim/detection/12/50/OpenLoRa/TnB/.initial=-41.8,
/ppcolssim/crcok/12/50/OpenLoRa/TnB/.initial=-37.2,
/ppcolssim/detection/12/50/CIC/CoRa/.initial=-77.3,
/ppcolssim/crcok/12/50/CIC/CoRa/.initial=-51.9,
/ppcolssim/detection/12/50/CIC/OpenLoRa/.initial=3.0,
/ppcolssim/crcok/12/50/CIC/OpenLoRa/.initial=3.3,
/ppcolssim/detection/12/50/CIC/Pyramid/.initial=0.2,
/ppcolssim/crcok/12/50/CIC/Pyramid/.initial=3.7,
/ppcolssim/detection/12/50/CIC/TnB/.initial=-38.8,
/ppcolssim/crcok/12/50/CIC/TnB/.initial=-33.9,
/ppcolssim/detection/12/50/Pyramid/CoRa/.initial=-77.5,
/ppcolssim/crcok/12/50/Pyramid/CoRa/.initial=-55.6,
/ppcolssim/detection/12/50/Pyramid/OpenLoRa/.initial=2.8,
/ppcolssim/crcok/12/50/Pyramid/OpenLoRa/.initial=-0.4,
/ppcolssim/detection/12/50/Pyramid/CIC/.initial=-0.2,
/ppcolssim/crcok/12/50/Pyramid/CIC/.initial=-3.7,
/ppcolssim/detection/12/50/Pyramid/TnB/.initial=-39.0,
/ppcolssim/crcok/12/50/Pyramid/TnB/.initial=-37.6,
/ppcolssim/detection/12/50/TnB/CoRa/.initial=-38.5,
/ppcolssim/crcok/12/50/TnB/CoRa/.initial=-18.1,
/ppcolssim/detection/12/50/TnB/OpenLoRa/.initial=41.8,
/ppcolssim/crcok/12/50/TnB/OpenLoRa/.initial=37.2,
/ppcolssim/detection/12/50/TnB/CIC/.initial=38.8,
/ppcolssim/crcok/12/50/TnB/CIC/.initial=33.9,
/ppcolssim/detection/12/50/TnB/Pyramid/.initial=39.0,
/ppcolssim/crcok/12/50/TnB/Pyramid/.initial=37.6,
/ppcolssim/detection/7/50/CoRa/OpenLoRa/.initial=63.4,
/ppcolssim/crcok/7/50/CoRa/OpenLoRa/.initial=48.9,
/ppcolssim/detection/7/50/CoRa/CIC/.initial=49.0,
/ppcolssim/crcok/7/50/CoRa/CIC/.initial=32.2,
/ppcolssim/detection/7/50/CoRa/Pyramid/.initial=64.6,
/ppcolssim/crcok/7/50/CoRa/Pyramid/.initial=56.9,
/ppcolssim/detection/7/50/CoRa/TnB/.initial=14.8,
/ppcolssim/crcok/7/50/CoRa/TnB/.initial=6.4,
/ppcolssim/detection/7/50/OpenLoRa/CoRa/.initial=-63.4,
/ppcolssim/crcok/7/50/OpenLoRa/CoRa/.initial=-48.9,
/ppcolssim/detection/7/50/OpenLoRa/CIC/.initial=-14.4,
/ppcolssim/crcok/7/50/OpenLoRa/CIC/.initial=-16.7,
/ppcolssim/detection/7/50/OpenLoRa/Pyramid/.initial=1.3,
/ppcolssim/crcok/7/50/OpenLoRa/Pyramid/.initial=8.0,
/ppcolssim/detection/7/50/OpenLoRa/TnB/.initial=-48.6,
/ppcolssim/crcok/7/50/OpenLoRa/TnB/.initial=-42.5,
/ppcolssim/detection/7/50/CIC/CoRa/.initial=-49.0,
/ppcolssim/crcok/7/50/CIC/CoRa/.initial=-32.2,
/ppcolssim/detection/7/50/CIC/OpenLoRa/.initial=14.4,
/ppcolssim/crcok/7/50/CIC/OpenLoRa/.initial=16.7,
/ppcolssim/detection/7/50/CIC/Pyramid/.initial=15.7,
/ppcolssim/crcok/7/50/CIC/Pyramid/.initial=24.7,
/ppcolssim/detection/7/50/CIC/TnB/.initial=-34.2,
/ppcolssim/crcok/7/50/CIC/TnB/.initial=-25.8,
/ppcolssim/detection/7/50/Pyramid/CoRa/.initial=-64.6,
/ppcolssim/crcok/7/50/Pyramid/CoRa/.initial=-56.9,
/ppcolssim/detection/7/50/Pyramid/OpenLoRa/.initial=-1.3,
/ppcolssim/crcok/7/50/Pyramid/OpenLoRa/.initial=-8.0,
/ppcolssim/detection/7/50/Pyramid/CIC/.initial=-15.7,
/ppcolssim/crcok/7/50/Pyramid/CIC/.initial=-24.7,
/ppcolssim/detection/7/50/Pyramid/TnB/.initial=-49.9,
/ppcolssim/crcok/7/50/Pyramid/TnB/.initial=-50.4,
/ppcolssim/detection/7/50/TnB/CoRa/.initial=-14.8,
/ppcolssim/crcok/7/50/TnB/CoRa/.initial=-6.4,
/ppcolssim/detection/7/50/TnB/OpenLoRa/.initial=48.6,
/ppcolssim/crcok/7/50/TnB/OpenLoRa/.initial=42.5,
/ppcolssim/detection/7/50/TnB/CIC/.initial=34.2,
/ppcolssim/crcok/7/50/TnB/CIC/.initial=25.8,
/ppcolssim/detection/7/50/TnB/Pyramid/.initial=49.9,
/ppcolssim/crcok/7/50/TnB/Pyramid/.initial=50.4,
/ppcolssim/detection/7/40/CoRa/OpenLoRa/.initial=61.9,
/ppcolssim/crcok/7/40/CoRa/OpenLoRa/.initial=49.3,
/ppcolssim/detection/7/40/CoRa/CIC/.initial=49.3,
/ppcolssim/crcok/7/40/CoRa/CIC/.initial=33.4,
/ppcolssim/detection/7/40/CoRa/Pyramid/.initial=64.4,
/ppcolssim/crcok/7/40/CoRa/Pyramid/.initial=58.1,
/ppcolssim/detection/7/40/CoRa/TnB/.initial=14.8,
/ppcolssim/crcok/7/40/CoRa/TnB/.initial=6.9,
/ppcolssim/detection/7/40/OpenLoRa/CoRa/.initial=-61.9,
/ppcolssim/crcok/7/40/OpenLoRa/CoRa/.initial=-49.3,
/ppcolssim/detection/7/40/OpenLoRa/CIC/.initial=-12.6,
/ppcolssim/crcok/7/40/OpenLoRa/CIC/.initial=-15.8,
/ppcolssim/detection/7/40/OpenLoRa/Pyramid/.initial=2.5,
/ppcolssim/crcok/7/40/OpenLoRa/Pyramid/.initial=8.8,
/ppcolssim/detection/7/40/OpenLoRa/TnB/.initial=-47.1,
/ppcolssim/crcok/7/40/OpenLoRa/TnB/.initial=-42.4,
/ppcolssim/detection/7/40/CIC/CoRa/.initial=-49.3,
/ppcolssim/crcok/7/40/CIC/CoRa/.initial=-33.4,
/ppcolssim/detection/7/40/CIC/OpenLoRa/.initial=12.6,
/ppcolssim/crcok/7/40/CIC/OpenLoRa/.initial=15.8,
/ppcolssim/detection/7/40/CIC/Pyramid/.initial=15.2,
/ppcolssim/crcok/7/40/CIC/Pyramid/.initial=24.7,
/ppcolssim/detection/7/40/CIC/TnB/.initial=-34.5,
/ppcolssim/crcok/7/40/CIC/TnB/.initial=-26.6,
/ppcolssim/detection/7/40/Pyramid/CoRa/.initial=-64.4,
/ppcolssim/crcok/7/40/Pyramid/CoRa/.initial=-58.1,
/ppcolssim/detection/7/40/Pyramid/OpenLoRa/.initial=-2.5,
/ppcolssim/crcok/7/40/Pyramid/OpenLoRa/.initial=-8.8,
/ppcolssim/detection/7/40/Pyramid/CIC/.initial=-15.2,
/ppcolssim/crcok/7/40/Pyramid/CIC/.initial=-24.7,
/ppcolssim/detection/7/40/Pyramid/TnB/.initial=-49.7,
/ppcolssim/crcok/7/40/Pyramid/TnB/.initial=-51.2,
/ppcolssim/detection/7/40/TnB/CoRa/.initial=-14.8,
/ppcolssim/crcok/7/40/TnB/CoRa/.initial=-6.9,
/ppcolssim/detection/7/40/TnB/OpenLoRa/.initial=47.1,
/ppcolssim/crcok/7/40/TnB/OpenLoRa/.initial=42.4,
/ppcolssim/detection/7/40/TnB/CIC/.initial=34.5,
/ppcolssim/crcok/7/40/TnB/CIC/.initial=26.6,
/ppcolssim/detection/7/40/TnB/Pyramid/.initial=49.7,
/ppcolssim/crcok/7/40/TnB/Pyramid/.initial=51.2,
/ppcolssim/detection/12/10/CoRa/OpenLoRa/.initial=88.7,
/ppcolssim/crcok/12/10/CoRa/OpenLoRa/.initial=85.1,
/ppcolssim/detection/12/10/CoRa/CIC/.initial=79.7,
/ppcolssim/crcok/12/10/CoRa/CIC/.initial=74.8,
/ppcolssim/detection/12/10/CoRa/Pyramid/.initial=91.8,
/ppcolssim/crcok/12/10/CoRa/Pyramid/.initial=87.8,
/ppcolssim/detection/12/10/CoRa/TnB/.initial=19.4,
/ppcolssim/crcok/12/10/CoRa/TnB/.initial=19.5,
/ppcolssim/detection/12/10/OpenLoRa/CoRa/.initial=-88.7,
/ppcolssim/crcok/12/10/OpenLoRa/CoRa/.initial=-85.1,
/ppcolssim/detection/12/10/OpenLoRa/CIC/.initial=-9.0,
/ppcolssim/crcok/12/10/OpenLoRa/CIC/.initial=-10.3,
/ppcolssim/detection/12/10/OpenLoRa/Pyramid/.initial=3.1,
/ppcolssim/crcok/12/10/OpenLoRa/Pyramid/.initial=2.7,
/ppcolssim/detection/12/10/OpenLoRa/TnB/.initial=-69.3,
/ppcolssim/crcok/12/10/OpenLoRa/TnB/.initial=-65.6,
/ppcolssim/detection/12/10/CIC/CoRa/.initial=-79.7,
/ppcolssim/crcok/12/10/CIC/CoRa/.initial=-74.8,
/ppcolssim/detection/12/10/CIC/OpenLoRa/.initial=9.0,
/ppcolssim/crcok/12/10/CIC/OpenLoRa/.initial=10.3,
/ppcolssim/detection/12/10/CIC/Pyramid/.initial=12.1,
/ppcolssim/crcok/12/10/CIC/Pyramid/.initial=13.0,
/ppcolssim/detection/12/10/CIC/TnB/.initial=-60.3,
/ppcolssim/crcok/12/10/CIC/TnB/.initial=-55.3,
/ppcolssim/detection/12/10/Pyramid/CoRa/.initial=-91.8,
/ppcolssim/crcok/12/10/Pyramid/CoRa/.initial=-87.8,
/ppcolssim/detection/12/10/Pyramid/OpenLoRa/.initial=-3.1,
/ppcolssim/crcok/12/10/Pyramid/OpenLoRa/.initial=-2.7,
/ppcolssim/detection/12/10/Pyramid/CIC/.initial=-12.1,
/ppcolssim/crcok/12/10/Pyramid/CIC/.initial=-13.0,
/ppcolssim/detection/12/10/Pyramid/TnB/.initial=-72.4,
/ppcolssim/crcok/12/10/Pyramid/TnB/.initial=-68.2,
/ppcolssim/detection/12/10/TnB/CoRa/.initial=-19.4,
/ppcolssim/crcok/12/10/TnB/CoRa/.initial=-19.5,
/ppcolssim/detection/12/10/TnB/OpenLoRa/.initial=69.3,
/ppcolssim/crcok/12/10/TnB/OpenLoRa/.initial=65.6,
/ppcolssim/detection/12/10/TnB/CIC/.initial=60.3,
/ppcolssim/crcok/12/10/TnB/CIC/.initial=55.3,
/ppcolssim/detection/12/10/TnB/Pyramid/.initial=72.4,
/ppcolssim/crcok/12/10/TnB/Pyramid/.initial=68.2,
/ppcolssim/detection/7/15/CoRa/OpenLoRa/.initial=47.9,
/ppcolssim/crcok/7/15/CoRa/OpenLoRa/.initial=45.6,
/ppcolssim/detection/7/15/CoRa/CIC/.initial=41.1,
/ppcolssim/crcok/7/15/CoRa/CIC/.initial=34.9,
/ppcolssim/detection/7/15/CoRa/Pyramid/.initial=61.7,
/ppcolssim/crcok/7/15/CoRa/Pyramid/.initial=68.4,
/ppcolssim/detection/7/15/CoRa/TnB/.initial=7.0,
/ppcolssim/crcok/7/15/CoRa/TnB/.initial=5.9,
/ppcolssim/detection/7/15/OpenLoRa/CoRa/.initial=-47.9,
/ppcolssim/crcok/7/15/OpenLoRa/CoRa/.initial=-45.6,
/ppcolssim/detection/7/15/OpenLoRa/CIC/.initial=-6.8,
/ppcolssim/crcok/7/15/OpenLoRa/CIC/.initial=-10.8,
/ppcolssim/detection/7/15/OpenLoRa/Pyramid/.initial=13.9,
/ppcolssim/crcok/7/15/OpenLoRa/Pyramid/.initial=22.8,
/ppcolssim/detection/7/15/OpenLoRa/TnB/.initial=-40.9,
/ppcolssim/crcok/7/15/OpenLoRa/TnB/.initial=-39.8,
/ppcolssim/detection/7/15/CIC/CoRa/.initial=-41.1,
/ppcolssim/crcok/7/15/CIC/CoRa/.initial=-34.9,
/ppcolssim/detection/7/15/CIC/OpenLoRa/.initial=6.8,
/ppcolssim/crcok/7/15/CIC/OpenLoRa/.initial=10.8,
/ppcolssim/detection/7/15/CIC/Pyramid/.initial=20.7,
/ppcolssim/crcok/7/15/CIC/Pyramid/.initial=33.6,
/ppcolssim/detection/7/15/CIC/TnB/.initial=-34.0,
/ppcolssim/crcok/7/15/CIC/TnB/.initial=-29.0,
/ppcolssim/detection/7/15/Pyramid/CoRa/.initial=-61.7,
/ppcolssim/crcok/7/15/Pyramid/CoRa/.initial=-68.4,
/ppcolssim/detection/7/15/Pyramid/OpenLoRa/.initial=-13.9,
/ppcolssim/crcok/7/15/Pyramid/OpenLoRa/.initial=-22.8,
/ppcolssim/detection/7/15/Pyramid/CIC/.initial=-20.7,
/ppcolssim/crcok/7/15/Pyramid/CIC/.initial=-33.6,
/ppcolssim/detection/7/15/Pyramid/TnB/.initial=-54.7,
/ppcolssim/crcok/7/15/Pyramid/TnB/.initial=-62.6,
/ppcolssim/detection/7/15/TnB/CoRa/.initial=-7.0,
/ppcolssim/crcok/7/15/TnB/CoRa/.initial=-5.9,
/ppcolssim/detection/7/15/TnB/OpenLoRa/.initial=40.9,
/ppcolssim/crcok/7/15/TnB/OpenLoRa/.initial=39.8,
/ppcolssim/detection/7/15/TnB/CIC/.initial=34.0,
/ppcolssim/crcok/7/15/TnB/CIC/.initial=29.0,
/ppcolssim/detection/7/15/TnB/Pyramid/.initial=54.7,
/ppcolssim/crcok/7/15/TnB/Pyramid/.initial=62.6,
/ppcolssim/detection/7/60/CoRa/OpenLoRa/.initial=64.8,
/ppcolssim/crcok/7/60/CoRa/OpenLoRa/.initial=47.5,
/ppcolssim/detection/7/60/CoRa/CIC/.initial=50.1,
/ppcolssim/crcok/7/60/CoRa/CIC/.initial=31.1,
/ppcolssim/detection/7/60/CoRa/Pyramid/.initial=64.3,
/ppcolssim/crcok/7/60/CoRa/Pyramid/.initial=52.9,
/ppcolssim/detection/7/60/CoRa/TnB/.initial=16.8,
/ppcolssim/crcok/7/60/CoRa/TnB/.initial=6.5,
/ppcolssim/detection/7/60/OpenLoRa/CoRa/.initial=-64.8,
/ppcolssim/crcok/7/60/OpenLoRa/CoRa/.initial=-47.5,
/ppcolssim/detection/7/60/OpenLoRa/CIC/.initial=-14.7,
/ppcolssim/crcok/7/60/OpenLoRa/CIC/.initial=-16.4,
/ppcolssim/detection/7/60/OpenLoRa/Pyramid/.initial=-0.5,
/ppcolssim/crcok/7/60/OpenLoRa/Pyramid/.initial=5.4,
/ppcolssim/detection/7/60/OpenLoRa/TnB/.initial=-48.0,
/ppcolssim/crcok/7/60/OpenLoRa/TnB/.initial=-41.0,
/ppcolssim/detection/7/60/CIC/CoRa/.initial=-50.1,
/ppcolssim/crcok/7/60/CIC/CoRa/.initial=-31.1,
/ppcolssim/detection/7/60/CIC/OpenLoRa/.initial=14.7,
/ppcolssim/crcok/7/60/CIC/OpenLoRa/.initial=16.4,
/ppcolssim/detection/7/60/CIC/Pyramid/.initial=14.1,
/ppcolssim/crcok/7/60/CIC/Pyramid/.initial=21.8,
/ppcolssim/detection/7/60/CIC/TnB/.initial=-33.4,
/ppcolssim/crcok/7/60/CIC/TnB/.initial=-24.6,
/ppcolssim/detection/7/60/Pyramid/CoRa/.initial=-64.3,
/ppcolssim/crcok/7/60/Pyramid/CoRa/.initial=-52.9,
/ppcolssim/detection/7/60/Pyramid/OpenLoRa/.initial=0.5,
/ppcolssim/crcok/7/60/Pyramid/OpenLoRa/.initial=-5.4,
/ppcolssim/detection/7/60/Pyramid/CIC/.initial=-14.1,
/ppcolssim/crcok/7/60/Pyramid/CIC/.initial=-21.8,
/ppcolssim/detection/7/60/Pyramid/TnB/.initial=-47.5,
/ppcolssim/crcok/7/60/Pyramid/TnB/.initial=-46.4,
/ppcolssim/detection/7/60/TnB/CoRa/.initial=-16.8,
/ppcolssim/crcok/7/60/TnB/CoRa/.initial=-6.5,
/ppcolssim/detection/7/60/TnB/OpenLoRa/.initial=48.0,
/ppcolssim/crcok/7/60/TnB/OpenLoRa/.initial=41.0,
/ppcolssim/detection/7/60/TnB/CIC/.initial=33.4,
/ppcolssim/crcok/7/60/TnB/CIC/.initial=24.6,
/ppcolssim/detection/7/60/TnB/Pyramid/.initial=47.5,
/ppcolssim/crcok/7/60/TnB/Pyramid/.initial=46.4,
/ppcolssim/detection/7/35/CoRa/OpenLoRa/.initial=61.1,
/ppcolssim/crcok/7/35/CoRa/OpenLoRa/.initial=51.1,
/ppcolssim/detection/7/35/CoRa/CIC/.initial=48.8,
/ppcolssim/crcok/7/35/CoRa/CIC/.initial=35.4,
/ppcolssim/detection/7/35/CoRa/Pyramid/.initial=65.7,
/ppcolssim/crcok/7/35/CoRa/Pyramid/.initial=62.5,
/ppcolssim/detection/7/35/CoRa/TnB/.initial=12.6,
/ppcolssim/crcok/7/35/CoRa/TnB/.initial=7.4,
/ppcolssim/detection/7/35/OpenLoRa/CoRa/.initial=-61.1,
/ppcolssim/crcok/7/35/OpenLoRa/CoRa/.initial=-51.1,
/ppcolssim/detection/7/35/OpenLoRa/CIC/.initial=-12.3,
/ppcolssim/crcok/7/35/OpenLoRa/CIC/.initial=-15.7,
/ppcolssim/detection/7/35/OpenLoRa/Pyramid/.initial=4.6,
/ppcolssim/crcok/7/35/OpenLoRa/Pyramid/.initial=11.4,
/ppcolssim/detection/7/35/OpenLoRa/TnB/.initial=-48.5,
/ppcolssim/crcok/7/35/OpenLoRa/TnB/.initial=-43.7,
/ppcolssim/detection/7/35/CIC/CoRa/.initial=-48.8,
/ppcolssim/crcok/7/35/CIC/CoRa/.initial=-35.4,
/ppcolssim/detection/7/35/CIC/OpenLoRa/.initial=12.3,
/ppcolssim/crcok/7/35/CIC/OpenLoRa/.initial=15.7,
/ppcolssim/detection/7/35/CIC/Pyramid/.initial=16.9,
/ppcolssim/crcok/7/35/CIC/Pyramid/.initial=27.1,
/ppcolssim/detection/7/35/CIC/TnB/.initial=-36.2,
/ppcolssim/crcok/7/35/CIC/TnB/.initial=-28.0,
/ppcolssim/detection/7/35/Pyramid/CoRa/.initial=-65.7,
/ppcolssim/crcok/7/35/Pyramid/CoRa/.initial=-62.5,
/ppcolssim/detection/7/35/Pyramid/OpenLoRa/.initial=-4.6,
/ppcolssim/crcok/7/35/Pyramid/OpenLoRa/.initial=-11.4,
/ppcolssim/detection/7/35/Pyramid/CIC/.initial=-16.9,
/ppcolssim/crcok/7/35/Pyramid/CIC/.initial=-27.1,
/ppcolssim/detection/7/35/Pyramid/TnB/.initial=-53.1,
/ppcolssim/crcok/7/35/Pyramid/TnB/.initial=-55.1,
/ppcolssim/detection/7/35/TnB/CoRa/.initial=-12.6,
/ppcolssim/crcok/7/35/TnB/CoRa/.initial=-7.4,
/ppcolssim/detection/7/35/TnB/OpenLoRa/.initial=48.5,
/ppcolssim/crcok/7/35/TnB/OpenLoRa/.initial=43.7,
/ppcolssim/detection/7/35/TnB/CIC/.initial=36.2,
/ppcolssim/crcok/7/35/TnB/CIC/.initial=28.0,
/ppcolssim/detection/7/35/TnB/Pyramid/.initial=53.1,
/ppcolssim/crcok/7/35/TnB/Pyramid/.initial=55.1
}

\pgfkeys{
/model/collisions/analytical/7/1/.initial=0.13,
/model/collisions/analytical/7/2/.initial=-6.14,
/model/collisions/analytical/7/3/.initial=-15.98,
/model/collisions/analytical/7/4/.initial=-21.36,
/model/collisions/analytical/7/5/.initial=-24.01,
/model/collisions/analytical/7/6/.initial=-25.46,
/model/collisions/analytical/7/7/.initial=-26.26,
}
\pgfkeys{
/model/collisions/analytical/12/1/.initial=-0.26,
/model/collisions/analytical/12/2/.initial=-17.87,
/model/collisions/analytical/12/3/.initial=-35.65,
/model/collisions/analytical/12/4/.initial=-44.07,
/model/collisions/analytical/12/5/.initial=-48.27,
/model/collisions/analytical/12/6/.initial=-50.58,
/model/collisions/analytical/12/7/.initial=-51.90,
}
\pgfkeys{
/model/collisions/mc/7/1/.initial=-0.21,
/model/collisions/mc/7/2/.initial=-3.65,
/model/collisions/mc/7/3/.initial=-7.96,
/model/collisions/mc/7/4/.initial=-10.28,
/model/collisions/mc/7/5/.initial=-17.69,
/model/collisions/mc/7/6/.initial=-20.07,
/model/collisions/mc/7/7/.initial=-21.32,
}
\pgfkeys{
/model/collisions/mc/12/1/.initial=-0.28,
/model/collisions/mc/12/2/.initial=-6.26,
/model/collisions/mc/12/3/.initial=-25.59,
/model/collisions/mc/12/4/.initial=-33.35,
/model/collisions/mc/12/5/.initial=-45.61,
/model/collisions/mc/12/6/.initial=-48.51,
/model/collisions/mc/12/7/.initial=-50.45,
}
\pgfkeys{
/model/collisions/analytical/7/1/.initial=0.13,
/model/collisions/analytical/7/2/.initial=-6.14,
/model/collisions/analytical/7/3/.initial=-15.98,
/model/collisions/analytical/7/4/.initial=-21.36,
/model/collisions/analytical/7/5/.initial=-24.01,
/model/collisions/analytical/7/6/.initial=-25.46,
/model/collisions/analytical/7/7/.initial=-26.26,
}
\pgfkeys{
/model/collisions/analytical/12/1/.initial=-0.26,
/model/collisions/analytical/12/2/.initial=-17.87,
/model/collisions/analytical/12/3/.initial=-35.65,
/model/collisions/analytical/12/4/.initial=-44.07,
/model/collisions/analytical/12/5/.initial=-48.27,
/model/collisions/analytical/12/6/.initial=-50.58,
/model/collisions/analytical/12/7/.initial=-51.90,
}
\pgfkeys{
/model/collisions/mc/7/1/.initial=-0.21,
/model/collisions/mc/7/2/.initial=-3.65,
/model/collisions/mc/7/3/.initial=-7.96,
/model/collisions/mc/7/4/.initial=-10.28,
/model/collisions/mc/7/5/.initial=-17.69,
/model/collisions/mc/7/6/.initial=-20.07,
/model/collisions/mc/7/7/.initial=-21.32,
}
\pgfkeys{
/model/collisions/mc/12/1/.initial=-0.28,
/model/collisions/mc/12/2/.initial=-6.26,
/model/collisions/mc/12/3/.initial=-25.59,
/model/collisions/mc/12/4/.initial=-33.35,
/model/collisions/mc/12/5/.initial=-45.61,
/model/collisions/mc/12/6/.initial=-48.51,
/model/collisions/mc/12/7/.initial=-50.45,
}

\definecolor{mpl_blue}{HTML}{1f77b4}
\definecolor{mpl_orange}{HTML}{ff7f0e}
\definecolor{mpl_green}{HTML}{2ca02c}
\definecolor{mpl_red}{HTML}{d62728}
\definecolor{mpl_purple}{HTML}{9467bd}
\begin{document}

\title{\our: Robust LoRa Frame Synchronization Under Frame Collisions and Ultra-Low SNR Conditions}

\author{Jos{\'e} {\'A}lamos\orcid{0000-0001-5681-2591},
        Thomas C. Schmidt\orcid{0000-0002-0956-7885},~\IEEEmembership{Member,~IEEE,}
        and~Matthias W\"ahlisch\orcid{0000-0002-3825-2807},~\IEEEmembership{Member,~IEEE}
\IEEEcompsocitemizethanks{
\IEEEcompsocthanksitem Jos\'e \'Alamos and Thomas C. Schmidt are with the Faculty of Computer Science at HAW Hamburg, Berliner Tor 7, 20099 Hamburg, Germany.\protect\\
E-mail: \{\texttt{jose.alamos, t.schmidt}\}\texttt{@haw-hamburg.de}.
\IEEEcompsocthanksitem Matthias W{\"a}hlisch is with the Faculty of Computer Science, TU Dresden, Helmholtzstr. 10, 01069 Dresden, and also with the Barkhausen Institut, 01187 Dresden,  Germany. \protect
E-mail: \texttt{m.waehlisch@tu-dresden.de}
\IEEEcompsocthanksitem  This work was supported by the German Federal Ministry of Research, Technology and Space (BMFTR) within the projects RESCUE-MATE (Grant \# 13N16843), and C-ray4edge (Grant \#s 16KIS1695, 16KIS1694K).}.
}
 


\maketitle

\setlength{\TPHorizModule}{\textwidth}
\setlength{\TPVertModule}{\paperheight}
\TPMargin{5pt}
\begin{textblock}{1}(.1,0.01)
\noindent
\footnotesize
If you refer to this paper, please cite the peer-reviewed publication:
J. Alamos, T. C. Schmidt, M. W{\"a}hlisch.\\
LZn: Robust LoRa Frame Synchronization Under Frame Collisions and Ultra-Low SNR Conditions.\\
    \emph{IEEE Transactions on Wireless Communications}, vol. 25, pp. 26041-26056, 2026,
    doi: 10.1109/TWC.2026.3747053.
\end{textblock}

\begin{abstract}
    LoRa has become a widely adopted wireless modulation scheme in
    LPWANs due to its low cost, long range, and minimal
    transmission power. However, collisions between frames of the same
    spreading factor -- common in dense LoRa deployments -- prevent
    conventional LoRa receivers from detecting and correctly decoding frames. 
    Recent work has introduced methods to improve recovery, yet
    their detection stage degrades sharply under low signal-to-noise ratio
    (SNR) and high collision rates. In this work, we introduce
    \our, a synchronization scheme driven by a minimum-based spectral
	intersection across consecutive symbols. While preamble search incurs a modest overhead, \our reduces the cost of fine synchronization by two to three orders of magnitude compared to existing collision-tolerant receivers. Our method  enables robust frame synchronization even under
    multiple packet overlaps or extremely low SNR conditions. We evaluate \our on
    simulations and three independent, real-world LoRa datasets. 
    \our improves detection sensitivity by up to 10\,dB
    and increases detection probability by up to $1.54\times$.
    In real-world datasets, \our
    improves decoding by $3.46\times$ in the most
    challenging single-user scenario and up to $1.22\times$ in collision
    scenarios compared to the second best collision-tolerant
    scheme (TnB).
    These results demonstrate that \our substantially improves the frame
    recovery of LoRa receivers, while remaining compatible with real-time
    requirements.
\end{abstract}


\section{Introduction}

LoRa, with its open specification and low-cost hardware, has gained significant
attention as a key technology for Low-Power Wide Area Networks (LPWANs).
LoRa networks commonly employ the LoRaWAN protocol, which
relies on an unslotted ALOHA medium-access scheme. End devices
transmit whenever data is available, without preset time slots or any prior
synchronization. This lack of coordination makes LoRaWAN networks highly
susceptible to frame collisions~\cite{f-cplal-17,ofjg-cccal-19}.

LoRa receivers can decode colliding frames with different chirp rates (\ie
different spreading factors or bandwidths), which are quasi-orthogonal. For
identical chirp rates, standard receivers can typically decode only the
strongest frame if its power exceeds that of interfering signals by at least
6\,dB, a phenomenon known as the capture effect. While multi-gateway
deployments increase the likelihood of successful capture, this effect is
insufficient in dense networks, where Adaptive Data Rate (ADR) often leads to
similar transmission power levels.

Recent studies explore demodulating LoRa symbols under same-chirp-rate frame collisions,
increasing throughput by more than an order of
magnitude~\cite{spcbk-cicdm-21,rz-trclb-22,xzg-fpdlt-19,txw-cemrl-20,gmgcg-iclgi-21,asw-cclsd-25}.
They introduce new frame synchronization strategies, since traditional LoRa
preamble detection does not work reliably under frame collisions. The 
proposed methods encounter three main limitations. \one Many approaches result
in degraded detection
sensitivity~\cite{spcbk-cicdm-21,rz-trclb-22,wtxx-rcltb-24}, hindering frame
detection at low signal-to-noise ratios (SNR), a common challenge in LoRaWAN
deployments; 
\two existing techniques deteriorate under high collision
rates~\cite{xzg-fpdlt-19, txw-cemrl-20, gmgcg-iclgi-21}, common in dense urban
scenarios~\cite{afsw-lcrr-25}; 
\three some methods require high computational complexity for frame
detection~\cite{spcbk-cicdm-21, xzg-fpdlt-19, txw-cemrl-20}, challenging 
 live frame detection. 

To address these challenges, we present and validate \our (short for \textit{LoRa Sync}), a frame
synchronization algorithm that operates even at low SNR levels and under
heavy collision scenarios, while keeping the computational
complexity compatible with live synchronization requirements.
Unlike prior spectral intersection approaches, which use multiplicative aggregation,
\our employs a minimum-based spectral intersection across consecutive preamble symbols.
This preserves persistent signal peaks while suppressing collision-induced artifacts and avoids 
the peak masking problem of multiplicative approaches.

The proposed method is designed for gateway-side operation with concurrent
uncoordinated transmissions. Its complexity matches conventional FFT-based LoRa
demodulators, with a slightly higher cost during preamble search that is offset by a substantially cheaper fine synchronization stage. This allows integration as a synchronization front-end without
modifying the demodulator.
Our key contributions read:

\begin{enumerate}
    \item We design a frame synchronization algorithm
        that achieves accurate detection in collision scenarios and ultra-low
        SNR conditions at a computation cost comparable to conventional FFT-based LoRa demodulators.
    \item We evaluate it against existing methods using simulated and real-world LoRa captures,
        assessing frame detection and impact on demodulation performance.
\end{enumerate}

The remainder of this paper is structured as follows. \autoref{sec:background}
provides  background on synchronization of LoRa frames and the
effect of collisions at the same chirp rate, as well as related work. \autoref{sec:core} describes our 
\our frame synchronization algorithm, for which  
 we contribute a  theoretical framework that 
derives analytical bounds in \autoref{sec:analytical}.
We evaluate the performance 
in~\autoref{sec:evaluation} and validate results in \autoref{sec:assessment}
 using real-world traces. \autoref{sec:complexity}
analyzes the computational complexity of LZn. We discuss  implications 
in \autoref{sec:discussion} and conclude  with an outlook in~\autoref{sec:conclusion}.

%

\section{Background and related work}\label{sec:background}


\subsection{LoRa Symbol Encoding}\label{subsec:lora_demod}

LoRa uses Chirp Spread Spectrum (CSS) modulation, generating each symbol
from a complex reference chirp. The symbol time and
chirp rate are determined by two key parameters: the bandwidth $B$ and
the spreading factor $S$. Specifically, the symbol duration is $T_{sym}=\frac{2^S}{B}$,
and the chirp rate is $\mu=\frac{B^2}{2^S}$. The chirp can either be an upchirp
(positive chirp rate) or a downchirp (negative chirp rate).

To encode data, LoRa cyclically shifts the reference upchirp by $\frac{m}{B}$,
where $m$ is the symbol value and can range from 0 to $M-1$, where $M=2^S$ is the
number of distinct symbols (\ie symbol cardinality). This time shift embeds the symbol information in the
signal. The instantaneous frequency of the modulated symbol is defined as:

$$
f_m(t) = \begin{cases}
    \frac{B^2}{M} t + \frac{B}{2} - \frac{B m}{M} & \text{if } 0 \leq t < \frac{m}{B} \\
    \frac{B^2}{M} t - \frac{B}{2} - \frac{B m}{M} & \text{if } \frac{m}{B} \leq t < T_{sym}
\end{cases}
$$

The baseband representation of the transmitted symbol is a complex signal
defined as $Y_m(t) = e^{j 2 \pi \int_{x=0}^t f_m(t) dt}$. When $m=0$, the
signal corresponds to the reference upchirp $Y_0(t)$. 

\subsection{LoRa Demodulation}\label{sec:lora_demodulation}

Dechirping the received signal $Y_m(t)$ by mixing it with the complex conjugate of
the reference chirp $Y_0^*(t)$ yields a signal $X(t)$ consisting of two clipped waveforms

\begin{align}
X(t) &= Y_m(t) \cdot Y_0^{*}(t) \notag \\
     &=e^{j 2 \pi \int_{x=0}^t \left( f_m(t) - f_0(t)\right) dt} \notag \\
     &= 
\begin{cases}
    e^{j 2 \pi B t \frac{M - m}{M}}  & \text{if } t < \frac{m}{B} \\
    e^{j 2 \pi m \frac{M - B t}{M}} & \text{if } t \geq \frac{m}{B}
\end{cases}\label{eq:dechirp}
\end{align}

Sampling the signal at the Nyquist rate $f_s=B$ aliases the continuous-time
waveform $X(t)$, causing the clipped waveforms to fold onto one another.
The resulting discrete-time sequence resembles a complex waveform $x[n]=e^{j 2
\pi \frac{m n}{M}}$, where $n \in {0..M-1}$ indexes the discrete time samples
and $m$ is the symbol value. This results in an M-ary FSK signal.

Therefore, the problem reduces to demodulating an FSK signal, for which several
well-established strategies exist~\cite{mmp-itpdt-19}. Coherent demodulation
correlates the received signal with the reference waveform (\eg using
DFT) to perform maximum-likelihood symbol detection under AWGN conditions,
achieving optimal performance but requiring precise carrier phase alignment.
Non-coherent demodulation avoids phase sensitivity by
computing the magnitude spectrum of the dechirped signal (via DFT) and
selecting the symbol corresponding to the peak energy bin (\ie \textit{argmax} of
the magnitude spectrum), trading some noise resilience for robustness against
phase distortion and channel impairments. Finally, quadrature detection offers
a computationally efficient alternative by approximating frequency demodulation
through phase differentiation, effectively transforming the FSK problem into an
ASK-like demodulation task, albeit with suboptimal performance due to the resulting
non-Gaussian noise~\cite{mmp-itpdt-19}.

We focus on non-coherent demodulation because collision scenarios in LoRa networks
introduce unpredictable phase distortions (\eg from overlapping chirps), which
severely degrade the performance of coherent demodulation (due to its reliance
on phase alignment) and quadrature demodulation (due to its sensitivity to phase
discontinuities).

\subsection{Frame Detection and Synchronization}\label{subsec:frame_detection}

The LoRa receiver performs frame detection using the preamble, which consists
of repeating upchirps, a synchronization word and a short downchirp segment.
The upchirp sequence is primarily used for packet detection and coarse
synchronization, while the downchirp sequence provides additional structure for compensation of
sample time offset (STO) and carrier frequency offset (CFO).

Carrier frequency offset (CFO) arises from oscillator mismatches between
transmitter and receiver. With a typical oscillator tolerance of
10\,ppm~\cite{s-anlmc-17} at 868\,MHz, CFO in practical LoRa systems is in the
order of several kHz, up to approximately 8.68\,kHz.

Frame detection is based on the property that dechirped upchirps produce a
stable spectral peak even under moderate misalignment of the observation
window. As a result, consecutive demodulation windows over the preamble yield
consistent peak locations, which vary linearly with the combined effect of
sample time offset (STO) and carrier frequency offset (CFO). This consistency
enables reliable frame detection through repeated peak observations. 

In the absence of CFO and STO, upchirp-induced ($s_{up}$) and downchirp-induced peaks ($s_{down}$) align
at zero. STO shifts the peaks in opposite directions, while CFO shifts both in the
same direction. This enables STO and CFO estimation as:
$$
\text{CFO} = \frac{s_{up} + s_{down}}{2},\quad \text{STO} = \frac{s_{up} - s_{down}}{2}
$$

This information is jointly used to align the demodulation window in time and frequency
with the frame symbols, enabling subsequent demodulation.

Fractional CFO and fractional STO components introduce spectral leakage (\ie scalloping loss)
and peak distortion due to phase discrepancies in the clipped complex waveforms~(\autoref{eq:dechirp}),
which reduce the effective symbol signal-to-noise ratio (SNR) and impair demodulation performance.

\subsection{Frame Detection Under Collisions}\label{subsec:collisions}

Simultaneous transmissions of multiple LoRa frames with identical
spreading factors and bandwidth cause temporal overlaps of symbols, \ie 
packet collisions.

The spectrum of a dechirped symbol affected by a collision comprises a complete
waveform representing the desired symbol, superimposed with one or more clipped
waveforms generated by the interfering symbols. Therefore, the discrete Fourier
transform (DFT) of this signal yields a sharp peak at the true symbol value
together with sinc-shaped peaks (Dirichlet kernels) from collision artifacts, and when those
artifacts carry enough energy they can surpass the true peak magnitude, leading
to erroneous symbol classification.

This behavior significantly impacts packet detection. Because frame detection
relies on the consistent identification of identical symbol indices across consecutive
demodulation windows, the presence of interference peaks disrupts
this consistency. Consequently, detection may fail entirely, rendering the
standard LoRa synchronization method ineffective in collision scenarios.

\subsection{Related Work}

LoRa frame detection under collisions has been extensively studied.
Cross-correlation methods detect preambles using reference chirps but degrade
under dense collision because of interference-induced correlation
peaks~\cite{xzg-fpdlt-19, txw-cemrl-20,zlj-pcsoe-25,cwh-mlfhi-24}.
CIC~\cite{spcbk-cicdm-21} mitigates this problem by employing cross-correlation
using the shorter downchirp segment of the LoRa frame for detection (\ie the
conjugate chirp segment), but its effectiveness is limited by residual
interference~\cite{sgiv-udano-22} that is not fully suppressed in the
transformed domain.

Peak tracking approaches, such as TnB~\cite{rz-trclb-22} and Pyramid~\cite{wtxx-rcltb-24},
exploit spectral consistency across symbols but become unreliable under low SNR
and strong interference. 

Successive interference cancellation~\cite{gmgcg-iclgi-21,pb-imdgd-22} improves robustness by iteratively
removing dominant signals, yet depends on correct prior demodulation and is
sensitive to early errors.

Xhonneux~\etal~\cite{xtbba-mtrlc-22} propose a two-user receiver that
multiplies consecutive symbol spectra to suppress collision artifacts and
relies on a finite-state machine to detect collisions. However, this design is
limited to two concurrent transmissions, restricting applicability in dense
LoRa deployments.

In addition, protocol-level approaches focus on collision avoidance and medium
access scheduling at the MAC layer~\cite{da-dspsl-24}, whereas this work targets physical-layer
synchronization under concurrent transmissions.

Existing methods address synchronization under low SNR conditions.
Grid-search~\cite{armt-lssde-22} of synchronization parameters improves
robustness in adverse conditions but typically incur high computational
complexity due to exhaustive hypothesis evaluation.
Savaux~\etal~\cite{sds-tslsa-22} reduce this cost through structured search
strategies that limit the explored parameter space while maintaining
near-optimal performance. Motivated by these ideas, we adopt a restricted
grid-search–style refinement only in the initial synchronization stage to
improve robustness without incurring full exhaustive-search complexity.

Hou~\etal~\cite{hxz-dmwpb-22} introduce MALoRa, which coherently
combines weak signals across multiple antennas via antenna diversity,
improving reception under severe channel conditions. In parallel, several works
focus on low-complexity receiver
design~\cite{wwz-pdpll-24,bdd-lclfs-20,xabl-llsar-22,vc-spdsl-22}.

\section{LZn design}\label{sec:core}

\subsection{Overview}

\begin{figure}[htbp]
    \centering
    \begin{tikzpicture}[
    box/.style={
        draw,
        rounded corners,
        align=center,
        font=\small,
        inner sep=3pt,
        minimum width=2cm,
        minimum height=1cm,
        fill=#1
    },
    boxsmall/.style={
        draw,
        rounded corners,
        align=center,
        font=\small,
        inner sep=3pt,
        minimum width=2.5cm,
        fill=#1
    },
    arrow/.style={->, thick}
]

\node[boxsmall=white] (input) {Received signal $r[n]$};
\node[box=blue!15, below=0.6cm of input] (coarse) {Coarse sync};
\node[box=blue!30, right=0.6cm of coarse] (fine) {Fine sync};
\node[box=cyan!20, right=0.6cm of fine] (valid) {Sync Word\\Validation};
\node[boxsmall=white, below=1cm of valid] (out) {Frame demodulation};

\draw[arrow] (input) -- (coarse);
\draw[arrow] (coarse) -- (fine);
\draw[arrow] (fine) -- (valid);
\draw[arrow] (valid) -- (out) node [pos=0.5, xshift=0.7cm, yshift=-0.1cm] {$(\hat{T_p}, \hat{F_p})$};
\node[draw, dashed, rounded corners, inner sep=6pt,
      fit=(coarse)(fine)(valid),
      label={[xshift=2cm] above:{\small \textbf{Proposed Method}}}] (methodbox) {};

\end{tikzpicture}
    \caption{Block diagram of the proposed pipeline: coarse synchronization
    via spectral intersection over a grid of time-and-frequency-aligned hypotheses,
    followed by hypotheses refinement (fine synchronization) and sync word
    validation.}
    \label{fig:sys_arch}
\end{figure}

Let $r[n]$ denote the received complex discrete-time baseband signal
sampled at the Nyquist rate $f_s=B$, with $B$ the LoRa bandwidth. The received signal consists
of $P$ potentially overlapping frames. We assume additive white Gaussian noise
(AWGN), which serves as the baseline noise model for the subsequent analytical
development~(\autoref{sec:analytical}).

\begin{equation}
r[n] = \sum_{p=0}^{P-1} s_{p}[n  - T_p] e^{2 \pi F_{p} \frac{n}{M}} + w[n],
\end{equation}

where:
\begin{itemize}
    \item $s_p[n]$ is the $p$-th transmitted frame
    \item $T_p$ is the start index of frame $p$ (potentially fractional)
    \item $F_p$ is the carrier frequency offset of frame $p$ in bins.
    \item $M=2^{\textit{SF}}$ is the number of samples per symbol
    \item $w[n]$ is additive white Gaussian noise (AWGN)
\end{itemize}

The synchronization problem consists of estimating the set of frame indices
and carrier frequency offsets (CFO)
$$
\left\{\left(\hat{T_p}, \hat{F_p} \right)\right\}_{p=0}^{P-1} .
$$

This formulation accommodates concurrent transmissions, as the sum allows
frames to partially overlap.
In practice, the signal can be sampled at a higher rate, then
filtered and decimated to reduce quantization noise, enhancing SNR. However, all
subsequent computations and synchronization operations are performed at the
Nyquist sampling frequency, which suffices to capture the essential signal
structure.

We propose the \our synchronization pipeline~(\autoref{fig:sys_arch}), which
continuously processes the received signal $r[n]$ to detect and demodulate
LoRa frames in a streamlined fashion.
This pipeline enables real-time processing, while
achieving reliable frame detection even in challenging conditions such
as very low Signal-to-Noise Ratio (SNR) or high collision rates.

The process is structured as follows. The received signal is processed
incrementally, advancing $M$ samples, where $M=2^{\text{SF}}$ denotes the number of
samples per symbol and SF is the spreading factor.
During the coarse synchronization
stage, the method generates initial
hypotheses for a potential frame start $T_p$ and a carrier frequency offset $F_p$.
This step leverages spectral intersection across multiple demodulation window
alignments and frequency offsets -- inspired by prior work~\cite{xtbba-mtrlc-22,armt-lssde-22}.
This technique effectively reduces the collision-induced
peaks and mitigates the effect of fractional STO and fractional CFO. Next, the
hypotheses undergo a collision-aware fine synchronization
step, which corrects STO and CFO to retrieve the synchronization parameters
$T_p$ and $f_p$. This step enables reliable synchronization even in challenging
scenarios such as low Signal-to-Noise Ratio (SNR) conditions or high traffic
scenarios. Finally, the refined hypotheses are validated using their sync word,
ensuring both false-positive rejection and enabling multiplexing capabilities
-- such as those required by private or public LoRaWAN networks. Validated
frames are then demodulated using the estimated parameter $(\hat{T_p}, \hat{F_p})$ to recover the transmitted data.

\subsection{Coarse Synchronization}\label{sec:coarse_sync}

Traditional LoRa preamble detection relies on symbol-by-symbol demodulation to
identify sequences of identical symbol values. Each symbol is demodulated
independently by selecting the maximum spectral peak via an \textit{argmax}
operation. While simple, this approach inherently assumes a single dominant
candidate peak per symbol and thus cannot distinguish multiple overlapping
preambles. Under collisions, weaker preamble peaks may be masked
by stronger interference peaks or be discarded during the single-peak selection~process. 

To overcome these limitations, our method exploits the temporal
consistency of preamble peaks. Instead of independently selecting one candidate
per symbol, we construct a spectral representation by performing spectral
intersection across consecutive symbols, preserving multiple peaks that remain
consistent over time.
This operation reduces the influence of collision-induced distortions by
retaining peaks that persist over time while suppressing inconsistent
interference components. By identifying each persistent peak independently,
\our can detect multiple overlapping preambles within the same observation
window. The capability of identifying multiple simultaneous preamble peaks from
this representation is evaluated in~\autoref{subsec:timing}.

Prior spectral intersection approaches based on
multiplication~\cite{xtbba-mtrlc-22} suffer from masking. When two or more
peaks have similar magnitudes (\eg on sub-symbol-offset collisions),
one peak can dominate and mask the others. This leads to detection misses. 
Our spectral intersection scheme addresses this problem by taking the
element-wise minimum across $N_p-1$ consecutive symbols, with $N_p$ the number
of reference upchirps in the preamble. This operation
preserves the true peak (which remains consistent across symbols) while
suppressing noise or collision-induced peaks, which rarely align across multiple
symbols. This approach is theoretically justified in \autoref{sec:analytical},
in which the filter not only effectively suppresses collision artifacts, but also
enhances the signal-to-noise ratio (SNR) by attenuating uncorrelated
interference.

We use $N_p-1$ symbols (rather than $N_p$) to ensure
full alignment of the preamble peaks, but avoid partial symbols at the edges of
misaligned frames, which enhances robustness to collisions and noise.
To further improve the effectiveness of the spectral intersection, we adjust
the demodulation windows using a fractional grid for timing and frequency offsets

\[
\Delta = \left\{ \left\{ \frac{k}{2 N_{\delta}} \right\} \;\middle|\; k \in \{0, \dots, N_{\delta}-1\} \right\}\\
\]
\[
\mathcal{F} = \left\{ \frac{k}{2 N_{f}} \;\middle|\; k \in \{0, \dots, N_{\mathcal{F}}-1\} \right\}
\]

These grids allow for window adjustments by up to half a bin -- \ie worst
case alignment -- in both delay and frequency, mitigating residual fractional
delay and fractional frequency, such as scalloping loss or phase distortion.

\begin{figure}[htbp]
    \centering
    \begin{tikzpicture}
\begin{groupplot}[
group style={
  group size=2 by 2,
  horizontal sep=0.35cm,
  vertical sep=0.3cm,
  ylabels at=edge left,
  xlabels at=edge bottom,
  yticklabels at=edge left,
  xticklabels at=edge bottom,
},
height=0.15\textwidth,
width=0.25\textwidth,
axis on top,
xmin=0, ymin=0,
xmax = 256, ymax=1.2,
ytick = {0,0.5, 1},
xlabel = {Frequency bin [\#]},
legend cell align={left},
legend style={
  fill opacity=0.8,
  draw opacity=1,
  text opacity=1,
  at={(0.97,0.05)},
  nodes={scale=0.8, transform shape},
  anchor=south east,
  draw=white!80!black,
},
tick align=outside,
tick pos=left,
x grid style={white!69.0196078431373!black},
xtick style={color=black},
xtick={0,128,256},
y grid style={white!69.0196078431373!black},
ytick style={color=black},
cycle list name=exotic,
]
\nextgroupplot[title={$f=0$}]
\addplot [thick] table [col sep=comma, x index=0, y index=1] {data/spec_leak_mit.csv};
\nextgroupplot[title={$f=\frac{1}{4}$}]
\addplot [thick] table [col sep=comma, x index=0, y index=2] {data/spec_leak_mit.csv};
\nextgroupplot[]
\addplot [thick] table [col sep=comma, x index=0, y index=3] {data/spec_leak_mit.csv};
\nextgroupplot[]
\addplot [thick] table [col sep=comma, x index=0, y index=4] {data/spec_leak_mit.csv};
\end{groupplot}
    \node[anchor=west] at ([xshift=0.1cm]group c2r1.east) {\scriptsize $\delta=0$};
    \node[anchor=west] at ([xshift=0.1cm]group c2r2.east) {\scriptsize $\delta=\frac{1}{4}$};

\path (group c1r1.north west) -- (group c1r2.south west) node[pos=0.5, xshift=-1cm, rotate=90, anchor=south] {Magnitude [--]};
\end{tikzpicture}
    \caption{Spectral intersection result ($Z_{\delta,f}[k]$) for a grid with
    $N_{\delta}=2$ and $N_f=2$ (four windows), illustrating how window alignment
    affects peak prominence. In this example, the peak
    achieves its maximum at $\delta=f=\frac{1}{4}$.}
    \label{fig:spectral_int}
\end{figure}

Formally, if $\textbf{Y}_{\delta,f}$ denotes the \textit{i-th} symbol window ($i \in
\{0,\dots,N_p-2\}$) extracted from the received signal -- shifted by a
small fractional delay $\delta$ and frequency-compensated by a small fractional
CFO $f$ -- the spectral intersection of $L$ consecutive symbols is given by  

\begin{equation}\label{eq:spectral_int}
    Z_{\delta,f} = \min_{i=0}^{L-1} \left| \text{DFT}(\mathbf{Y_{\delta,f}}) \right|.
\end{equation}

The fractional parameters $\delta$ and $f$ adjust the alignment of the demodulation
window to compensate for spectral leakage caused by misalignment due to
asynchronous frame transmission. While the preamble peak may occasionally align
by chance, most frames arrive with unknown timing and frequency offset. By
sweeping $\delta$ and $f$ over a small range, the method actively sharpens the
preamble peak after spectral intersection and ensures its prominence even when
misaligned. As illustrated in \autoref{fig:spectral_int}, the parameter sweep
significantly improves peak clarity. This approach enhances the reliability of
frame detection. However, it introduces a trade-off in computational complexity due
to the increased number of demodulations~(evaluated in~\autoref{sec:evaluation}
and further analyzed in~\autoref{sec:complexity}). Note that fractional delays
and frequency offsets are not interchangeable. For example,
$(\delta, f) = (0,\frac{1}{2})$ induces different phase distortions than
$(\delta, f) = (\frac{1}{2},0)$. Hence, both dimensions are swept
independently.

For each combination of fractional delay $\delta$ and fractional frequency $f$,
we generate preamble hypotheses by detecting peaks in the corresponding signal
$Z_{\delta, f}$. Peaks corresponding to the same physical preamble may appear
multiple times due to neighbouring $(\delta, f)$ hypotheses. Therefore, we
consolidate only these redundant detections by retaining the $(\delta, f)$
configuration that yields the maximum magnitude for each peak candidate. This
step only merges redundant representations of the same spectral peak, allowing
multiple simultaneous preambles to remain as separate candidates. The resulting
set is further refined using Non-Maximum Suppression (NMS) with a 3-bin window.
NMS is applied solely to suppress local side lobes around each detected peak
and is not intended to separate preamble peaks, of which the main lobes overlap within
the suppression window. Since NMS operates strictly locally, stronger peaks
only affect candidates within the suppression window. This prevents global
suppression of weaker candidates and allows preambles of unequal received
powers to remain as independent candidates.

In the final hypothesis pruning step, a short signal around each detected unique
 peak is extracted from the $Z_{\delta, f}$ at the location of the
maximum amplitude. Each segment is then independently cross-correlated with a
bank of preamble peak templates. Cross correlation is selected as a validation
metric due to its robustness in additive white Gaussian noise (AWGN). It
effectively reduces to a matched-filter detector, which is optimal in AWGN
scenarios~\cite{ps-dc-01}, and therefore enables reliable discrimination of
hypotheses. Since each candidate is evaluated independently, the correlation
score of a peak is unaffected by the magnitude or presence of other peaks,
allowing weaker candidates to be retained under power imbalance. Only peaks
whose correlation with at least one of the templates exceeds a predefined
threshold are retained, ensuring consistency with the expected preamble peak
structure.
Each template is the magnitude of a Dirichlet kernel with width $M$, which captures the DFT peak
structure at an arbitrary (including fractional) frequency bin~\cite{wtxx-rcltb-24}.
$$D_M\left(\frac{s}{4 \cdot \max(N_\delta, N_f)} - \frac{n}{M}\right), n \in
\left\{-\left\lfloor \frac{K}{2}\right\rfloor,\dots,\left\lfloor \frac{K}{2}\right\rfloor\right\},$$ where $K=15$ samples and $
{-2 \leq s \leq 2}$ introduce fractional shifts.
We construct five templates centered at the nominal peak ($s=0$). These templates are
symmetrically shifted to account for residual misalignments not resolved by the
discrete $(\delta, f)$ grid.
These shifts correspond to $\pm \frac{s}{2 \cdot \max(N_\delta, N_f)}$ and
capture sub-grid offsets in both time and frequency.
Empirical analysis of the Dirichlet kernel autocorrelation reveals
that shifts below $\frac{1}{8}$ -- achievable with $N_\delta \geq 2$ or
$N_f \geq 2$ -- reduce the correlation score by less than $0.03$ with $K=15$.
Therefore, five templates are sufficient to cover relevant misalignments without
significant computational overhead. 

\autoref{alg:coarse_sync} summarizes the proposed coarse detection.
For readability, we define the function
$\mathrm{DechirpBlock}(r, \delta, f, L)$, which extracts $L$ consecutive
demodulation windows from the received signal $r$, starting at a (possible)
fractional index $\delta$, applies frequency correction $f$ via complex phasor
multiplication, and performs per-symbol dechirping. Here, DFT denotes the
computation of the full spectrum, while $DFT_{\mathcal{K}}$ restricts computation
to the bins  in the set $\mathcal{K}$.

The $\mathrm{PeakDetection}$ function uses a Modified Z-Score-based 
outlier selection scheme~\cite{ih-hdho-93} (MAD threshold $\sigma=3.5$) to identify prominent peaks
and reject noise. To minimize complexity, we first check if
$Z_{\delta, f}$ contains a prominent spectral peak. A window is deemed
prominent if the max-to-median ratio exceeds a threshold $\rho = 5$. This
threshold is derived analytically in~\autoref{sec:evaluation} and chosen
conservatively to discard  noise-dominated windows while preserving valid
preamble windows across different communication scenarios. Peak detection is
then performed on the remaining windows by identifying all local maxima and
minima (zero crossing), computing peak prominences using adjacent valleys,
and filtering outliers.

\begin{algorithm}
\caption{Coarse Synchronization}\label{alg:coarse_sync}
\begin{algorithmic}[1]

\Require Received signal $r$,delay set $\Delta$, frequency set $\mathcal{F}$, samples/symbol $M$, template bank $\tau$, current sample $j$,
\Ensure Hypothesis set $\mathcal{H}$

\State $\mathcal{C} \gets \emptyset$, $\mathcal{U} \gets \emptyset$
\State $K \gets 15,\quad K_c \gets \lfloor K/2 \rfloor$
\For{each $\delta \in \Delta$}
\For{each $f \in \mathcal{F}$}
    \State $\tilde{\delta} = j + \delta - (N_p - 1) \cdot M$ \label{line:spec_start}
    \State $\mathbf{Y}_{\delta,f} \gets \text{DechirpBlock}(r, \tilde{\delta}, f, N_p-1)$
    \State $\mathbf{X}_{\delta,f} \gets \textsc{DFT}(\mathbf{Y}_{\delta,f})$
    \State $\mathbf{A}_{\delta,f} \gets |\mathbf{X}_{\delta,f}|$ \label{line:spec_end}
    \Statex \textit{Note: $\mathbf{A}_{\delta,f}$ can be cached for reuse across iterations}
    \State $Z_{\delta,f}[k] \gets \min_{i=0}^{N_p-2} A_{\delta,f}[i,k]$ \label{line:spec_int}

    \State $\mathcal{P}_{\delta,f} \gets \text{PeakDetection}(\mathbf{Z}_{\delta,f})$ \label{line:peak_detection}

    \For{each $p \in \mathcal{P}_{\delta,f}$}
        \State $\mathcal{C} \gets \mathcal{C} \cup \{(p,\delta,f)\}$ \label{line:add_candidates}
    \EndFor

\EndFor
\EndFor

    \For{each unique peak index $p$}\label{line:nms_start}
    \State $(\delta^\ast,f^\ast) \gets \arg\max_{\delta,f} Z_{\delta,f}[p]$ \label{line:argmax}
    \State $\mathcal{U} \gets \mathcal{U} \cup \{(p, \delta^\ast, f^\ast)\}$ \Comment{Add peak and its coordinates to the set}
\EndFor

    \State $\mathcal{U} \gets \text{NonMaxSuppression}(\mathcal{U}, 3)$ \Comment{Filter sidelobes with 3-bin NMS}\label{line:nms}

\For{each unique peak index $p\in \mathcal{U}$}
    \State $\mathbf{z}_p[n] \gets Z_{\delta^\ast, f^\ast}[(p + n - K_c)\ \mathrm{mod}\ M]$ \label{line:extract}
    \Statex \hspace{\algorithmicindent} \hspace{\algorithmicindent} \hspace{\algorithmicindent}$n = 0,\dots,K-1$

    \State $\rho_{\text{cc}} \gets \displaystyle \max_{t \in \mathcal{T}} \big| \langle \mathbf{z}_p, t \rangle \big|$ \label{line:xcorr}

    \If{Correlation score exceeds threshold}
        \State $\hat{\ell}_p \gets \tilde{\delta} + p + \delta^\ast - M$ \label{line:loc_calc}
        \State $\mathcal{H} \gets \mathcal{H} \cup \{(\hat{\ell}_p, f^\ast)\}$ \label{line:emit_coarse}
    \EndIf

\EndFor

\State \Return $\mathcal{H}$

\end{algorithmic}
\end{algorithm}

The algorithm starts by generating candidate hypotheses for potential preamble
start indices and frequency offsets. For each combination of fractional delay
$\delta$ and fractional frequency $f$ from the predefined grid, the magnitude
spectrum of the last $N_p-1$ is computed
(\autoref{line:spec_start}--\autoref{line:spec_end}) and then
intersected~(\autoref{line:spec_int}). Peaks detected using the max-to-median
ratio and Modified Z-score outlier selection~(\autoref{line:peak_detection}) with
coordinates $(\delta, f)$ are added to the candidate set $\mathcal{C}$ (\autoref{line:add_candidates}).

After merging redundant peaks and applying Non-Maximum Suppression~(\autoref{line:nms_start} -- \autoref{line:nms}), the most prominent peaks are cross-correlated with templates~(\autoref{line:xcorr}).

Validated peaks, along with the fractional frequency $f$, are used to derive coarse estimations
of $T_p$ and $f_p$. The frame start is estimated relative to the demodulation window
coordinates and the detected peak location, with an adjustment for the spectral
intersection step. The initial CFO is taken from the $f$ parameter associated to the peak.
The resulting hypotheses are added to the set $\mathcal{H}$~\autoref{line:emit_coarse} for
further refinement in the fine synchronization stage.

\subsection{Fine Synchronization}\label{subsec:fine_sync}

The fine synchronization stage -- implemented in \autoref{alg:fine_sync} -- refines the estimates of $T_p$ and $f_p$
generated during coarse synchronization through four key steps: \one coarse
refinement, which performs coarse estimation of STO and CFO using the
upchirp-induced and downchirp-induced peak locations detected via spectral
intersection; \two fractional CFO compensation achieved by analyzing
phase variations between upchirp peaks across symbols; \three fractional STO
correction performed using a high resolution spectrum; \four a final residual
integer correction step to address any residual drifts from earlier steps.
Unlike prior methods that require expensive, iterative fine synchronization
(\eg exhaustive search or full padded-DFTs for each symbol), our approach
reduces complexity by restricting processing on narrow frequency regions using
the Goertzel algorithm~\cite{g-eagwm-69} for fractional CFO compensation and a
single Zoom FFT~\cite{rsr-cza-69} per frame for STO refinement. In addition, redundant spectral
computations are avoided by coherently summing identical symbols in the time
domain, followed by a one-shot spectrum evaluation.

\paragraph{Coarse Refinement} After coarse synchronization, the integer STO and CFO
components remain uncompensated. This can cause an offset of the true start of the packet
by up to \mbox{$f_{max} = M \cdot \frac{\text{MaxCFO}}{B}$} samples
(up to 7\% of symbol duration at 125\,kHz bandwidth in the EU868 band).

We advance the demodulation windows by $\frac{M}{8}$ samples~(\autoref{line:isi}) to
account for this deviation, ensuring at least two consecutive demodulation
windows fully overlap the downchirp segment (2.25 symbols) and mitigating inter-symbol interference (ISI).
We then compensate for STO and CFO using
the relative locations of the upchirp-induced peak -- placed at $\frac{M}{8}$
after the advancement -- and the downchirp-induced peak
(see~\autoref{subsec:frame_detection}).

\begin{algorithm}
\caption{Fine Synchronization}\label{alg:fine_sync}
\begin{algorithmic}[1]
    \Require Hypothesis set $\mathcal{H}$, received signal $y$, samples per symbol $M$, preamble length $N_p$
    \Ensure Synchronization parameters $T_p$ and $f_p$
	\State $\mathcal{K}_{\text{downchirp}} \gets \{n, -\frac{M}{8}-2 f_{max} \leq n \leq -\frac{M}{8} + 2 f_{max}\}$\label{line:downchirp_search}
\State $\mathcal{K}_0 \gets \{n, -2 \leq n \leq 2\}$
    \State $\mathcal{K}_{\text{highres}} \gets \{\frac{n}{8}, -16 \leq n \leq 16 \}$ 
    \State $\mathcal{K}_{\text{corr}} \gets \{n, -15 \leq n \leq 15 \}$
    \State $\text{SW}_\text{align} \gets [0, SW_{\text{LSB}} - 1,  SW_{\text{MSB - 1}}]$

\For{each hypothesis $h \in \mathcal{H}$}
    \Statex \textbf{Step I: Coarse Refinement}
    \State $\tilde{\delta} \gets h.\hat{\ell} + (N_p + 2) \cdot M + \frac{M}{8}$ \label{line:isi}
    \State $\mathbf{Y}_{h} \gets \text{DechirpBlock}(r, \tilde{\delta}, h.\hat{f}, 2)$ \label{line:dc_start}
	\State $\mathbf{X}_{h} \gets \text{DFT}_{\mathcal{K}_{\text{downchirp}}}(\mathbf{Y}_h)$
    \State $Z_{h}[k] \gets \min_{i=0}^{1} |Z_{h}[i,k]|$ \label{line:spec_dc_int}
    \State $s_{\text{up}} \gets \frac{M}{8}$, $s_{\text{down}}^\ast \gets \arg\max_{k \in K_{\text{downchirp}}}Z_h[k]$ 
    \State Update $h.\hat{\ell}$ and $h.\hat{f}$ based on peak locations~(\S~\ref{subsec:frame_detection})

    \Statex \textbf{Step II: Fractional CFO Compensation}
    \State $\mathbf{Y}_{p} \gets \text{DechirpBlock}(r, h.\hat{\ell}, h.\hat{f}, N_p)$
    \State $\mathbf{X}_{p} \gets \text{DFT}_{\mathcal{K}_0}(\mathbf{Y}_p)$ \Comment Goertzel\label{line:goertzel_1}
    \State $Z_{p}[k] \gets \min_{i=0}^{N_p-1} |Z_{p}[i,k]|$ 
    \State $s_{up}^\ast \gets \arg\max_{k \in \mathcal{K}_0} \mathbf{Z}_p$

    \State $X^\ast[i] \gets X_p[i, s_{up}^\ast], \quad i=0,\dots,N_p$ 
    \State $\Delta_{\phi}[i] \gets \angle\!\left(X^\ast[i+1]\;{X^\ast}^{*}[i]\right)$, \quad $i=0,\dots,N_p-1$ \label{line:phase_diff}
    \State $\mathcal{C}_\phi \gets \text{OutlierRejection}(\mathbf{\Delta_\phi})$ \Comment By modified Z-Score
    \State $f_{\text{frac}} \gets \frac{1}{2\pi} \text{mean}(\mathcal{C}_\phi)$
    \State $\mathbf{X}_{p}[i] \gets \mathbf{X}_{p}[i] \cdot e^{-j2\pi f_{\text{frac}} i}$, \quad $i=0,\dots,N_p-1$ \label{line:mean}
    \State $h.\hat{f} \gets h.\hat{f} - f_{\text{frac}}$

    \Statex \textbf{Step III: Fractional STO Compensation}
    \State $X_{\text{agg}} \gets \sum_{i=0}^{N_p} X_p[i,k]$ \Comment Coherent sum\label{line:agg_2}
    \State $\mathbf{X_{\text{zoom}}} \gets \text{DFT}_{\mathcal{K}_{\text{highres}}}(X_{\text{agg}})$ \Comment Zoom FFT\label{line:zoom_fft}
	\State $\delta_{frac}^\ast \gets \arg\max |\mathbf{X_{\text{zoom}}}|$
    \State $h.\hat{\ell} \gets h.\hat{\ell} - \delta_{\text{frac}}^\ast $ \Comment{Fractional STO,  CFO correction}

    \Statex \textbf{Step IV: Residual Integer STO/CFO Compensation}
    \State $\mathbf{Y}_{p} \gets \text{DechirpBlock}(y, h.\hat{\ell}, h.\hat{f}, N_p)$
    \State $\mathbf{Y}_{d} \gets \text{DechirpBlock}(y, h.\hat{\ell} + 10 \cdot M, h.\hat{f}, 2)$
    \State $X_{\text{agg,p}} \gets \sum_{i=0}^{N_p-1} X_p[i,k]$ \Comment Coherent sum
    \State $X_{\text{agg,d}} \gets \sum_{i=0}^{1} X_d[i,k]$ \Comment Coherent sum
    \State $Z_{p}[k] \gets |\textsc{DFT}_{K_0}(\mathbf{Y}_p)[k]|$
    \State $Z_{d}[k] \gets |\textsc{DFT}_{K_0}(\mathbf{Y}_d)[k]|$
    \State $s_{\text{up}}^\ast \gets \arg\max_{k} Z_p[k]$, $s_{\text{down}}^\ast \gets \arg\max_{k} Z_d[k]$
    \State Update $h.\hat{\ell}$ and $h.\hat{f}$ based on peak locations~(\S~\ref{subsec:frame_detection}\label{line:integer_sto_cfo})
    \State \Return $(h.\hat{\ell}, h.\hat{f})$
    \EndFor

\end{algorithmic}
\end{algorithm}

Spectral intersection across the two windows is applied to sharpen the
peak~(\autoref{line:dc_start} -- \autoref{line:spec_dc_int}) and the method
finds the downchirp-induced peaks using simple search (\textit{argmax}).
Because the duration of the downchirp segment is short, collisions with other
downchirp segments are unlikely, which motivates this search strategy.
To further mitigate interference, we restrict the search to the region where
the peak is expected to lie (\autoref{line:downchirp_search}), as established
in previous work~\cite{rz-trclb-22}.

However, other overlapping chirps may still distort the downchirp-induced peak
within the search region, potentially leading to incorrect CFO and STO estimates whose residual misalignment reduces the effective SNR during demodulation.
After compensation, the resulting frame undergoes fractional CFO and STO refinement.
The method first compensates the fractional carrier frequency offset, enabling
coherent summation of the signal. This allows computing the spectrum 
from the aggregated signal -- effectively cancelling out noise -- without requiring
a separate DFT for each preamble symbol.

\paragraph{Fractional CFO Compensation}
The spectrum of the compensated frame is first computed using the Goertzel
algorithm~(\autoref{line:goertzel_1}), which efficiently evaluates only the bins around
zero ($\{-2,\cdots,2\}$). This region is chosen to account for any potential
drifts that may arise during coarse refinement. By focusing on this narrow
range, the method significantly reduces the computational complexity compared
to a full FFT.

Following an approach first introduced in~\cite{bdd-lclfs-20}, we compensate
fractional CFO using the phase difference between consecutive upchirp-induced
peaks~(\autoref{line:phase_diff}), which is linearly related to
the CFO.
While averaging these phase differences yields an accurate estimate in noisy
conditions, collision artifacts can introduce outliers that distort this
estimation. To mitigate this, potential outliers are discarded using
Modified Z-Score prior to averaging~(\autoref{line:mean}).

\paragraph{Fractional STO Compensation} This method assumes that any remaining
sub-bin shift in the peak, which manifests as spectral leakage, is solely
due to the fractional STO, as fractional CFO has already been compensated.

The aggregated preamble spectrum~(\autoref{line:agg_2}), is analyzed at a
high resolution ($8\times$) using Zoom FFT around the
expected peak location~(\autoref{line:zoom_fft}).
This focused approach avoids the computation cost of a full padded-DFT
calculation, while precisely identifying the sub-sample location of the peak
maximum. The result is then mapped to the fractional sample time offset \ie the
shift required to optimally align the preamble with its true start.

\paragraph{Residual STO/CFO Compensation}
Any residual STO and CFO is corrected with the location of the upchirp-induced
and downchirp-induced peaks~(see~\autoref{subsec:frame_detection}), which are
identified from the spectrum of the aggregated signal computed with Goertzel
around the origin~(\autoref{line:integer_sto_cfo}).

\subsection{Sync Word Validation}

To validate the
sync word, the algorithm first aligns the peaks of the sync word symbols
(typically 8 and 16 for public LoRaWAN networks) to the reference upchirp
peaks, which are centered at zero after fine synchronization. This alignment --
achieved via a complex phasor multiplication, or equivalently by rotating the
signal in the frequency domain in practical implementations -- enables the
reuse of the spectral intersection step to suppress collision artifacts and
isolate the true peaks. Frames are retained only if the cross-correlation of
the signal segment around zero with the bank of templates $\tau$ -- consistent
with the coarse synchronization step -- exceeds the correlation
threshold. This
ensures that false positives or symbols with incorrect sync word patterns are
discarded, as their peaks would fail to align at the origin.



\section{Formal Analysis and Theoretical Guarantees}\label{sec:analytical}

Although the proposed pipeline comprises multiple stages, we develop a generic
model of the spectral intersection step -- the key mechanism for suppressing
inconsistent peaks across symbols. This model provides insight into the
achievable interference suppression under idealized assumptions, rather than  
the end-to-end performance of the full synchronization pipeline. As such, the model 
provides an optimistic prediction of the full pipeline (upper bound).
We first formalize the spectral intersection step using a generic signal model,
then specialize it to the single- and multi-user cases.

Let $L$ denote the number of contiguous symbols used for preamble detection, with $L \leq N_p-1$.
Consider $L$ consecutive dechirped LoRa symbols with spectrum $Y_i[k]$ ($i \in
\left\{0,\dots,L-1\right\},\, k \in \left\{0,\dots,M-1\right\}$), where $M=2^\text{SF}$ is
the number of frequency bins per symbol and SF the spreading factor. All symbols
carry the same preamble chirp and are observed under additive white Gaussian
noise with symbol energy $E_s$ and normalized noise energy $E_n=1$.
Unless stated otherwise, we assume perfect synchronization (\ie zero carrier frequency
offset and zero sample time offset) to establish an upper bound in performance.

The standard LoRa receiver generates a preamble hypothesis when the \textit{argmax}
is identical across all $L$ symbols, denoted $m$ for the proposed model.
Our method instead computes the intersected spectrum $Z[k] = \min_{i=0}^{L-1} Y_i[k]$ and identifies
the peak $m$ via peak detection. For analytical tractability, we approximate this
by an \textit{argmax} over the intersected spectrum.

Let $\mathcal{S}$ denote the event that the correct peak is successfully identified,
\ie $Z[m] > \max_{k \neq m} Z[k]$.
We model  the probability of successfully identifying the peak $m$
 by
\begin{equation}\label{eq:signal_model}
P\bigl( \mathcal{S} \bigr)
= \int_{0}^{\infty}
\underbrace{
\mathbb{P}\!\left( \max_{k \ne m} Z[k] < z \,\middle|\, Z[m] = z \right)
}_{\substack{\text{conditional} \\ \text{success probability}}}
\\
\cdot
\underbrace{
f_{Z[m]}(z)
    }_{\substack{\text{Probability Density} \\ \text{Function of } Z[m]}}
\,dz
\end{equation}

The evaluation of the success probability $P(\mathcal{S})$ requires characterizing the distribution of $Z[k]$, which
depends on the considered scenario (single-user or multi-user).

\subsection{Single-user Scenario (Collision Free)}\label{subsec:single_user}

In the single-user scenario, the aligned bin $Y_i[m]$ contains both the signal
and noise, and thus follows a Rician distribution~\cite{er-calmb-18} with
signal amplitude $\nu=\sqrt{M E_s}$, under circularly symmetric complex
Gaussian noise with variance $\frac{1}{2}$ per real dimension.
In contrast, non-aligned bins $Y_i[k], k \neq m$ contain noise only and therefore
follow Rayleigh distributions with identical variance~\cite{er-calmb-18}. The samples $Y_i[k]$
are i.i.d. (independent and identically distributed) across symbol indices $i$ for each frequency bin $k$. Moreover,
noise components are also i.i.d. across frequency bins, \ie for $k \neq m$.

Given the characterization above, $Z[k]$ is the minimum of $L$ i.i.d. realizations of $Y_i[k]$
The cumulative distribution function (CDF) of $Z[k]$ is then given by
\begin{equation}\label{eq:model_gen}
F_{Z[k]}(z)=1-\bigl(1-F_{Y_i[k]}(z)\bigr)^L
\end{equation}

Since the distribution of $Y_i[k]$ depends on whether $k=m$ or $k\neq m$, we
distinguish between signal and noise bins. Accordingly, $F_{Z[k]}(z)$ takes
different forms for the two cases.
For the noise bins (\ie $k \neq m$), $Y_i[k] \sim \mathrm{Rayleigh}(\sigma^2 = \frac{1}{2})$, yielding
$F_{Y_i[k]}(z)=1-e^{-z^2}$, whereas for the signal bin (\ie $k=m$), $Y_i[m] \sim
\mathrm{Rician}(\nu=\sqrt{M E_s}, \sigma^2 = \frac{1}{2})$, whose CDF is expressed using the Marcum $Q$-function~\cite{m-sttdp-60}, hence
\begin{equation}\label{eq:zk_cdf}
F_{Z[k]}(z)=
\begin{cases}
    F_{Z_I}(z) = 1 - e^{-L z^2}, & k \neq m,\\[6pt]
    F_{Z_m}(z) = 1 - \left[ Q_1\!\left(\sqrt{2 M E_s}, \sqrt{2}\, z \right) \right]^L, & k = m,
\end{cases}
\end{equation}
where $F_{Z_I}(z)$ and $F_{Z_m}(z)$ denote the CDFs of noise and signal bins, respectively.

Since the noise samples $Y_i[k]$ are i.i.d. across $i$ and independent across
frequency bins, the variables $\{Z[k]\}_{k \neq m}$ are mutually independent
and independent of $Z[m]$. Therefore, the conditional probability in
\autoref{eq:signal_model} becomes
\begin{equation}\label{eq:cond}
\mathbb{P}\!\left(\max_{k \ne m} Z[k] < z \,\middle|\, Z[m] = z \right)
= \prod_{k \ne m} F_{Z_I}(z)
= \left[F_{Z_I}(z)\right]^{M-1}
\end{equation}
Combining the above expressions for the noise and signal bins with \autoref{eq:signal_model}, the success probability of identifying the preamble peak becomes
\begin{equation}\label{eq:ps_final}
P(\mathcal{S})
= \int_0^\infty 
\left[F_{Z_I}(z)\right]^{M-1}
f_{Z_m}(z)
\,dz,
\end{equation}
where $f_{Z_m}(z) = \frac{d}{dz} F_{Z_m}(z)$ is the probability density function (PDF) of $Z[m]$.

\begin{figure}[ht]
    \centering
    \begin{tikzpicture}
\definecolor{L7}{RGB}{68,1,84}
\definecolor{L6}{RGB}{59,82,139}
\definecolor{L5}{RGB}{33,145,140}
\definecolor{L4}{RGB}{94,201,98}
\definecolor{L3}{RGB}{173,220,48}
\definecolor{L2}{RGB}{190,170,50}
\definecolor{L1}{RGB}{210,200,60}

\pgfplotsset{
    no_xlabels/.style={
        xticklabels={},
        xlabel={}
    },
    no_ylabels/.style={
        yticklabels={},
        ylabel={}
    },
}
\begin{groupplot}[
height=0.35\linewidth,
width=0.5\linewidth,
group style={
  group size=2 by 2,
  horizontal sep=0.35cm,
  vertical sep=0.4cm,
},
axis on top,
ymax=1, ymin=0.01,
xmax = -25, xmax=-10,
legend cell align={left},
legend style={
  fill opacity=0.8,
  draw opacity=1,
  text opacity=1,
  at={(0.05,-1.4)},
  nodes={scale=0.8, transform shape},
  anchor=south west,
  draw=white!80!black,
  legend columns=4,
},
cycle list={
    {thick}, 
},
ymode=log,
tick align=outside,
tick pos=left,
x grid style={white!69.0196078431373!black},
xtick style={color=black},
y grid style={white!69.0196078431373!black},
ytick style={color=black},
ytick={0.01,0.1,1},
yticklabels={1, 10, 100},
]
\nextgroupplot[title=\textbf{SF8}, xmin=-20, xmax=-11, no_xlabels]
\addplot+ [L1] table [x index=0, y index=1, col sep=comma] {data/model_one_user_8.csv};
\addplot+ [L2] table [x index=0, y index=2, col sep=comma] {data/model_one_user_8.csv};
\addplot+ [L3] table [x index=0, y index=3, col sep=comma] {data/model_one_user_8.csv};
\addplot+ [L4] table [x index=0, y index=4, col sep=comma] {data/model_one_user_8.csv};
\addplot+ [L5] table [x index=0, y index=5, col sep=comma] {data/model_one_user_8.csv};
\addplot+ [L6] table [x index=0, y index=6, col sep=comma] {data/model_one_user_8.csv};
\addplot+ [L7] table [x index=0, y index=7, col sep=comma] {data/model_one_user_8.csv};
\nextgroupplot[title=\textbf{SF10}, no_xlabels, no_ylabels, xmin=-27, xmax=-16]
\addplot+ [L1] table [x index=0, y index=1, col sep=comma] {data/model_one_user_10.csv};
\addplot+ [L2] table [x index=0, y index=2, col sep=comma] {data/model_one_user_10.csv};
\addplot+ [L3] table [x index=0, y index=3, col sep=comma] {data/model_one_user_10.csv};
\addplot+ [L4] table [x index=0, y index=4, col sep=comma] {data/model_one_user_10.csv};
\addplot+ [L5] table [x index=0, y index=5, col sep=comma] {data/model_one_user_10.csv};
\addplot+ [L6] table [x index=0, y index=6, col sep=comma] {data/model_one_user_10.csv};
\addplot+ [L7] table [x index=0, y index=7, col sep=comma] {data/model_one_user_10.csv};
\nextgroupplot[xmin=-20, xmax=-11]
\addplot+ [L1] table [x index=0, y index=1, col sep=comma] {data/model_analytical_mc_8.csv};
\addlegendentry{L=1}
\addplot+ [L2] table [x index=0, y index=2, col sep=comma] {data/model_analytical_mc_8.csv};
\addlegendentry{L=2}
\addplot+ [L3] table [x index=0, y index=3, col sep=comma] {data/model_analytical_mc_8.csv};
\addlegendentry{L=3}
\addplot+ [L4] table [x index=0, y index=4, col sep=comma] {data/model_analytical_mc_8.csv};
\addlegendentry{L=4}
\addplot+ [L5] table [x index=0, y index=5, col sep=comma] {data/model_analytical_mc_8.csv};
\addlegendentry{L=5}
\addplot+ [L6] table [x index=0, y index=6, col sep=comma] {data/model_analytical_mc_8.csv};
\addlegendentry{L=6}
\addplot+ [L7] table [x index=0, y index=7, col sep=comma] {data/model_analytical_mc_8.csv};
\addlegendentry{L=7}
\nextgroupplot[no_ylabels, xmin=-27, xmax=-16]
\addplot+ [L1] table [x index=0, y index=1, col sep=comma] {data/model_analytical_mc_10.csv};
\addplot+ [L2] table [x index=0, y index=2, col sep=comma] {data/model_analytical_mc_10.csv};
\addplot+ [L3] table [x index=0, y index=3, col sep=comma] {data/model_analytical_mc_10.csv};
\addplot+ [L4] table [x index=0, y index=4, col sep=comma] {data/model_analytical_mc_10.csv};
\addplot+ [L5] table [x index=0, y index=5, col sep=comma] {data/model_analytical_mc_10.csv};
\addplot+ [L6] table [x index=0, y index=6, col sep=comma] {data/model_analytical_mc_10.csv};
\addplot+ [L7] table [x index=0, y index=7, col sep=comma] {data/model_analytical_mc_10.csv};
\end{groupplot}

\node[anchor=base,rotate=-90,yshift=0.275cm] at (group c2r1.east) {\textbf{Model}};
\node[anchor=base,rotate=-90,yshift=0.275cm] at (group c2r2.east) {\textbf{Simulation}};
\path (group c1r1.north west) -- (group c1r2.south west) node[pos=0.5, xshift=-1cm, rotate=90, anchor=south] {Detection Miss Rate [\%]};
\path (group c1r2.south east) -- (group c2r2.south west) node[pos=0.5, yshift=-1cm] {SNR [dB]};
\end{tikzpicture}
    \caption{Detection miss rate versus SNR, for varying spectral intersection depth $L$
    in single-user scenarios with spreading factors 8 and 10.
    The top and bottom row show analytical predictions and Monte Carlo simulation results, respectively.
    }

    \label{fig:model_single_user}
\end{figure}

We evaluate \autoref{eq:ps_final} numerically by setting $E_s=10^{\frac{\gamma}{10}}$ and $M=2^{SF}$, where
$\gamma$ denotes the SNR. The results are shown in \autoref{fig:model_single_user} (top)
for two spreading factor configurations (8 and 10) and $L \in \left\{1,\dots,7\right\}$.
We also validate the model via Monte Carlo simulations
(\autoref{fig:model_single_user}, bottom) by generating dechirped symbols under
additive white Gaussian noise (AWGN), observing close agreement between
simulated and theoretical peak-preservation probabilities.

Our results indicate that increasing the number of intersection steps ($L$)
significantly improves the probability of preserving the correct preamble peak.
For example, achieving a 99\% peak-selection probability requires a lower SNR
as $L$ increases. Without spectral intersection ($L=1$), the required SNR for
correct peak selection is approximately $-11.7$\,dB~(\autoref{fig:model_single_user}, top left) for SF8 and
$-17.22$\,dB for SF10~(\autoref{fig:model_single_user}, top right). By intersecting $L$ windows, the effective noise is
reduced, lowering the required SNR to $-15$\,dB (SF8) and $-20.5$\,dB (SF10),
at full intersection ($L=7$).

To benchmark the proposed intersection step against the standard LoRa detector,
we consider a baseline detector that processes the $L$ consecutive symbols
independently without spectral intersection. Let \mbox{$P_d \triangleq
P(\mathcal{S})\big|_{L=1}$} denote the single-symbol demodulation probability.
Under the assumption of independent noise realizations, the probability that
all $L$ individual symbol decisions are correct is given by $(P_d)^L$. This
yields exponential decay in overall detection as $L$ increases.

Our method, in contrast, replaces independent decisions with a single-shot
detection on an aggregate signal, improving effective SNR compared to a single
symbol.

In practice, the end-to-end detection performance of \our is affected by stages
beyond the analytical model, including CFO/STO estimation during fine
synchronization. This stage operates
with a reduced intersection depth of $L=2$, resulting in lower sensitivity
compared to the idealized spectral intersection model. Under the most favorable
conditions, the analytical model predicts a 1\% detection error threshold of
approximately $-13.5$\,dB for SF8 and $-18.5$\,dB for
SF10~(\autoref{fig:model_single_user}, top left/right, $L=2$).
Correlation-based peak detection further degrades sensitivity, slightly
underperforming simple peak search (\textit{argmax}) as noise distorts the
correlation peak.

The standard LoRa detector requires higher SNR to compete.
Using the best-case single-symbol threshold (\ie $-11.7$\,dB for SF8), independent
decisions succeed in only 96\% of cases for $L=4$ symbols and $93.2$\% for $L=7$.

\subsection{Two-user Scenario}

Interfering symbols with the same spreading factor as the desired signal
produce sinc-shaped components (\ie Dirichlet kernels) in the spectrum of the dechirped
demodulation window (see \autoref{subsec:collisions}). 
This arises from the fact that the FFT is applied over a finite-length observation
window, effectively corresponding to the Fourier transform of time-limited (clipped)
complex exponentials~\cite{os-dsp-99}.
When a colliding frame is misaligned with respect to the target symbol, the
observation window captures partial contributions from adjacent interfering
symbols. The resulting received spectrum $Y[k]$ consists of the desired symbol
contribution at bin $m$ and interference from misaligned transmissions. The received spectrum can be expressed as
\begin{align}\label{eq:dirichlet}
Y[k] &= A_s \, \delta[k - m] + A_i \Big( Y_1[k] + Y_2[k] \Big)
\end{align}
where
\begin{align}
Y_1[k] &= e^{j \phi_1} D_T(\Delta_1), \nonumber \\
Y_2[k] &= e^{j (\phi_2 + 2 \pi T \Delta_2)} D_{M-T}(\Delta_2).
\end{align}

We assume the noise power is negligible compared to the
interference power and is therefore omitted.
In this formulation, the Kronecker delta models the desired symbol contribution, while the
second term captures interference (sinc-shaped peaks) from misaligned signal
segments. The amplitudes $A_s$ and $A_i$ are the target and interference
amplitudes, respectively. The function $D_T(f)$ denotes the Dirichlet kernel of
length $T$, and $\phi_{1,2}, \Delta_{1,2}$ and $T$ describe the phase offsets,
frequency offsets and segment durations, respectively.

The distributions of $Y[k]$ are identical for all noise bins ($k \neq m$) because the
interference peaks -- governed by random time and frequency offsets -- do not favor
any specific frequency bin. Each noise bin experiences interference contributions
from the same statistical process (Dirichlet kernels with random offsets), resulting
in identically distributed noise bins. For the signal bins ($k = m$), the distribution
differs because it includes both the desired signal (modeled as a deterministic
Kronecker delta) and interference. However, interference in the
signal bin follows the same statistical process as the noise bins,
ensuring identical distributions across symbols.

Despite the identical distribution, the bins $Y[k]$ are not independent across
frequency bins due to structure induced by finite-length observation of dechirped
signals (see previous discussion on the resulting Dirichlet kernel structure).
This induces spectral leakage across frequency bins, leading to correlation
between neighbouring bins. Across symbols, dependencies arise only locally when
interfering symbol overlaps two adjacent demodulation windows, contributing energy
to the same frequency bin in both observations and inducing correlation between
consecutive symbols. This effect is limited to adjacent windows, since each
symbol can span at most two windows and consecutive symbols are otherwise
independent due to LoRa symbol whitening.

As a result, $Z[k]$ -- obtained by taking the minimum across $L$ windows -- can
still be approximated using the standard order-statistics framework for the
minimum of $L$ independent realizations (\autoref{eq:model_gen}). This approximation is
justified because most symbols in the set of $L$ windows are effectively
uncorrelated, and the minimum operation naturally selects the least affected
window -- even if some adjacent windows are correlated. Nevertheless, this approximation
is optimistic: in reality, the two correlated adjacent windows may both yield
high interference values, effectively reducing the number of independent windows
competing for minimum. This raises the expected value of $Z[k]$ compared to the model
assuming $L$ fully independent windows, meaning the analysis provides a lower bound on
the true expected minimum.

Note that the minimum operation also helps decorrelate the resulting $Z[k]$ across
frequency bins, as the dominant interference contributions at each bin may
arise from different windows, most of which are independent. Additionally, the
distributions of $Z[k]$ are identically distributed across noise bins ($k \neq m$),
since the collision process -- driven by random time and frequency offsets -- does
not favor a specific frequency bin. The minimum operation preserves this
uniformity.

Following these arguments, we approximate the distributions of noise bins
$Z[k], k\neq m$ as i.i.d, and similarly for the signal bin ($k=m$). This
enables the computation of success probability using \autoref{eq:ps_final}.
Since the exact distribution of $Y[k]$ under the correlated model
is intractable, we evaluate it numerically via Monte Carlo simulations by sweeping
parameters in \autoref{eq:dirichlet}.

\begin{figure}[ht]
    \centering
    \begin{tikzpicture}
\definecolor{L7}{RGB}{68,1,84}
\definecolor{L6}{RGB}{59,82,139}
\definecolor{L5}{RGB}{33,145,140}
\definecolor{L4}{RGB}{94,201,98}
\definecolor{L3}{RGB}{173,220,48}
\definecolor{L2}{RGB}{190,170,50}
\definecolor{L1}{RGB}{210,200,60}

\pgfplotsset{
    no_xlabels/.style={
        xticklabels={},
        xlabel={}
    },
    no_ylabels/.style={
        yticklabels={},
        ylabel={}
    },
}
\begin{groupplot}[
height=0.35\linewidth,
width=0.5\linewidth,
group style={
  group size=2 by 2,
  horizontal sep=0.35cm,
  vertical sep=0.3cm,
},
axis on top,
ymax=1, ymin=0.01,
legend cell align={left},
legend style={
  fill opacity=0.8,
  draw opacity=1,
  text opacity=1,
  at={(0.05,-1.4)},
  nodes={scale=0.8, transform shape},
  anchor=south west,
  draw=white!80!black,
  legend columns=4,
},
cycle list={
    {thick}, 
},
ymode=log,
tick align=outside,
tick pos=left,
x grid style={white!69.0196078431373!black},
xtick style={color=black},
y grid style={white!69.0196078431373!black},
ytick style={color=black},
ytick={0.01,0.1,1},
yticklabels={1,10,100},
]
\nextgroupplot[title=\textbf{SF7}, xmin=-40, xmax=1, no_xlabels]
\addplot+ [L1] table [x index=0, y index=1, col sep=comma] {data/collisions.analytical.7.csv};
\addplot+ [L2] table [x index=0, y index=2, col sep=comma] {data/collisions.analytical.7.csv};
\addplot+ [L3] table [x index=0, y index=3, col sep=comma] {data/collisions.analytical.7.csv};
\addplot+ [L4] table [x index=0, y index=4, col sep=comma] {data/collisions.analytical.7.csv};
\addplot+ [L5] table [x index=0, y index=5, col sep=comma] {data/collisions.analytical.7.csv};
\addplot+ [L6] table [x index=0, y index=6, col sep=comma] {data/collisions.analytical.7.csv};
\addplot+ [L7] table [x index=0, y index=7, col sep=comma] {data/collisions.analytical.7.csv};
\nextgroupplot[title=\textbf{SF12}, xmin=-55, no_ylabels, no_xlabels]
\addplot+ [L1] table [x index=0, y index=1, col sep=comma] {data/collisions.analytical.12.csv};
\addplot+ [L2] table [x index=0, y index=2, col sep=comma] {data/collisions.analytical.12.csv};
\addplot+ [L3] table [x index=0, y index=3, col sep=comma] {data/collisions.analytical.12.csv};
\addplot+ [L4] table [x index=0, y index=4, col sep=comma] {data/collisions.analytical.12.csv};
\addplot+ [L5] table [x index=0, y index=5, col sep=comma] {data/collisions.analytical.12.csv};
\addplot+ [L6] table [x index=0, y index=6, col sep=comma] {data/collisions.analytical.12.csv};
\addplot+ [L7] table [x index=0, y index=7, col sep=comma] {data/collisions.analytical.12.csv};
\nextgroupplot[xmin=-40, xmax=1]
\addplot+ [L1] table [x index=0, y index=1, col sep=comma] {data/collisions.mc.7.csv};
\addlegendentry{L=1}
\addplot+ [L2] table [x index=0, y index=2, col sep=comma] {data/collisions.mc.7.csv};
\addlegendentry{L=2}
\addplot+ [L3] table [x index=0, y index=3, col sep=comma] {data/collisions.mc.7.csv};
\addlegendentry{L=3}
\addplot+ [L4] table [x index=0, y index=4, col sep=comma] {data/collisions.mc.7.csv};
\addlegendentry{L=4}
\addplot+ [L5] table [x index=0, y index=5, col sep=comma] {data/collisions.mc.7.csv};
\addlegendentry{L=5}
\addplot+ [L6] table [x index=0, y index=6, col sep=comma] {data/collisions.mc.7.csv};
\addlegendentry{L=6}
\addplot+ [L7] table [x index=0, y index=7, col sep=comma] {data/collisions.mc.7.csv};
\addlegendentry{L=7}
\nextgroupplot[no_ylabels, xmin=-55]
\addplot+ [L1] table [x index=0, y index=1, col sep=comma] {data/collisions.mc.12.csv};
\addplot+ [L2] table [x index=0, y index=2, col sep=comma] {data/collisions.mc.12.csv};
\addplot+ [L3] table [x index=0, y index=3, col sep=comma] {data/collisions.mc.12.csv};
\addplot+ [L4] table [x index=0, y index=4, col sep=comma] {data/collisions.mc.12.csv};
\addplot+ [L5] table [x index=0, y index=5, col sep=comma] {data/collisions.mc.12.csv};
\addplot+ [L6] table [x index=0, y index=6, col sep=comma] {data/collisions.mc.12.csv};
\addplot+ [L7] table [x index=0, y index=7, col sep=comma] {data/collisions.mc.12.csv};
\end{groupplot}
\node[anchor=base,rotate=-90,yshift=0.275cm] at (group c2r1.east) {\textbf{Model}};
\node[anchor=base,rotate=-90,yshift=0.275cm] at (group c2r2.east) {\textbf{Simulation}};
\path (group c1r1.north west) -- (group c1r2.south west) node[pos=0.5, xshift=-1cm, rotate=90, anchor=south] {Detection Miss Rate [\%]};
\path (group c1r2.south east) -- (group c2r2.south west) node[pos=0.5, yshift=-1cm] {SIR [dB]};
\end{tikzpicture}
    \caption{Detection miss rate versus Signal-to-Interference Ratio (SIR) for
	varying spectral intersection depth $L$ in two-user scenarios with
	spreading factors 7 and 12. The top and bottom rows show analytical
	predictions and Monte Carlo simulation results, respectively.
    }
    \label{fig:model_col}
\end{figure}

To evaluate the effect of inter-symbol dependencies caused by collisions, we additionally
simulate the dechirped preamble of $N_p - 1$ consecutive symbols, with $N_p=8$ the
number of reference upchirp symbols in the preamble.
In the dechirped domain, the collision is modeled as the superposition of a target
tone and an interfering FSK-like signal with random sample offset and
frequency offsets. Consequently, components corresponding to the frame
frequency appear in consecutive frequency bins due to the temporal overlap of
symbols.

We present the results in \autoref{fig:model_col}.
The model provides a lower bound of detection sensitivity (\ie 1\% detection error)
compared to the simulations. For $L \leq 2$,
the spectral intersection step cannot fully suppress interference peaks that persist
across consecutive dechirped windows, raising the noise floor and causing the
observed deviation in the simulation. For  $L > 3$, simulations converge
towards the model, though slight differences remain due to these dependencies.

The results indicate that increasing the spectral intersection depth $L$
reduces the SIR threshold for preserving the desired peak. This demonstrates the
interference suppression capability of the spectral intersection operation in
collision scenarios. According to the idealized model, the spectral
intersection step achieves a peak-preservation threshold as low as
approximately $\FetchKey{/model/collisions/analytical/7/7}$\,dB
(\autoref{fig:model_col}, top, SF7) and
$\FetchKey{/model/collisions/analytical/12/7}$\,dB for $L=7$
(\autoref{fig:model_col}, top, SF12).

As in the single-user case, the end-to-end detection sensitivity of \our is
further constrained by the STO/CFO estimation step during fine synchronization,
which employs $L=2$.
In the absence of opposite-chirp interference -- \ie for colliding signals with
the same spreading factor and identical chirp rate sign -- the simulations
indicate an SNR limit of approximately $\FetchKey{/model/collisions/mc/7/2}$\,dB (\autoref{fig:model_col}, SF7,
bottom) and $\FetchKey{/model/collisions/mc/12/2}$\,dB (\autoref{fig:model_col}, SF12, bottom) at $L=2$. Given
the short duration of the segment compared to the full frame, collisions with
downchirp sections of other frames are less likely, and most collisions involve
interfering upchirps.

Dechirping a signal with the opposite chirp rate spreads its energy across the
spectrum, so the resulting components are not fully orthogonal~\cite{sgiv-udano-22}. Specifically, dechirping a
symbol with chirp rate of opposite chirp slope produces a quadratic-phase
signal with an effective doubled chirp rate. Modelling colliding symbols as
ideal linear chirps of opposite slope yields therefore a sequence with period
$\frac{M}{2}$, the DFT of which is supported only on every second frequency bin with
equal magnitude across the active bins.

Since the total energy is normalized to $E_i=1$, Parseval's theorem implies
that this energy of a single symbol is distributed over $\frac{M}{2}$ active
bins. This yields an amplitude scaling of $\sqrt{2}$ per bin.
Assuming worst-case destructive interference at the signal bin, the minimum
detectable peak in this idealized model becomes $2 \sqrt{2}$. This corresponds to an SNR of
\mbox{$\gamma = 10\, log_{10}(\frac{(2 \sqrt{2})^2}{M})$}, which evaluates to
$\UDColThresh{7}$\,dB for SF7 and $\UDColThresh{12}$\,dB for SF12. The spectral intersection step can
further mitigate this interference. 

In the idealized model, the interference
primarily occupies even bins, while temporal misalignment across symbols
decorrelates its contribution between observations. As a result, although the
target peak may experience slight attenuation, the interference does not
combine coherently at the signal bin. Consequently, the signal peak is
preserved relative to the interference-dominated bins. The minimum operation
therefore selectively suppresses misaligned interference, explaining the
improved peak-preservation behavior observed in the analytical model compared
to single-symbol processing.

\subsection{Multi-user Scenario}

We consider a dense regime. Under standard stochastic geometry
assumptions~\cite{h-sgwn-12}, interference is modeled as independently
distributed in time and frequency, with independent fading and finite-variance
channel gains.

Due to path-loss attenuation and spatial dispersion of transmitters, 
individual interferers contribute roughly similar fractions of the aggregate
interference power~\cite{h-sgwn-12}. Consequently, the interference accumulates 
many comparable contributions of limited variance -- without dominance of any
single term. In this regime, the aggregate interference can be approximated as
a complex Gaussian random variable based on the central limit theorem, since it
arises from the sum of many independent, roughly similar contributions.

This justifies the single-user analysis in \autoref{fig:model_single_user} as 
a useful approximation for the spectral intersection operation under dense
interference conditions, where the aggregated interference approaches Gaussian
behavior. However, this analysis does not capture additional effects caused by
dominant interferers, synchronization errors or other non-Gaussian interference
structures.

\section{Evaluation}\label{sec:evaluation}

We evaluate \our against four baselines (TnB, CIC, Pyramid, and
OpenLoRa~\cite{mkskc-ovlie-23}) using simulated LoRa captures. These traces
exhibit carrier frequency offset (CFO) up to 4.88\,kHz, carry a 12-byte
payload, and employ coding rate~$\frac{4}{8}$.

TnB, CIC, and Pyramid provide collision-tolerant detection. OpenLoRa, in contrast,
represents conventional LoRa detection and is not designed to resolve overlapping
frames.

Our method is implemented in \textit{Python} and released within this work. The
implementation takes discrete baseband LoRa captures as input and outputs
packet arrival time (in samples) and CFO estimates. These estimates are used to
align demodulation windows for subsequent symbol demodulation.

\paragraph{Selection of operational parameters}
The noise rejection parameter  $\rho$ (see~\autoref{sec:coarse_sync}) is defined as the ratio of the 20th percentile of the signal-bin distribution to the median of the noise-bin distribution. We evaluate this ratio at the SNR where the full spectral intersection fails, which corresponds to $-15$\,dB for SF8. Both distributions are derived in~\autoref{sec:analytical}. Assuming AWGN, this results in a conservative threshold of $\rho \approx 5$, which avoids rejecting valid frames across the considered SNR range.
Peak and sync-word validation use a correlation threshold of 0.7, selected
empirically and stable within the range 0.65--0.75.

\subsection{Sensitivity versus Computational Complexity}\label{subsec:deltas}

We study the effect of the number of fractional delay hypotheses~$N_{\delta}$ and fractional frequency
shift hypotheses~$N_{f}$~(\autoref{sec:coarse_sync}) on the detection sensitivity.
Increasing these parameters improves the sample and frequency resolution, at the
cost of $N_{\delta} \cdot N_{f}$ FFTs per demodulation window.
To quantify this trade-off, we evaluate the detection sensitivity (\ie
99\% detection probability) using traces with spreading factor 10; other
spreading factors yield comparable trends. We summarize the results in
\autoref{fig:deltas_sf10}.

The baseline configuration $(N_{\delta}=1,N_{f}=1)$ achieves a sensitivity of 
$-15.2$\,dB. Expanding to $(1,2)$ or $(2,1)$ improves the
sensitivity by approximately 1.1--1.3\,dB, highlighting the benefits of finer resolution.
However, gains beyond $(2,2)$, which reaches $-17.2$\,dB, diminish rapidly 
($0.3$\,dB improvement at four times the computational cost).
While the minimal configuration minimizes overhead, it sacrifices $1.5$\,dB relative to $(2,2)$.
Hence, we balance  sensitivity
($-17.2$\,dB) with efficiency  and  select $(2,2)$ (four FFTs per window) as the operation point. 
All subsequent experiments use this $(2,2)$ configuration.

\begin{figure}[ht]
    \centering
    \begin{tikzpicture}
\begin{axis}[%
  enlargelimits=false,
    colormap={col_mod}{
rgb(0cm)=(1,1,1); rgb(1cm)=(0.8,0,0)
  },
  colormap name=col_mod,
  width=0.3\textwidth,
  height=0.2\textwidth,
  colorbar,
  colorbar style={
    width=0.5em,
    font=\small,
     ylabel={SNR [dB]},
     ylabel style={
         yshift=-3cm
     }
},
  xlabel={$N_{\delta }$ [\#]},
  ylabel={$N_{f}$ [\#]},
  xtick={1,2,3,4},
  ytick={1,2,3,4},
  xmin=0.5,xmax=4.5,
  ymin=0.5,ymax=4.5,
  point meta min=-20,
  point meta max=-10,
  nodes near coords={\pgfmathfloatifflags{\pgfplotspointmeta}{0}{}{\pgfmathprintnumber{\pgfplotspointmeta}}},
  nodes near coords style={
      /pgf/number format/fixed,
      font=\small,
  },
  every node near coord/.append style={xshift=0pt,yshift=6pt, black},
    tick label style={font=\small},
    label style={font=\small},
]

\addplot[
matrix plot,
mesh/cols=4,
mesh/rows=4,
point meta=explicit]
table[x=t, y=f, meta=rx_sens, col sep=comma] {data/deltas_sf10.csv};
\end{axis}
\end{tikzpicture}
    \caption{Detection sensitivity (99\% frame detection) for varying number of
	delay hypotheses ($N_\delta$) and frequency shift hypotheses ($N_f$), for spreading factor 10.}
    \label{fig:deltas_sf10}
\end{figure}

\subsection{Synchronization Performance in Single-User Scenarios}\label{subsec:rx_sensitivity}
\begin{figure}
    \centering
    \begin{tikzpicture}
\pgfplotsset{
    no_xlabels/.style={
        xticklabels={},
        xlabel={}
    },
    no_ylabels/.style={
        yticklabels={},
        ylabel={}
    },
}
\begin{groupplot}[
cycle list={
    {thick,mark=none, mpl_blue}, 
    {thick,mark=none, mpl_orange}, 
    {thick,mark=none, mpl_green}, 
    {thick,mark=none, mpl_red}, 
    {thick,mark=none, mpl_purple}, 
},
group style={
  group size=2 by 2,
  horizontal sep=0.15cm,
  vertical sep=0.2cm,
},
width=0.26\textwidth,
height=0.165\textwidth,
axis on top,
xmin=-25,xmax=12,
ymin=0.01,
ymode=log,
xlabel= {},
ylabel= {},
legend cell align={left},
legend style={
  fill opacity=1,
  draw opacity=1,
  text opacity=1,
  at={(-1.3,-1)},
  legend image code/.code={
    \draw[#1] (0cm,-0.0cm) -- (0.3cm,-0.0cm);
  }
  nodes={scale=0.5, transform shape},
  anchor=north west,
  draw=white!80!black,
  legend columns=5,
},
tick align=outside,
tick pos=left,
x grid style={white!69.0196078431373!black},
xtick style={color=black},
y grid style={white!69.0196078431373!black},
ytick style={color=black},
ytick={0.01,0.1,1},
yticklabels={1, 10, 100},
]

\nextgroupplot[title=\textbf{SF8}, no_xlabels, ylabel={MFR [\%]}]
    \addplot+ [] table [x index=0, y index=1, col sep=comma] {data/rxsens_detected_CoRa.csv};
    \addplot+ [] table [x index=0, y index=1, col sep=comma] {data/rxsens_detected_OpenLoRa.csv};
    \addplot+ [] table [x index=0, y index=1, col sep=comma] {data/rxsens_detected_TnB.csv};
    \addplot+ [] table [x index=0, y index=1, col sep=comma] {data/rxsens_detected_CIC.csv};
    \addplot+ [] table [x index=0, y index=1, col sep=comma] {data/rxsens_detected_Pyramid.csv};

\nextgroupplot[title=\textbf{SF10}, no_xlabels, no_ylabels]
    \addplot+ [] table [x index=0, y index=2, col sep=comma] {data/rxsens_detected_CoRa.csv};
    \addplot+ [] table [x index=0, y index=2, col sep=comma] {data/rxsens_detected_OpenLoRa.csv};
    \addplot+ [] table [x index=0, y index=2, col sep=comma] {data/rxsens_detected_TnB.csv};
    \addplot+ [] table [x index=0, y index=2, col sep=comma] {data/rxsens_detected_CIC.csv};
    \addplot+ [] table [x index=0, y index=2, col sep=comma] {data/rxsens_detected_Pyramid.csv};

    \nextgroupplot[ylabel={PER [\%]}]
    \addplot+ [] table [x index=0, y index=1, col sep=comma] {data/rxsens_prr_CoRa.csv};
    \addplot+ [] table [x index=0, y index=1, col sep=comma] {data/rxsens_prr_OpenLoRa.csv};
    \addplot+ [] table [x index=0, y index=1, col sep=comma] {data/rxsens_prr_TnB.csv};
    \addplot+ [] table [x index=0, y index=1, col sep=comma] {data/rxsens_prr_CIC.csv};
    \addplot+ [] table [x index=0, y index=1, col sep=comma] {data/rxsens_prr_Pyramid.csv};

\nextgroupplot[no_ylabels]
    \addlegendimage{thick, color=mpl_blue,
      legend image code/.code={
        \draw[thick] (0cm,-0.0cm) -- (0.3cm,-0.0cm);
      }
        }
    \addlegendentry{\our}
    \addlegendimage{thick, color=mpl_orange,
      legend image code/.code={
        \draw[thick] (0cm,-0.0cm) -- (0.3cm,-0.0cm);
      }
    }
    \addlegendentry{OpenLoRa}

    \addlegendimage{thick, color=mpl_green,
      legend image code/.code={
        \draw[thick] (0cm,-0.0cm) -- (0.3cm,-0.0cm);
      }
        }
    \addlegendentry{TnB}
    \addlegendimage{thick, color=mpl_red,
      legend image code/.code={
        \draw[thick] (0cm,-0.0cm) -- (0.3cm,-0.0cm);
      }
    }
    \addlegendentry{CIC}
    \addlegendimage{thick, color=mpl_purple,
      legend image code/.code={
        \draw[thick] (0cm,-0.0cm) -- (0.3cm,-0.0cm);
      }
    }
    \addlegendentry{Pyramid}
    \addplot+ [] table [x index=0, y index=2, col sep=comma] {data/rxsens_prr_CoRa.csv};
    \addplot+ [] table [x index=0, y index=2, col sep=comma] {data/rxsens_prr_OpenLoRa.csv};
    \addplot+ [] table [x index=0, y index=2, col sep=comma] {data/rxsens_prr_TnB.csv};
    \addplot+ [] table [x index=0, y index=2, col sep=comma] {data/rxsens_prr_CIC.csv};
    \addplot+ [] table [x index=0, y index=2, col sep=comma] {data/rxsens_prr_Pyramid.csv};

\end{groupplot}
\path (group c1r2.south east) -- (group c2r2.south west) node[pos=0.5, yshift=-1cm] {SNR [dB]};
\end{tikzpicture}
    \caption{Missed Frame Ratio (MFR) and Packet Error Ratio (PER) of \our
    and the baseline methods in the single-user (collision-free) scenario 
    for two spreading factors (8 and 10) and varying SNR levels (-25\,dB to
    12\,dB). }
\label{fig:rx_sensitivity_sim}
\end{figure}

\autoref{fig:rx_sensitivity_sim} (top row) shows the Missed Frame Ratio (MFR)
versus SNR for simulated LoRa captures, across two spreading factors (8 and
10), for \our and the baseline approaches.

While all methods approach 100\% detection miss ratio at the lowest SNR ($-25$\,dB), \our consistently
outperforms the evaluated baselines across both spreading factors within the
evaluated SNR range. In particular, \our achieves 99\% frame detection at
approximately $-11.2$\,dB (SF8) and $-17.2$\,dB (SF10), corresponding to gains of 
approximately 5\,dB over OpenLoRa (SF10), 10\,dB over TnB (SF8), 18\,dB over CIC (SF10) and over
20\,dB over Pyramid. 
While superior detection is a prerequisite, it does not guarantee successful
data recovery. To verify whether these superior frame detection capabilities translate into
end-to-end decoding performance, we demodulate frames using the standard LoRa
demodulation (\textit{argmax}) and decode them using the standard FEC decoder.
For Pyramid, which combines detection and demodulation within a single
pipeline, we use the method as provided without modifications.
As shown in \autoref{fig:rx_sensitivity_sim} (bottom row), \our exhibits a
maximum gap of 0.2\,dB between detection and decoding thresholds at 99\%
detection ratio. In particular at SF10, decoding succeeds down to $-17.2$\,dB,
while signal detection remains possible down to $-17.4$\,dB.
This represents a decoding improvement of  $\approx 5$\,dB over the best
performing baseline (OpenLoRa, SF10) and  $\approx 10$\,dB over the best
performing collision-tolerant synchronization method (TnB, SF8). 
Most baselines
(OpenLoRa, CIC and TnB) maintain a tight coupling between detection and
decoding performance, with decoding on a slightly higher decoding threshold
than detection. In stark contrast, Pyramid reports detections at low SNR that
rarely result in decoding. For example, at $10$\,dB, Pyramid exhibits 21\%
decoding error for SF8, despite failing to detect only 2\% of transmitted frames.
This discrepancy stems from the peak tracking strategy of Pyramid. This method
estimates peak locations (\ie the symbol value) via linear interpolation over
tracked peak amplitudes across successive demodulation windows. This estimation
is sensitive to tracking errors, as missing or noisy peak observations can
shift the inferred intersection by one or two frequency bins, which propagates
into decoding errors. Moreover, peak tracking terminates when
peak detection fails, preventing further symbol demodulation.

Overall, \our dominates the evaluated landscape, achieving the highest
detection and decoding performance across all single-user scenarios.
Crucially, this performance translates seamlessly to real-world single-user
conditions, as demonstrated by our real-world trace assessment in
\autoref{sec:assessment}.

\subsection{Performance of Synchronization in Multi-User Scenarios}

\begin{figure}
    \centering
    \newcommand{\DrawCols}[2]{
    \addplot [mpl_blue] table [x index=0, y index=1, col sep=comma] {data/cols_#1_#2.csv};
    \addplot [mpl_green] table [x index=0, y index=3, col sep=comma] {data/cols_#1_#2.csv};
    \addplot [mpl_red] table [x index=0, y index=4, col sep=comma] {data/cols_#1_#2.csv};
    \addplot [mpl_orange] table [x index=0, y index=2, col sep=comma] {data/cols_#1_#2.csv};
    \addplot [mpl_purple] table [x index=0, y index=5, col sep=comma] {data/cols_#1_#2.csv};
}

\newcommand{\DrawColsPackets}[2]{
    \addplot [mpl_blue] table [x index=0, y index=1, col sep=comma] {data/cols_crc_ok_#1_#2.csv};
    \addplot [mpl_green] table [x index=0, y index=3, col sep=comma] {data/cols_crc_ok_#1_#2.csv};
    \addplot [mpl_red] table [x index=0, y index=4, col sep=comma] {data/cols_crc_ok_#1_#2.csv};
    \addplot [mpl_orange] table [x index=0, y index=2, col sep=comma] {data/cols_crc_ok_#1_#2.csv};
    \addplot [mpl_purple] table [x index=0, y index=5, col sep=comma] {data/cols_crc_ok_#1_#2.csv};
}

\begin{tikzpicture}
\pgfplotsset{
    no_xlabels/.style={
        xticklabels={},
        xlabel={}
    },
    no_ylabels/.style={
        yticklabels={},
        ylabel={}
    },
}
\begin{groupplot}[
cycle list={
    {thick,mark=none, mpl_blue}, 
    {thick,mark=none, mpl_orange}, 
    {thick,mark=none, mpl_green}, 
    {thick,mark=none, mpl_red}, 
    {thick,mark=none, mpl_purple}, 
},
group style={
  group size=2 by 2,
  horizontal sep=0.15cm,
  vertical sep=0.2cm,
},
width=0.26\textwidth,
height=0.165\textwidth,
axis on top,
xmin=5,xmax=60,
xlabel= {},
legend cell align={left},
legend style={
  fill opacity=1,
  draw opacity=1,
  text opacity=1,
  at={(-1.3,-1)},
  legend image code/.code={
    \draw[#1] (0cm,-0.0cm) -- (0.3cm,-0.0cm);
  }
  nodes={scale=0.5, transform shape},
  anchor=north west,
  draw=white!80!black,
  legend columns=5,
},
tick align=outside,
tick pos=left,
x grid style={white!69.0196078431373!black},
xtick style={color=black},
y grid style={white!69.0196078431373!black},
ytick style={color=black},
ytick={0, 0.25,0.50,0.75, 1},
yticklabels={0, 25, 50, 75, 100},
]
    \nextgroupplot[ylabel={FDR[\%]}, title=\textbf{SF7}, no_xlabels]
    \DrawCols{high}{7}

\nextgroupplot[title=\textbf{SF12}, no_ylabels, no_xlabels]
    \DrawCols{high}{12}
\nextgroupplot[ylabel={PRR [\%]}]
    \DrawColsPackets{high}{7}
\nextgroupplot[no_ylabels]
    \addlegendimage{thick, color=mpl_blue}
    \addlegendentry{\our}
    \addlegendimage{thick, color=mpl_green}
    \addlegendentry{TnB}
    \addlegendimage{thick, color=mpl_orange}
    \addlegendentry{OpenLoRa}
    \addlegendimage{thick, color=mpl_red}
    \addlegendentry{CIC}
    \addlegendimage{thick, color=mpl_purple}
    \addlegendentry{Pyramid}
    \DrawColsPackets{high}{12}

\end{groupplot}
\path (group c1r2.south east) -- (group c2r2.south west) node[pos=0.5, yshift=-1cm] {TX Rate [\si{\pkt\per\second}]};
\end{tikzpicture}
    \caption{Frame Detection Ratio (FDR) and Packet Reception Ratio (PRR) of \our    and the baseline methods in a multi-user scenario
	for two spreading factors (7 and 12), varying  SNR values (0 to 30\,dB) and transmission rates (5--60\,\textit{pkt/s}).}
\label{fig:cols_sim_detection}
\end{figure}

\autoref{fig:cols_sim_detection} (top) shows the frame detection ratio across
varying transmission rates for two spreading factor configurations (7 and 12).
These configurations are selected as they respectively represent the shortest and longest time-on-air (ToA),
 thereby capturing the impact of packet duration on collision probability.
Packet arrivals follow a Poisson process. At the highest evaluated transmission
rate (60 \textit{pkt/s}), the expected number of overlapping packets during a preamble is \PoissonSFSevenTXRSixty{} for SF7 and \PoissonSFTwelveTXRSixty{} for SF12, regarding a preamble length of 12.25 symbols.

We observe that \our outperforms all baselines for every transmission rate and
spreading factor configuration. At the highest transmission rate
(60\,\textit{pkt/s}), our solution achieves a detection ratio of
\PercColsSim{detection}{7}{60}{CoRa} for SF7 and
\PercColsSim{detection}{12}{60}{CoRa} for SF12. These results translate into
significant gains over the baselines. In particular, up
to \PPColsSim{detection}{12}{60}{CoRa}{TnB} percentage points over TnB (SF12),
\PPColsSim{detection}{12}{60}{CoRa}{Pyramid} points over Pyramid (SF12),
\PPColsSim{detection}{12}{60}{CoRa}{CIC} points over CIC (SF12) and
\PPColsSim{detection}{12}{60}{CoRa}{OpenLoRa} points over OpenLoRa (SF12). At
low transmission rates (5\,\textit{pkt/s}), our method shows an improvement
over TnB of up to \PPColsSim{detection}{12}{5}{CoRa}{TnB} percentage point
(SF12), while the remaining methods lag behind by up to
\PPColsSim{detection}{12}{5}{CoRa}{OpenLoRa} points (OpenLoRa, SF12),
\PPColsSim{detection}{12}{5}{CoRa}{CIC} points (CIC, SF12) and
\PPColsSim{detection}{12}{5}{CoRa}{Pyramid} points (Pyramid, SF12).

Next we evaluate whether the superior detection of \our under multiple
interferers translates to higher frame decoding performance.
We apply all four decoding pipelines (CoRa, TnB, CIC and OpenLoRa) to the
frames reported by each synchronization algorithm (OpenLoRa, TnB, CIC and
\our), counting a frame as successfully decoded if at least one decoding
pipeline passes the payload CRC. This allows us to isolate the effect of
synchronization from that of the decoding pipeline. The fifth receiver,
Pyramid, integrates detection and demodulation within a single pipeline and
provides no separate frame timestamps or CFO estimates.
We therefore report only the frames decoded by its full pipeline.

\autoref{fig:cols_sim_detection} (bottom) presents the packet reception ratio for
each synchronization algorithm across varying transmission rates and spreading
factors (7 and 12). We observe that \our consistently outperforms all baselines.
At the highest transmission rate (60\,\textit{pkt/s}), \our
achieves gains up to \PPColsSim{crcok}{12}{60}{CoRa}{TnB} points more than TnB (SF12), \PPColsSim{crcok}{12}{60}{CoRa}{CIC} points over CIC (SF12), \PPColsSim{crcok}{12}{60}{CoRa}{Pyramid} points
over Pyramid (SF12) and \PPColsSim{crcok}{12}{60}{CoRa}{OpenLoRa} points over OpenLoRa (SF12).
These results indicate that \our exhibits its strongest advantage under high collision
conditions (\ie high transmission rate and high spreading factor), achieving the
highest performance gain over the baselines.
As observed before, Pyramid exhibits a gap between detection and decoding
performance. This discrepancy is most prominent at the lowest transmission
rates (5\, \textit{pkt/s}) and spreading factor 7, where Pyramid decodes only
\PercColsSim{crcok}{7}{5}{Pyramid} of frames despite detecting
\PercColsSim{detection}{7}{5}{Pyramid}.

\subsection{Impact of Timing Offset and Power Imbalance}\label{subsec:timing}

\begin{figure}
    \centering
    \begin{tikzpicture}
\begin{axis}[
    small,
    view={0}{90},
    colorbar,
    colormap name=viridis,
    height=3.5cm,
    xlabel={Offset [symbols]},
    ylabel={SIR [dB]},
    tick align=outside,
    tick pos=left,
    x grid style={white!69.0196078431373!black},
    xtick style={color=black},
    y grid style={white!69.0196078431373!black},
    ytick style={color=black},
    axis on top,
    xmin=-13,xmax=13,
    colorbar style={
        ylabel={FDR [\%]},
        ytick = {0,0.25,0.5,0.75,1},
        yticklabels = {0,25,50,75,100},
        ylabel style={
            yshift=-2.7cm
        }
    },
    ytick = {-40,-30,-20,-10,0,10},
]
\addplot3 [
surf,
shader=interp,
]
table[
    col sep=comma,
    x=offset,
    y=pw,
    z=prr,
] {data/cols_analysis.csv};

\end{axis}
\end{tikzpicture}
    \caption{Frame Detection Ratio (FDR) versus collision offsets (in symbols) for
	varying Signal-to-Interference Ratio (SIR) levels in a simulated two-user scenario (SF8).
    Frames experience carrier frequency offset up to $4.88$\,kHz. Positive offsets
    indicate that the interfering frame arrives after the target frame.}
\label{fig:collision_offset}
\end{figure}

We evaluate the detection performance of \our in a two-user scenario with
temporally overlapping transmissions, capturing the impact of
timing misalignment and power imbalance (\ie the near-far effect) on
synchronization performance. We vary the relative frame offset
by up to $13$ symbols in both directions, covering the complete preamble
duration. For offsets below the preamble length (12.25 symbols), the
preambles overlap, producing structured chirp
interference during synchronization. We summarize the results in
\autoref{fig:collision_offset}.

For signal-to-interference ratios above -15\,dB (\ie interference powers 
 up to $\approx 32 \times$ the desired signal power), \our achieves
near-perfect detection. However, degradation occurs around
integer symbol-aligned offsets, particularly at -1, 0 and +1 symbols,
where detection drops to around 50\% even under favorable power imbalance.
These regions represent overlapping target and
interference peaks in the dechirped signal during synchronization (assuming zero-centered CFO),
 which reduces the desired peak prominence.
Additionally, multiple downchirp-induced peaks during the initial STO/CFO
estimation~(\autoref{subsec:fine_sync}) may cause incorrect peak association,
which further degrades synchronization. 

For fractional offsets below one symbol, the reduced upchirp-induced
interference during initial CFO estimation yields low interference floor in the
downchirp segment and enables detection down to $\approx -30$\,dB. The notch
around $-2.25$ symbols is caused by degraded sync-word detection due to
interference between the target sync-word and the interfering downchirp
segment. For positive offsets above 11 symbols, detection approaches 100\% even
at $-40$\,dB, as the first downchirp symbol remains sufficiently unaffected for
reliable CFO estimation.

Overall, \our remains robust against strong near-far conditions while resolving
sub-symbol collision offsets, where multiple preamble peaks appear within the
demodulation window after spectral intersection. By identifying individual
peaks, \our separates overlapping frames in scenarios where the \textit{argmax}
approach can only select the strongest peak, demonstrating the effectiveness of
our proposed strategy.

\section{Assessment using Real-World Traces}\label{sec:assessment}

The following results validate our findings in \autoref{sec:evaluation} using
three real-world datasets as characterized in Table \ref{tab:tx_params}.

\begin{table}
\normalsize
\centering
\begin{tabular}{lccc}
\toprule
\textbf{Parameter} & \textbf{CIC} & \textbf{TnB} & \textbf{\our} \\
\midrule
SF & 8 & 8,10 & 7--10 \\
BW [kHz] & 250 & 125 & 125 \\
CR & 4/5 & 4/5--4/8 & 4/5 \\

Rate [pkt/s] & 5--100 & 20,25 & 2 \\
TX Dev. [\#] & 20 & 20 & 1 \\

\midrule
\multirow{2}{*}{SNR [dB]}
& \textit{D1}: $30$--$42$
& \textit{Ind}.: $-9$--$23$
& \textit{WM5}: $-7$--$14$ \\

& \textit{D4}: $-17$--$5$
& \textit{Out}.: $-17$--$12$
& \textit{WM4}: $-17$--$6$ \\

\bottomrule
\end{tabular}
\caption{Configuration parameters of LoRa devices and scenario characteristics for the captured datasets.}
\label{tab:tx_params}
\end{table}
\subsection{Description of Datasets} These datasets are complementary, each
capturing different operational challenges. The \textit{CIC} dataset evaluates dense
networks with many concurrent devices under varying SNR and traffic conditions. 
The \textit{TnB} dataset extends this evaluation to heterogeneous PHY configurations
with varying spreading factors and coding rates.

\begin{wrapfigure}{r}{0.15\textwidth}
    \centering
    \includegraphics[width=\linewidth]{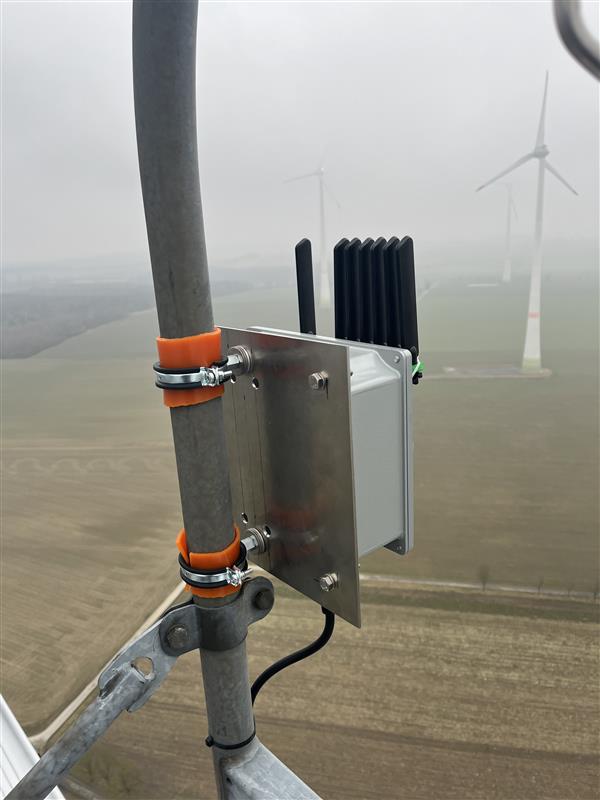}
    \caption{Gateway deployment for the \our dataset.}
    \label{fig:wm5}
\end{wrapfigure}

Finally, the \textit{\our} dataset provides recordings from multiple LoRa
transmitters, captured individually over distinct propagation paths and
power levels for evaluating synchronization robustness under realistic channel
conditions.

We generate the \our dataset with programmable multi-radio outdoor
gateways~\cite{wrapf-appmo-25}, each combining a low-cost software-defined
radio platform (\textit{BladeRF x40}) with a COTS
Nucleo-WL55JC board embedding an SX1262 transceiver. Three gateways are deployed
at an operational wind turbine site in Paderborn, Germany~(\autoref{fig:wm5}), with
one SDR receiver and two LoRa transmitters.

The receiver sits in the tower base at approximately 3 meters height.
One transmitter (\textit{WM4}) is placed on top of the same turbine (150\,m
distance, no line-of-sight), creating a strongly multi-path dominated link,
while the second transmitter (\textit{WM5}) is located on a separate wind
turbine (430\,m distance) with partially obstructed line-of-sight.
On each transmission, the device randomly selects a transmission power from the
fixed set ${2, 6, 10, 14\,dBm}$. For each capture, we record 300\,s of wireless
traffic at a constant spreading factor between 7~and~10.

All three datasets were sampled with an oversampling factor (OSF) of 8, yielding sampling
rates according to the configured bandwidth.

\begin{figure}
    \centering
    \newcommand{\DrawWMFigure}[2]{%
    \pgfplotsforeachungrouped \file/\style in {
        CoRa/mpl_blue,
	Baseline/mpl_orange,
	TnB/mpl_green,
	CIC/mpl_red,
	Pyramid/mpl_purple
    }{

        \edef\temp{
            \noexpand\addplot [\style, name path global=pathLower] table [
                x index=0, y index=2, col sep=comma
            ] {data/LZn_#1_#2_\file.csv};
            \noexpand\addplot [draw=none, name path global=pathLower] table [
                x index=0, y index=5, col sep=comma
            ] {data/LZn_#1_#2_\file.csv};
            \noexpand\addplot [draw=none, name path global=pathUpper] table [
                x index=0, y index=6, col sep=comma
            ] {data/LZn_#1_#2_\file.csv};
        }
        \temp

	\edef\asd{\noexpand\addplot [\style, opacity=0.3]
	    fill between[of=pathLower and pathUpper];}
	\asd
    }
}
\begin{tikzpicture}
\pgfplotsset{
    no_xlabels/.style={
        xticklabels={},
        xlabel={}
    },
    no_ylabels/.style={
        yticklabels={},
        ylabel={}
    },
}
\begin{groupplot}[
cycle list={
    {thick,mark=none, mpl_blue}, 
    {thick,mark=none, mpl_orange}, 
    {thick,mark=none, mpl_green}, 
    {thick,mark=none, mpl_red}, 
},
group style={
  group size=2 by 4,
  horizontal sep=0.45cm,
  vertical sep=0.2cm,
},
width=0.26\textwidth,
height=0.165\textwidth,
axis on top,
xmin=2,xmax=14,
ymin=0,
xlabel= {},
ylabel= {},
legend cell align={left},
legend style={
  fill opacity=1,
  draw opacity=1,
  text opacity=1,
  at={(-1.3,-1)},
  legend image code/.code={
    \draw[#1] (0cm,-0.0cm) -- (0.3cm,-0.0cm);
  }
  nodes={scale=0.5, transform shape},
  anchor=north west,
  draw=white!80!black,
  legend columns=5,
},
tick align=outside,
tick pos=left,
x grid style={white!69.0196078431373!black},
xtick style={color=black},
y grid style={white!69.0196078431373!black},
ytick style={color=black},
ytick={0,0.25,0.5,0.75,1},
yticklabels={0, 25, 50, 75, 100},
xtick={2,6,10,14}
]
\nextgroupplot[title=\textbf{WM5}, no_xlabels]
    \DrawWMFigure{WM5}{7}

\nextgroupplot[title=\textbf{WM4}, no_xlabels,  no_ylabels, no_xlabels]
    \DrawWMFigure{WM4}{7}

\nextgroupplot[no_xlabels]
    \DrawWMFigure{WM5}{8}

\nextgroupplot[ no_ylabels, no_xlabels]
    \DrawWMFigure{WM4}{8}

\nextgroupplot[no_xlabels]
    \DrawWMFigure{WM5}{9}
\nextgroupplot[ no_ylabels, no_xlabels]
    \DrawWMFigure{WM4}{9}

\nextgroupplot[]
    \DrawWMFigure{WM5}{10}
\nextgroupplot[ no_ylabels]
    \addlegendimage{thick, color=mpl_blue,
      legend image code/.code={
        \draw[thick] (0cm,-0.0cm) -- (0.3cm,-0.0cm);
      }
        }
    \addlegendentry{\our}
    \addlegendimage{thick, color=mpl_orange,
      legend image code/.code={
        \draw[thick] (0cm,-0.0cm) -- (0.3cm,-0.0cm);
      }
    }
    \addlegendentry{OpenLoRa}

    \addlegendimage{thick, color=mpl_green,
      legend image code/.code={
        \draw[thick] (0cm,-0.0cm) -- (0.3cm,-0.0cm);
      }
        }
    \addlegendentry{TnB}
    \addlegendimage{thick, color=mpl_red,
      legend image code/.code={
        \draw[thick] (0cm,-0.0cm) -- (0.3cm,-0.0cm);
      }
    }
    \addlegendentry{CIC}
    \addlegendimage{thick, color=mpl_purple,
      legend image code/.code={
        \draw[thick] (0cm,-0.0cm) -- (0.3cm,-0.0cm);
      }
    }
    \addlegendentry{Pyramid}
    \DrawWMFigure{WM4}{10}

\end{groupplot}
\node[anchor=base,rotate=-90,yshift=0.275cm] at (group c2r1.east) {\textbf{SF7}};
\node[anchor=base,rotate=-90,yshift=0.275cm] at (group c2r2.east) {\textbf{SF8}};
\node[anchor=base,rotate=-90,yshift=0.275cm] at (group c2r3.east) {\textbf{SF9}};
\node[anchor=base,rotate=-90,yshift=0.275cm] at (group c2r4.east) {\textbf{SF10}};
\path (group c1r1.north west) -- (group c1r4.south west) node[pos=0.5, xshift=-1cm, rotate=90, anchor=south] {PDR [\%]};
\path (group c1r4.south east) -- (group c2r4.south west) node[pos=0.5, yshift=-1cm] {TX Power [dBm]};
\end{tikzpicture}
    \caption{Packet Detection Ratio (PDR) of \our and the baseline methods for
    the \our dataset (collision-free transmissions), evaluated under high
    (\textit{WM5}) and low (\textit{WM4}) SNR scenarios, across four spreading
    factors (7, 8, 9, and 10) and four transmission power levels (2, 6, 10 or
    14\,dBm). Shaded areas represent the 95\% confidence intervals.}
\label{fig:rx_sensitivity}
\end{figure}

\subsection{Validation of Single-User Scenario}\label{subsec:ev_single}
We validate our single-user simulation findings (\autoref{subsec:rx_sensitivity}) using real-world SDR captures from the \textit{WM4} (low SNR) and \textit{WM5} (high SNR) scenarios.

As in simulations, all methods achieve near-perfect performance under high SNR
(\autoref{fig:rx_sensitivity}, \textit{WM5}) with only minor deviations
(approximately 5\%) for CIC and Pyramid under low-link-budget settings (SF7,
2\,dBm). At higher link budgets within the low SNR scenario (\autoref{fig:rx_sensitivity}, \textit{WM4}), \our performs comparably to
TnB, reaching up to 97\% at 14\,dBm.

In the low SNR regime (\autoref{fig:rx_sensitivity}, \textit{WM4}), \our
consistently outperforms all baselines across low link-budget configurations.
The performance gains of our solution are most pronounced at SF7 and 2\,dBm,
where our solution attains 90\% detection. Under the same low link-budget
conditions, the evaluated baselines fall short, exhibiting 70\% (OpenLoRa),
26\% (TnB), 5\% (CIC) and 4\% (Pyramid).

\begin{figure}
    \centering
    \newcommand{\DrawWMFigure}[2]{%
    \pgfplotsforeachungrouped \file/\style in {
        CoRa/mpl_blue,
	Baseline/mpl_orange,
	TnB/mpl_green,
	CIC/mpl_red,
	Pyramid/mpl_purple
    }{

        \edef\temp{
            \noexpand\addplot [\style, name path global=pathLower] table [
                x index=0, y index=1, col sep=comma
            ] {data/LZn_#1_#2_\file.csv};
            \noexpand\addplot [draw=none, name path global=pathLower] table [
                x index=0, y index=3, col sep=comma
            ] {data/LZn_#1_#2_\file.csv};
            \noexpand\addplot [draw=none, name path global=pathUpper] table [
                x index=0, y index=4, col sep=comma
            ] {data/LZn_#1_#2_\file.csv};
        }
        \temp

	\edef\asd{\noexpand\addplot [\style, opacity=0.3]
	    fill between[of=pathLower and pathUpper];}
	\asd
    }
}
\begin{tikzpicture}
\pgfplotsset{
    no_xlabels/.style={
        xticklabels={},
        xlabel={}
    },
    no_ylabels/.style={
        yticklabels={},
        ylabel={}
    },
}
\begin{groupplot}[
cycle list={
    {thick,mark=none, mpl_blue}, 
    {thick,mark=none, mpl_orange}, 
    {thick,mark=none, mpl_green}, 
    {thick,mark=none, mpl_red}, 
},
group style={
  group size=2 by 4,
  horizontal sep=0.45cm,
  vertical sep=0.2cm,
},
width=0.26\textwidth,
height=0.165\textwidth,
axis on top,
xmin=2,xmax=14,
ymin=0,
xlabel= {},
ylabel= {},
legend cell align={left},
legend style={
  fill opacity=1,
  draw opacity=1,
  text opacity=1,
  at={(-1.3,-1)},
  legend image code/.code={
    \draw[#1] (0cm,-0.0cm) -- (0.3cm,-0.0cm);
  }
  nodes={scale=0.5, transform shape},
  anchor=north west,
  draw=white!80!black,
  legend columns=5,
},
tick align=outside,
tick pos=left,
x grid style={white!69.0196078431373!black},
xtick style={color=black},
y grid style={white!69.0196078431373!black},
ytick style={color=black},
ytick={0,0.25,0.5,0.75,1},
yticklabels={0, 25, 50, 75, 100},
xtick={2,6,10,14}
]
\nextgroupplot[title=\textbf{WM5}, no_xlabels]
    \DrawWMFigure{WM5}{7}

\nextgroupplot[title=\textbf{WM4}, no_xlabels,  no_ylabels, no_xlabels]
    \DrawWMFigure{WM4}{7}

\nextgroupplot[no_xlabels]
    \DrawWMFigure{WM5}{8}

\nextgroupplot[ no_ylabels, no_xlabels]
    \DrawWMFigure{WM4}{8}

\nextgroupplot[no_xlabels]
    \DrawWMFigure{WM5}{9}
\nextgroupplot[ no_ylabels, no_xlabels]
    \DrawWMFigure{WM4}{9}

\nextgroupplot[]
    \DrawWMFigure{WM5}{10}
\nextgroupplot[ no_ylabels]
    \addlegendimage{thick, color=mpl_blue,
      legend image code/.code={
        \draw[thick] (0cm,-0.0cm) -- (0.3cm,-0.0cm);
      }
        }
    \addlegendentry{\our}
    \addlegendimage{thick, color=mpl_orange,
      legend image code/.code={
        \draw[thick] (0cm,-0.0cm) -- (0.3cm,-0.0cm);
      }
    }
    \addlegendentry{OpenLoRa}

    \addlegendimage{thick, color=mpl_green,
      legend image code/.code={
        \draw[thick] (0cm,-0.0cm) -- (0.3cm,-0.0cm);
      }
        }
    \addlegendentry{TnB}
    \addlegendimage{thick, color=mpl_red,
      legend image code/.code={
        \draw[thick] (0cm,-0.0cm) -- (0.3cm,-0.0cm);
      }
    }
    \addlegendentry{CIC}
    \addlegendimage{thick, color=mpl_purple,
      legend image code/.code={
        \draw[thick] (0cm,-0.0cm) -- (0.3cm,-0.0cm);
      }
    }
    \addlegendentry{Pyramid}
    \DrawWMFigure{WM4}{10}

\end{groupplot}
\node[anchor=base,rotate=-90,yshift=0.275cm] at (group c2r1.east) {\textbf{SF7}};
\node[anchor=base,rotate=-90,yshift=0.275cm] at (group c2r2.east) {\textbf{SF8}};
\node[anchor=base,rotate=-90,yshift=0.275cm] at (group c2r3.east) {\textbf{SF9}};
\node[anchor=base,rotate=-90,yshift=0.275cm] at (group c2r4.east) {\textbf{SF10}};
\path (group c1r1.north west) -- (group c1r4.south west) node[pos=0.5, xshift=-1cm, rotate=90, anchor=south] {PRR [\%]};
\path (group c1r4.south east) -- (group c2r4.south west) node[pos=0.5, yshift=-1cm] {TX Power [dBm]};
\end{tikzpicture}
    \caption{Packet Reception Ratio (PRR) for packets demodulated using detections from \our and the
    baseline methods on the \our dataset. Performance is evaluated under high
    (\textit{WM5}) and low (\textit{WM4}) SNR conditions, across four
    spreading factors (7, 8, 9, and 10) and four transmission power levels (2, 6, 10 and 14\,dBm). Shaded areas represent the 95\% confidence intervals.}
\label{fig:wm_prr}
\end{figure}

We evaluate CRC correctness on the detected frames~(\autoref{fig:wm_prr}),
using standard LoRa demodulation.
All methods except Pyramid achieve near-perfect decoding in high SNR
(\textit{WM5}), with only SF7 showing a minor 3\% drop in average for these
demodulators.
As in simulation, Pyramid shows a pronounced decoupling
between detection and decoding performance with a maximum packet
reception ratio of 41\% at SF7
(\autoref{fig:wm_prr}, \textit{WM5}) despite near-perfect detection rates.

In the low SNR scenario (\autoref{fig:wm_prr}, \textit{WM4}), \our consistently
achieves the highest packet reception ratio across all spreading factors,
except at SF10 under high link-budget conditions, where it performs on par with
TnB at 87\% (\autoref{fig:wm_prr}, \textit{WM4}, SF10). In the most adverse
channel conditions (\autoref{fig:wm_prr}, \textit{WM4}, SF7, 2\,dBm), our
solution decodes 76\% of frames, outperforming OpenLoRa (36\%), TnB (20\%), CIC
(5\%) and Pyramid (0\%).

Overall, the real-world traces confirm the trends observed in the simulation-based evaluation,
with our solution excelling in low SNR scenarios.

\subsection{Validation of Multi-User Scenario}\label{subsec:ev_collisions}

\begin{figure}[htbp]
    \centering
    \newcommand{\DrawScatter}[1]{%
    \pgfplotsforeachungrouped \file/\style in {
        CoRa/mpl_blue,
	OpenLoRa/mpl_orange,
	TnB/mpl_green,
	CIC/mpl_red,
	Pyramid/mpl_purple
    }{

        \edef\temp{
            \noexpand\addplot [\style, name path global=pathLower] table [
                x index=0, y index=1, col sep=comma
            ] {data/cic_prr_#1_\file.csv};
            \noexpand\addplot [draw=none, name path global=pathLower] table [
                x index=0, y index=2, col sep=comma
            ] {data/cic_prr_#1_\file.csv};
            \noexpand\addplot [draw=none, name path global=pathUpper] table [
                x index=0, y index=3, col sep=comma
            ] {data/cic_prr_#1_\file.csv};
        }
        \temp

	\edef\asd{\noexpand\addplot [\style, opacity=0.15]
	    fill between[of=pathLower and pathUpper];}
	\asd
    }
}
\begin{tikzpicture}

\pgfplotsset{
    no_xlabels/.style={
        xticklabels={},
        xlabel={}
    },
    no_ylabels/.style={
        yticklabels={},
        ylabel={}
    },
}
\begin{groupplot}[
cycle list={
    {thick,mark=none, mpl_blue}, 
    {thick,mark=none, mpl_orange}, 
    {thick,mark=none, mpl_green}, 
    {thick,mark=none, mpl_red}, 
    {thick,mark=none, mpl_purple}, 
    {thick,mark=none, dashed, mpl_blue}, 
    {thick,mark=none, dashed, mpl_orange}, 
    {thick,mark=none, dashed, mpl_green}, 
    {thick,mark=none, dashed, mpl_red}, 
    {thick,mark=none, dashed, mpl_purple}, 
},
group style={
  group size=2 by 1,
  horizontal sep=0.35cm,
  vertical sep=1.3cm,
},
height=0.37\linewidth,
width=0.55\linewidth,
axis on top,
xtick = {0, 25, 50, 75, 100},
xmin=5,xmax=105,
ymin=-5,ymax=75,
xlabel= {},
    ylabel= {Throughput [\si{\pkt\per\second}]},
legend cell align={left},
legend style={
  fill opacity=1,
  draw opacity=1,
  text opacity=1,
  at={(-1.3,-0.8)},
  legend image code/.code={
    \draw[#1] (0cm,-0.0cm) -- (0.3cm,-0.0cm);
  }
  nodes={scale=0.5, transform shape},
  anchor=north west,
  draw=white!80!black,
  legend columns=5,
},
tick align=outside,
tick pos=left,
x grid style={white!69.0196078431373!black},
xtick style={color=black},
y grid style={white!69.0196078431373!black},
ytick style={color=black},
ytick = {0, 25, 50, 75, 100},
yticklabels = {0, 25, 50, 75, 100},
]
\nextgroupplot[title=\textbf{D1}]
\DrawScatter{high}
\nextgroupplot[title=\textbf{D4}, no_ylabels]
    \addlegendimage{thick, color=mpl_blue,
      legend image code/.code={
        \draw[thick] (0cm,-0.0cm) -- (0.3cm,-0.0cm);
      }
        }
    \addlegendentry{\our}
    \addlegendimage{thick, color=mpl_orange,
      legend image code/.code={
        \draw[thick] (0cm,-0.0cm) -- (0.3cm,-0.0cm);
      }
    }
    \addlegendentry{OpenLoRa}

    \addlegendimage{thick, color=mpl_green,
      legend image code/.code={
        \draw[thick] (0cm,-0.0cm) -- (0.3cm,-0.0cm);
      }
        }
    \addlegendentry{TnB}
    \addlegendimage{thick, color=mpl_red,
      legend image code/.code={
        \draw[thick] (0cm,-0.0cm) -- (0.3cm,-0.0cm);
      }
    }
    \addlegendentry{CIC}
    \addlegendimage{thick, color=mpl_purple,
      legend image code/.code={
        \draw[thick] (0cm,-0.0cm) -- (0.3cm,-0.0cm);
      }
    }
    \addlegendentry{Pyramid}
\DrawScatter{very_low}

\end{groupplot}
\path (group c1r1.south east) -- (group c2r1.south west) node[pos=0.5, yshift=-1cm] {TX Rate [\si{\pkt\per\second}]};
\end{tikzpicture}
    \caption{Throughput versus transmission rate (CIC dataset), for packets
    demodulated using detections from \our and the baseline methods. Results
    are shown for a high SNR scenario (\textit{D1}) and a low SNR scenario
    (\textit{D4}). Shaded areas (barely visible) represent the 95\% confidence intervals}
    \label{fig:cic_eval}
\end{figure}

We further validate our findings for multi-user scenarios using the
\textit{CIC} and \textit{TnB} datasets. Since these datasets do not provide
packet-level ground truth, detection results cannot be directly evaluated in
terms of missed detections or false positives. Therefore, and since false
positives may vary across methods, we focus on successfully decoded frames (CRC
OK) as an indicator of effective packet recovery. We use the same evaluation
methodology of~\autoref{sec:evaluation}, in which results are aggregated via a
union over multiple demodulators. Pyramid uses its original (unmodified)
pipeline.

On the CIC dataset, \our matches TnB under high SNR conditions
(\autoref{fig:cic_eval}, \textit{D1}) with up to 68\% PRR at 100 \textit{pkt/s}, 
while the other schemes exhibit  lower decoding ratios, \ie 
57\% (CIC), 11.6\% (OpenLoRa) and 2.17\% (Pyramid). In the low SNR regime
(\autoref{fig:cic_eval}, \textit{D4}), \our consistently outperforms baselines
with a maximum of 45.8\% at 100 \textit{pkt/s}. This represents gains of 9\% over
TnB (41.7\%), 57\% over CIC (28.51\%), $3.16\times$ over OpenLoRa and over $42\times$ over
Pyramid.

\begin{figure}
\captionsetup[subfigure]{skip=0pt}
    \centering

    \begin{subfigure}[t]{0.49\textwidth}
        \centering
        \newcommand{\DrawTnBScatter}[3]{
\addplot+ [
	   error bars/.cd,
	   y dir=both, y explicit,
	   ]
	table [
	   x=cr,
	   y=mean,
	   y error minus expr=\thisrow{mean} - \thisrow{ci_low},
	   y error plus expr=\thisrow{ci_high} - \thisrow{mean},
	   col sep=comma
   ] {data/tnb_prr_#1_#2_#3.csv};
}

\begin{tikzpicture}
\pgfplotsset{
    no_xlabels/.style={
        xticklabels={},
        xlabel={}
    },
    no_ylabels/.style={
        yticklabels={},
        ylabel={}
    },
}
\begin{groupplot}[
    group style={
      group size=1 by 2,
      horizontal sep=0.35cm,
      vertical sep=0.4cm,
    },
    legend cell align={left},
    legend style={
      fill opacity=1,
      draw opacity=1,
      text opacity=1,
      at={(0.2,-0.7)},
      nodes={scale=0.5, transform shape},
      anchor=north west,
      draw=white!80!black,
      legend columns=5,
    },
    ybar,
	symbolic x coords={1,2,3,4},
    tick align=outside,
    tick pos=left,
    enlarge x limits=0.25,
    ymin=0,ymax=17,
    xlabel={Group},
    ylabel={Value},
    axis on top,
    height=0.37\linewidth,
    width=1\linewidth,
    xtick = {1,2,3,4},
    xticklabels={4/5,4/6, 4/7,4/8},
	cycle list={
	    {thick,mark=none, fill=mpl_blue, draw=none}, 
	    {thick,mark=none, fill=mpl_orange, draw=none}, 
	    {thick,mark=none, fill=mpl_green, draw=none}, 
	    {thick,mark=none, fill=mpl_red, draw=none}, 
	    {thick,mark=none, fill=mpl_purple, draw=none}, 
	},
]

\nextgroupplot[bar shift auto, bar width=5pt, no_xlabels, ylabel={}]
\DrawTnBScatter{indoor}{8}{CoRa}
\DrawTnBScatter{indoor}{8}{OpenLoRa}
\DrawTnBScatter{indoor}{8}{TnB}
\DrawTnBScatter{indoor}{8}{CIC}
\DrawTnBScatter{indoor}{8}{Pyramid}

\nextgroupplot[bar width=5pt, xlabel={}, ylabel={}]

\DrawTnBScatter{outdoor1}{8}{CoRa}
\DrawTnBScatter{outdoor1}{8}{OpenLoRa}
\DrawTnBScatter{outdoor1}{8}{TnB}
\DrawTnBScatter{outdoor1}{8}{CIC}
\DrawTnBScatter{outdoor1}{8}{Pyramid}

\end{groupplot}
\node[anchor=south,rotate=-90,yshift=0.175cm] at (group c1r1.east) {\textbf{Indoor}};
\node[anchor=south,rotate=-90,yshift=0.175cm] at (group c1r2.east) {\textbf{Outdoor1}};
\path (group c1r1.south west) -- (group c1r2.north west) node[pos=0.5, anchor=south,rotate=90,yshift=0.675cm] {Throughput [\si{\pkt\per\second}]};
\path (group c1r2.south west) -- (group c1r2.south east) node[pos=0.5, yshift=-1cm] {Coding Rate [\#]};
\end{tikzpicture}
        \caption{Spreading Factor 8}\label{fig:tnb_sf8}
\vspace{0.5cm} 
    \end{subfigure}
    \begin{subfigure}[t]{0.49\textwidth}
        \centering
        \newcommand{\DrawTnBScatter}[3]{
\addplot+ [
	   error bars/.cd,
	   y dir=both, y explicit,
	   ]
	table [
	   x=cr,
	   y=mean,
	   y error minus expr=\thisrow{mean} - \thisrow{ci_low},
	   y error plus expr=\thisrow{ci_high} - \thisrow{mean},
	   col sep=comma
   ] {data/tnb_prr_#1_#2_#3.csv};
}

\begin{tikzpicture}
\pgfplotsset{
    no_xlabels/.style={
        xticklabels={},
        xlabel={}
    },
    no_ylabels/.style={
        yticklabels={},
        ylabel={}
    },
}
\begin{groupplot}[
    group style={
      group size=1 by 2,
      horizontal sep=0.35cm,
      vertical sep=0.4cm,
    },
    legend cell align={left},
    legend style={
      fill opacity=1,
      draw opacity=1,
      text opacity=1,
      at={(0,-0.8)},
      anchor=north west,
      draw=white!80!black,
      legend columns=5,
    },
    legend image code/.code={
        \draw[fill=\pgfplots@list@color,draw=none] 
            (0cm,-0.1cm) rectangle (0.3cm,0.2cm);
    },
    legend image post style={},
    ybar,
	symbolic x coords={1,2,3,4},
    tick align=outside,
    tick pos=left,
    enlarge x limits=0.25,
    ymin=0,ymax=17,
    xlabel={Group},
    ylabel={Value},
    axis on top,
    height=0.37\linewidth,
    width=1\linewidth,
    xtick = {1,2,3,4},
    xticklabels={4/5,4/6, 4/7,4/8},
	cycle list={
	    {thick,mark=none, fill=mpl_blue, draw=none}, 
	    {thick,mark=none, fill=mpl_orange, draw=none}, 
	    {thick,mark=none, fill=mpl_green, draw=none}, 
	    {thick,mark=none, fill=mpl_red, draw=none}, 
	    {thick,mark=none, fill=mpl_purple, draw=none}, 
	},
]

\nextgroupplot[bar width=5pt, no_xlabels, ylabel={}]
\DrawTnBScatter{indoor}{10}{CoRa}
\DrawTnBScatter{indoor}{10}{OpenLoRa}
\DrawTnBScatter{indoor}{10}{TnB}
\DrawTnBScatter{indoor}{10}{CIC}
\DrawTnBScatter{indoor}{10}{Pyramid}

\nextgroupplot[bar width=5pt, xlabel={}, ylabel={}]
    \addlegendimage{thick, color=mpl_blue,
      legend image code/.code={
        \draw[thick] (0cm,-0.0cm) -- (0.3cm,-0.0cm);
      }
        }
    \addlegendentry{\our}
    \addlegendimage{thick, color=mpl_orange,
      legend image code/.code={
        \draw[thick] (0cm,-0.0cm) -- (0.3cm,-0.0cm);
      }
    }
    \addlegendentry{OpenLoRa}

    \addlegendimage{thick, color=mpl_green,
      legend image code/.code={
        \draw[thick] (0cm,-0.0cm) -- (0.3cm,-0.0cm);
      }
        }
    \addlegendentry{TnB}
    \addlegendimage{thick, color=mpl_red,
      legend image code/.code={
        \draw[thick] (0cm,-0.0cm) -- (0.3cm,-0.0cm);
      }
    }
    \addlegendentry{CIC}
    \addlegendimage{thick, color=mpl_purple,
      legend image code/.code={
        \draw[thick] (0cm,-0.0cm) -- (0.3cm,-0.0cm);
      }
    }
    \addlegendentry{Pyramid}
\DrawTnBScatter{outdoor1}{10}{CoRa}
\DrawTnBScatter{outdoor1}{10}{OpenLoRa}
\DrawTnBScatter{outdoor1}{10}{TnB}
\DrawTnBScatter{outdoor1}{10}{CIC}
\DrawTnBScatter{outdoor1}{10}{Pyramid}

\end{groupplot}
\node[anchor=south,rotate=-90,yshift=0.175cm] at (group c1r1.east) {\textbf{Indoor}};
\node[anchor=south,rotate=-90,yshift=0.175cm] at (group c1r2.east) {\textbf{Outdoor1}};
\path (group c1r1.south west) -- (group c1r2.north west) node[pos=0.5, anchor=south,rotate=90,yshift=0.675cm] {Throughput [\si{\pkt\per\second}]};
\path (group c1r2.south west) -- (group c1r2.south east) node[pos=0.5, yshift=-1cm] {Coding Rate [\#]};
\end{tikzpicture}
        \caption{Spreading Factor 10}\label{fig:tnb_sf10}
    \end{subfigure}

    \caption{Throughput versus coding rate (TnB dataset) for packets demodulated using
    detections from \our and the baseline methods. Results are shown for a high SNR
    scenario (\textit{Indoor}) and a low SNR scenario (\textit{Outdoor1}), for two
    spreading factors (8 and 10). The error bars represent 95\% confidence intervals.\vspace{-10pt}}
    
    \label{fig:comparison}
\end{figure}

On the TnB dataset, our solution matches performance with TnB -- within error bars at SF8 and high SNR conditions --  
(\autoref{fig:tnb_sf8}, \textit{Indoor}), attaining a maximum of 15
\textit{pkt/s} at coding rate 4/7. We observe higher advantage of our solution in
the SF10 scenario, in which the time on air of frames is almost four times that
of SF8. Specifically, we observe up to $22\%$ gains over TnB
(\autoref{fig:tnb_sf10}, \textit{Indoor}, coding rate 4/8), up to $2.5\times$
that of CIC (coding rate 4/8), and up to $6.65\times$ that of OpenLoRa (coding
rate 4/7). Pyramid yields very few successfully decoded packets, leading to a
large relative gain (up to $400\times$) for \our at coding rate 4/5.

This trend also holds under low SNR
conditions~(\autoref{fig:tnb_sf10}, \textit{Outdoor1}), where 
\our  at SF10 continues to provide the largest gains, specifically $1.21\times$ over
TnB (CR 4/8), $3\times$ over CIC (CR 4/7), $7.8\times$ over OpenLoRa and over
$120\times$ over Pyramid (CR 4/7). At SF8, we also observe non-negligible
gains in certain configurations, reaching up to $21\%$ over TnB
(\autoref{fig:tnb_sf8}, \textit{Outdoor1}, CR 4/8).

Overall, these results are consistent with the simulation findings, in which \our improves decoding performance
under high collision rates and low SNR conditions.

\section{Computational Complexity and Feasibility of Near-Real-Time Operation}\label{sec:complexity}

We next analyze the computational complexity of our solution and compare with
the evaluated baselines.
The \our synchronization pipeline consists
of dechirping operations ($O(N)$), full FFTs ($O(N \log(N))$, including Zoom FFT),
Goertzel-based bin extraction ($O(N)$ for a fixed
number of bins), cross-correlation ($O(N)$ with fixed template length) and other
linear-time operations, such as median filtering and peak
detection.
Overall, the dominant cost is $O(N \log(N))$, driven by the FFTs.
Prior to spectral intersection,
our method performs $N_\delta \cdot N_f$ FFTs over the preamble search grid.
In our evaluation, this reduces to four FFTs
per demodulation window (stride $M$, with $M$ the samples per symbol).
For each detected frame hypothesis, fine synchronization applies a fixed sequence
of operations including processing of the final preamble symbol, synchronization
word and downchirp, along with one high-resolution Zoom FFT for fractional STO refinement. This sums up to 
 six FFTs per hypothesis.

\begin{table}[t]
\centering
\small
\begin{tabular}{l|c|m{1.6cm}|m{1.6cm}}
    \toprule
	\textbf{Method} & \textbf{Complexity} & \textbf{FFTs per window [\#]} &  \textbf{FFTs p. hypothesis [\#]} \\
    \midrule
    \our & $O (N \log N)$ & 4 & 6 \\
    CIC & $O (N^2)$ & -- & 2048  \\
    TnB & $O (N \log N)$ & 1 & 432  \\
    Pyramid & $O (N \log N)$ & 16 & --  \\
    OpenLoRa & $O (N \log N)$ & 1 & 2048  \\
    \bottomrule
\end{tabular}
    \caption{Computational complexity and FFT requirements of the evaluated methods.
    FFTs per window and per hypothesis denote the FFT operations during
    preamble search and for each detected hypothesis, respectively.
    Values are reported for SF8 using the default settings of each method.
    }\label{tab:complexity}
\end{table}

\paragraph{Comparison with baselines}
As shown in \autoref{tab:complexity}, all methods except CIC achieve
$O(N \log N)$ complexity, due to FFT-based preamble detection and
synchronization, whereas CIC relies on correlation-based frame detection with 
$O (N^2)$ complexity per symbol. However, asymptotic complexity alone does not capture
the practical cost, which is dominated by the number of FFTs required
for each method.

TnB, CIC and OpenLoRa incur substantial costs during fine
synchronization. TnB evaluates a grid of fractional CFO/STO hypotheses (36 configurations
with oversampling 8) across twelve symbols, requiring one FFT per evaluation.
CIC and OpenLoRa test all candidate alignments around the detected position,
resulting in a cost that scales with both oversampling factor and samples
per symbol (\eg 2048 for SF8 with oversampling factor 8).
Pyramid instead applies a sliding-window search with stride $Q=16$,
leading to multiple FFT evaluations per demodulation window while
largely integrating fine synchronization into this stage.

LZn incurs a modest $4\times$ per-window FFT increase compared to OpenLoRa,
CIC, and TnB but reduces per-hypothesis overhead to just 6 FFTs --
substantially lower than the up to 2048 FFTs required by CIC and OpenLoRa and
up to 432 required by TnB -- while exceeding the overhead of Pyramid,
which requires no additional FFTs at this stage.

To match the aggregate FFT workload of Pyramid, the \our system would require
unrealistically high traffic, with multiple packet
detections within a single symbol duration. For SF8 at 125\,kHz bandwidth (approximately
2\,ms per symbol), this regime (of $\approx$ 1000 \textit{pkt/s}) lies well beyond typical LoRa deployments. Since Pyramid has already been demonstrated to
operate in real-time, these results suggest that \our has a comparable
computational feasibility under typical traffic conditions. Furthermore, the
real-time feasibility has been evaluated for a prior demonstrator~\cite{afsw-lcrr-25}.
However, we note that the presented analysis is based on operation counts and
does not constitute a direct runtime measurement on a specific hardware
platform.

\section{Discussion}\label{sec:discussion}

The presented results consistently demonstrate that \our achieves the strongest
performance under challenging operating conditions, in particular low SNR and
high collision regimes. These scenarios emphasize  
synchronization robustness, as weak or overlapping packet signatures increase
the likelihood of timing and frequency estimation errors.

Simulation results indicate that our solution yields a 5\,dB improvement in detection
sensitivity compared to the evaluated standard LoRa detector (OpenLoRa) and
approximately 10\,dB improvement to the second best collision-tolerant baseline (TnB).
In high collision scenarios, we observe that \our 
detects 54\% more packets than the second best (TnB) -- 
 a significant advantage. We further observe that this improved detection
translates into higher frame detection. For example, the reception sensitivity
of our solution is only 0.2\,dB below its detection sensitivity, enabling
reliable decoding at very low SNR. This small gap between detection and decoding thresholds
suggests that detected frames are typically synchronized with sufficient
accuracy to permit successful demodulation, highlighting the effectiveness of
our synchronization strategy.

Our findings align with the results from real-world traces, with  28\% higher frame detection than the strongest baseline under very low SNR
conditions (SF7, 2\,dBm transmission power), corresponding to a relative improvement of
$2.11\times$. In collision
scenarios, we observe that our solution attains 9\% decoding gains over the
second best (TnB) in the low SNR scenario at high transmission rate (100  \textit{pkt/s}, CIC dataset). At high spreading factors,
the advantage becomes more pronounced, reaching up to
$22$\% gains over TnB at SF10 and coding rate 4/8. Since all frames are demodulated using the same receiver pipeline,
the observed performance differences primarily reflect improvements in synchronization and frame recovery rather than
differences in demodulation quality.

The analysis presented in~\autoref{sec:complexity} shows that our solution is
compatible with real-time requirements, achieving the same asymptotic complexity
as Pyramid, but requiring fewer FFTs in practical solutions. Since Pyramid has
previously been demonstrated in real-time settings, similar
real-time operation is expected for our method.

\section{Conclusions and Outlook}\label{sec:conclusion}

We presented LZn, a synchronization method for robust LoRa frame detection
under ultra-low SNR and equal-spreading-factor collision conditions.
By combining spectral intersection, template matching and outlier detection, \our achieves
reliable synchronization in challenging environments.

Simulation results demonstrate a $10$\,dB improvement of detection sensitivity and
$1.54 \times$ more detections than the second-best collision-tolerant baseline
(TnB). On real-world captures, \our enables decoding up to $3.46 \times$ more
packets in challenging single-user conditions (SF7, 2\,dBm) and achieves
$1.22\times$ higher throughput in collision scenarios. \our never
underperforms the standard LoRa receiver implementation (OpenLoRa) across the
evaluated scenarios, achieving up to twice as many successful decodings.
OpenLoRa provides a transparent and reproducible baseline despite not capturing
all aspects of commercial LoRa transceivers.

\our is designed for near-real-time operation, trading a modest increase in preamble search cost for a substantially cheaper fine synchronization stage,
which keeps its overall FFT cost below existing collision-tolerant approaches. Since a baseline
method (Pyramid) has already been demonstrated to operate in real time, the lower
computational cost of \our suggests comparable feasibility for practical
deployments.

In future work, we will implement \our on a Software-Defined Radio platform to assess online
detection performance in long-term dense-urban deployments. This will also allow us to
explore a wider range of practical insights.

\paragraph{Artifacts} Our artifacts, \ie implementations,
simulations, and data, are publicly available at \url{https://doi.org/10.5281/zenodo.23382502}.

\bibliographystyle{IEEEtran}
\bibliography{own,rfcs,ids,bibliography}
\vspace{-40pt}
\begin{IEEEbiography}[{\includegraphics[width=1in,height=1.25in,trim={0 50px 0 0},clip,keepaspectratio]{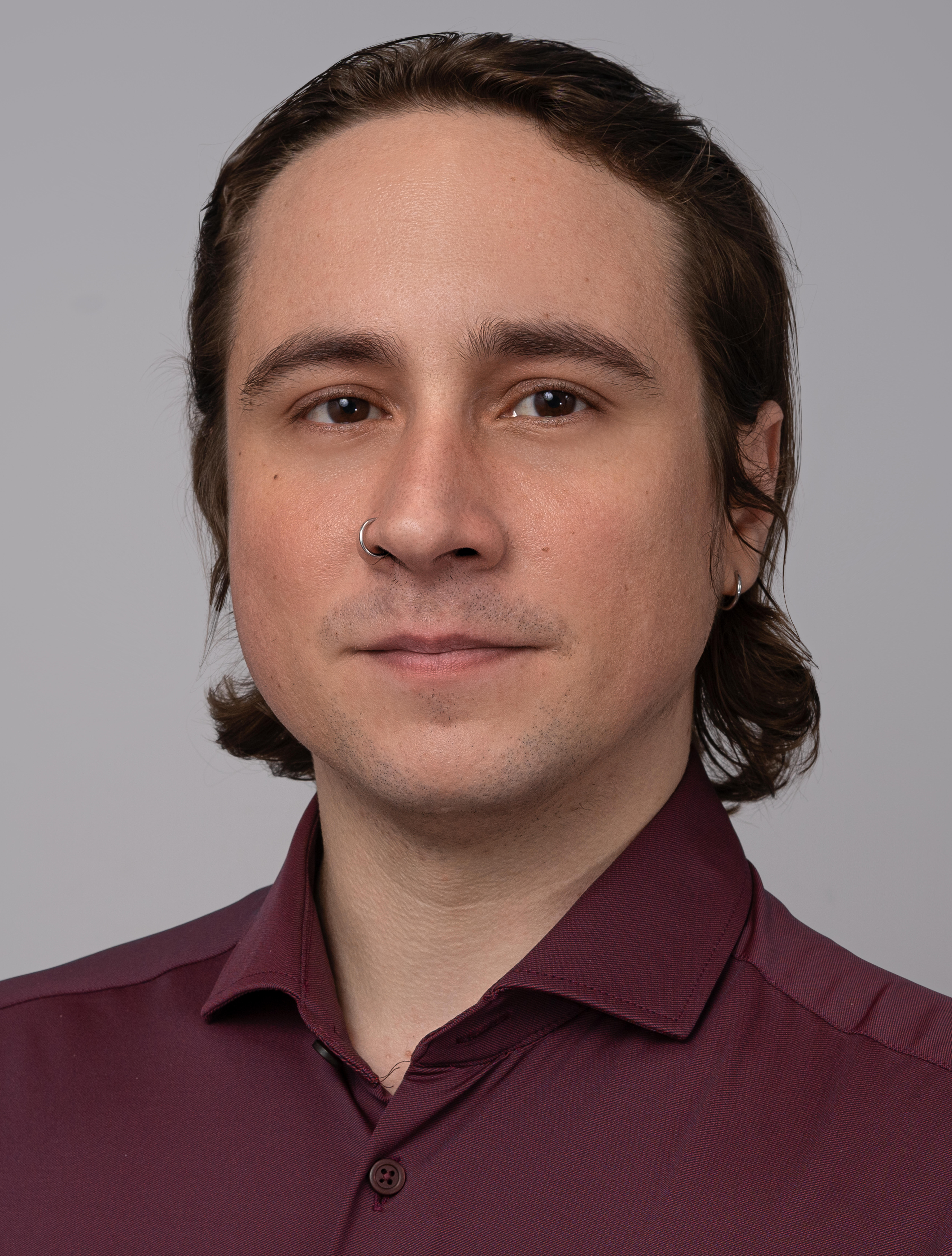}}]{Jos{\'e} {\'A}lamos}
	received the Civil Engineer of Industry and Diploma in Electrical Engineering degrees from the Pontificia Universidad Cat\'olica de
	Chile. He is a Ph.D. student in Computer Science and a Research Associate with the Hamburg University of Applied Sciences (HAW
	Hamburg). His research focuses on low-power IoT communications, particularly LoRa networking, physical-layer receiver design, and IPv6 connectivity for constrained devices. Since 2016, he has contributed to and maintained the RIOT operating system, developing MAC-layer protocols and radio abstraction layers for the constrained IoT.
\end{IEEEbiography}
\vspace{-35pt}
\begin{IEEEbiography}[{\includegraphics[width=1in,height=1.25in,clip,keepaspectratio]{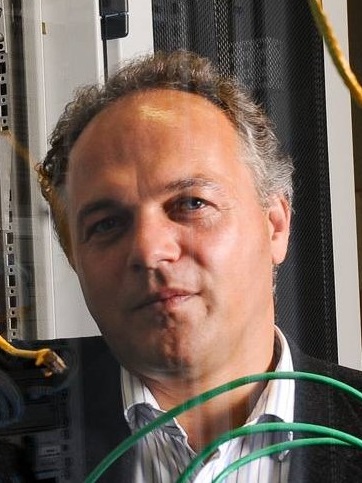}}]{Thomas C. Schmidt}
(Member, IEEE)  received the Ph.D. degree in mathematical physics from FU Berlin. He is a Professor of Computer Networks and Internet Technologies with the Hamburg University of Applied Sciences (HAW), where he heads the Internet Technologies Research Group.  His continued interests lie in the development, measurement, and analysis of the Internet.
He serves as co-editor and a technical expert in many occasions and is actively involved in the work of IETF and IRTF. Together with his group, he pioneered work on an information-centric Industrial IoT and the emerging data-centric Web of Things. He is a co-founder of several large open source projects and a coordinator of the community developing the RIOT operating system.
\end{IEEEbiography}
\vspace{-35pt}
\begin{IEEEbiography}[{\includegraphics[width=1in,height=1.25in,clip,keepaspectratio]{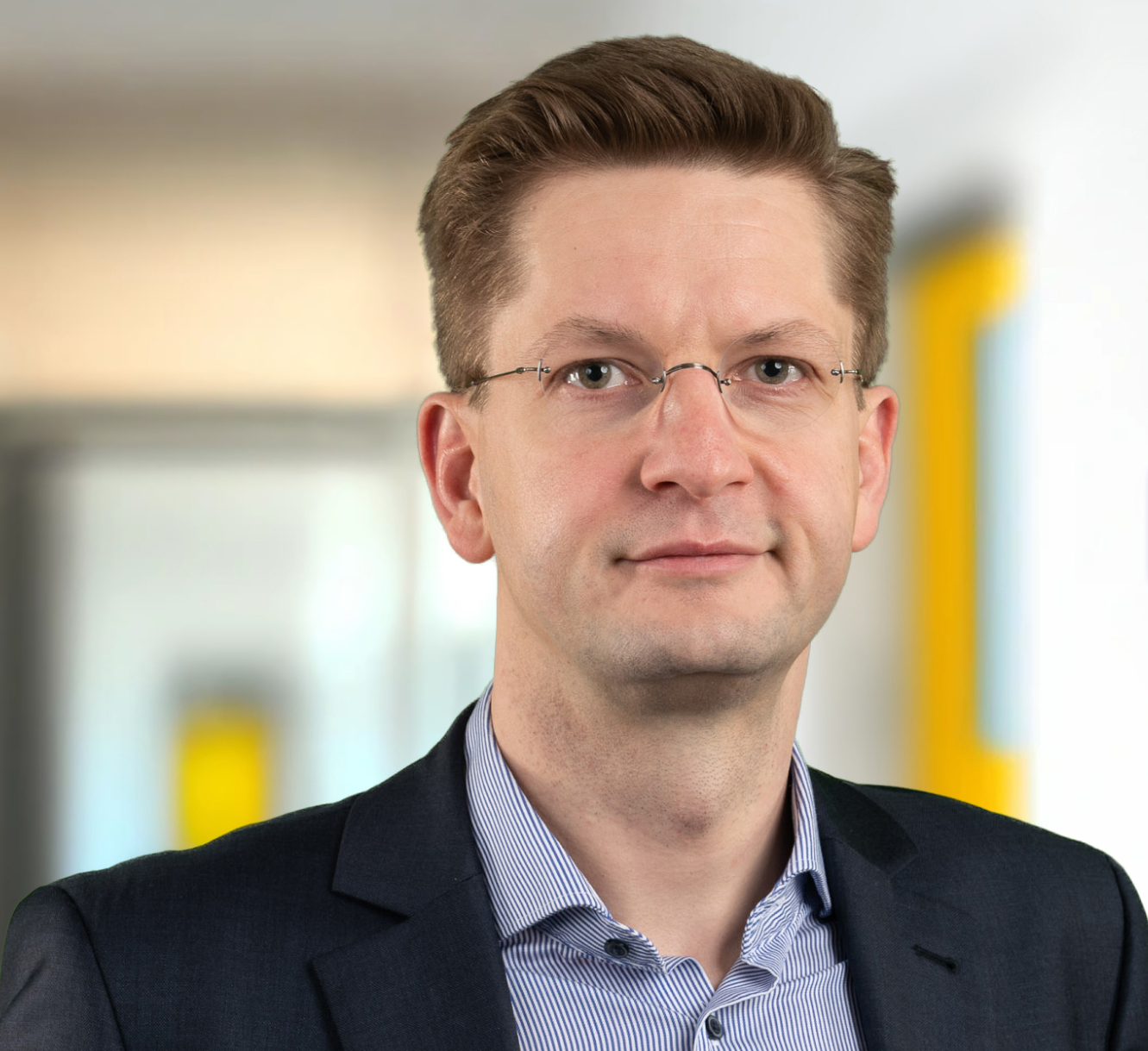}}]{Matthias W\"ahlisch}
(Member, IEEE) received the Ph.D. degree (Highest Hons.) in computer science from Freie Universität Berlin. He is a full professor and holds the Chair of Distributed and Networked Systems at the Faculty of Computer Science at TU~Dresden. He is also a Research Fellow of the Barkhausen Institut. His research and teaching focus on scalable, reliable, and secure Internet communication. This includes the design and evaluation of networking protocols and architectures, as well as Internet measurements and analysis.
 Since 2005, he has been active within IETF/IRTF. He published more than 160 peer-reviewed papers. He co-founded some successful open source projects, such as RIOT and RTRlib, where he is still responsible for the strategic development.
\end{IEEEbiography}

\end{document}